\documentclass[10pt, twoside, twocolumn]{article}

\usepackage[latin2]{inputenc}
\usepackage{natbib}
\usepackage{amsmath}
\usepackage{graphics}
\usepackage{graphicx}
\include{refek.bib}

\begin{document}

\begin{titlepage}

\vspace{10mm}
\center\Huge{Collective motion}
\vspace{20 mm}
\center\Large{Tam\'as Vicsek$^{1,2}$ \& Anna Zafeiris$^{1,3}$}
\vspace{10 mm}

\center\normalsize{$^1$ Department of Biological Physics, E\"otv\"os University - P\'azm\'any P\'eter stny. 1A, Budapest, Hungary H-1117}\\
\center\normalsize{$^2$ Statistical and Biological Physics Research Group of HAS - P\'azm\'any P\'eter stny. 1A, Budapest, Hungary H-1117}\\
\center\normalsize{$^3$ Department of Mathematics, National University of Athens - Panepistimioupolis 15784 Athens, Greece}

\vspace{15 mm}

\begin{abstract}
We review the observations and the basic laws describing the essential aspects of collective motion -- being one of the most common and spectacular manifestation of coordinated behavior. Our aim is to provide a balanced discussion of the various facets of this highly multidisciplinary field, including experiments, mathematical methods and models for simulations, so that readers with a variety of background could get both the basics and a broader, more detailed picture of the field. The observations we report on include systems consisting of units ranging from macromolecules through metallic rods and robots to groups of animals and people. Some emphasis is put on models that are simple and realistic enough to reproduce the numerous related observations and are useful for developing concepts for a better understanding of the complexity of systems consisting of many simultaneously moving entities. As such, these models allow the establishing of a few fundamental principles of flocking. In particular, it is demonstrated, that in spite of considerable differences, a number of deep analogies exist between equilibrium statistical physics systems and those made of self-propelled (in most cases living) units. In both cases only a few well defined macroscopic/collective states occur and the transitions between these states follow a similar scenario, involving discontinuity and algebraic divergences.
\end{abstract}

\end{titlepage}

\tableofcontents

\section{Introduction}\label{sec:Intro}

Most of us must have been fascinated by the eye-catching displays of collectively moving animals. Schools of fish can move in a rather orderly fashion or change direction amazingly abruptly. Under the pressure from a nearby predator the same fish can swirl like a vehemently stirred fluid. Flocks of hundreds of starlings can fly to the fields as a uniformly moving group, but then, after returning to their roosting site, produce turbulent, puzzling aerial displays. There are a huge number of further examples both from the living and the non-living world for the rich behavior in systems consisting of interacting, permanently moving units.\par

Although persistent motion is one of the conspicuous features of life, recently several physical and chemical systems have also been shown to possess interacting, ``self-propelled'' units. Examples include rods or disks of various kinds on a vibrating table \citep{BlairNKudrolliRods, PolarRodsKud, VibrPDisks, NematicsSw1, LongLivedFlucN, YamadaEtAl03, LightPowMicroMtrs}.\par

The concept of the present review is to on one hand introduce the readers to the field of flocking by discussing the most influential ``classic'' works on collective motion as well as providing an overview of the state of the art for those who consider doing research in this thriving multidisciplinary area. We have put a special stress on coherence and aimed at presenting a balanced account of the various experimental and theoretical approaches.\par

In addition to presenting the most appealing results from the quickly growing related literature we also deliver a critical discussion of the emerging picture and summarize our present understanding of flocking phenomena in the form of a systematic phenomenological description of the results obtained so far. In turn, such a description may become a good starting point for developing a unified theoretical treatment of the main laws of collective motion.\par

\subsection{The basic questions we address}

Are these observed motion patterns system specific? Such a conclusion would be quite common in biology. Or, alternatively, are there only a few typical classes which all of the collective motion patterns belong to? This would be a familiar thought for a statistical physicist dealing with systems of an enormous number of molecules in equilibrium. In fact, collective motion is one of the manifestations of a more general class of phenomena called collective behavior \citep{Vicsek01}. The studies of the latter have identified a few general laws related to how new, more complex qualitative features emerge as many simpler units are interacting. Throughout this review we consider collective motion as a phenomenon occurring in collections of similar, interacting units moving with about the same absolute velocity. In this interpretation the source of energy making the motion possible (the ways the units gain momentum) and the conditions ensuring the similar absolute velocities are not relevant.
\par

There is an amazing variety of systems made of such units bridging over many orders of magnitude in size. In addition, the nature of the entities in such systems can be purely physical, chemical as well as biological. Will they still exhibit the same motion patterns? If yes, what are these patterns and are there any underlying universal principles predicting that this has to be so (e.g., non-conservation of moments during interactions)?\par

\begin{figure}
\centerline{\includegraphics[angle=0,width=.95\columnwidth]{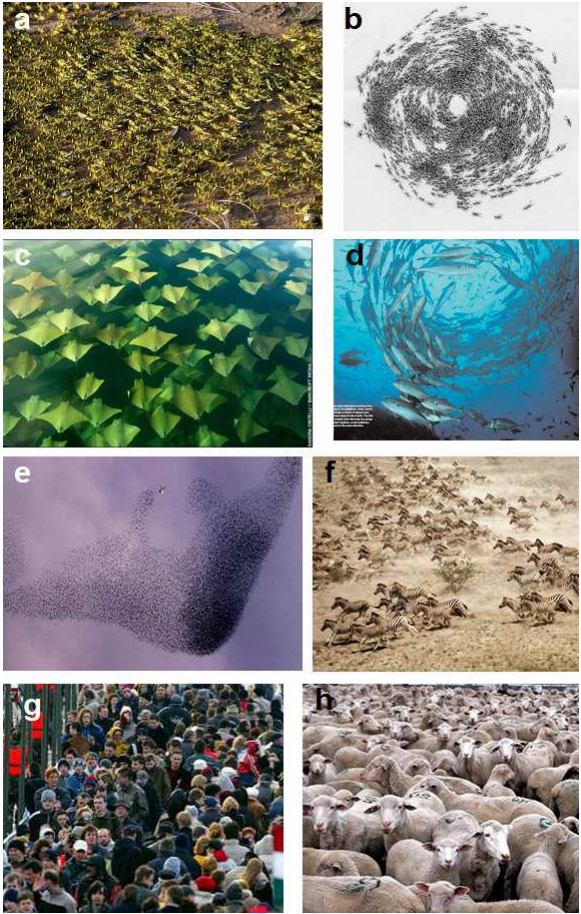}}
\caption{\label{figIntroMontazs}(Color) A gallery of images related to collective behavior. Among others, it illustrates the possible existence of very general behavioral patterns. (a) Wingless Locusts marching in the field. (b) A rotating colony of army ants. (c) A three-dimensional array of golden rays. (d) Fish are known to produce such vortices. (e) Before roosting, thousands of starlings producing a fascinating aerial display. They are also trying to avoid a predator bird close to the central,  finger-like structure. (f) A herd of zebra. (g) People spontaneously ordered into ``traffic lanes'' as they cross a pedestrian bridge in large numbers. (h) Although sheep are known to move very coherently, just as the corresponding theory predicts, when simply hanging around (no motion), well developed orientational patterns cannot emerge.}
\end{figure}

In Fig. \ref{figIntroMontazs}, we show a gallery of pictures representing a few of the many possible examples of the variety of collective motion patterns occurring in a highly diverse selection of biological systems.\par

\subsection{Collective behavior}

In a system consisting of many similar units (such as many molecules, but also, flocks of birds) the interactions between the units can be simple (attraction/repulsion) or more complex (combinations of simple interactions) and can occur between neighbors in space or in an underlying network. Under some conditions, transitions can occur during which the objects adopt a pattern of behavior almost completely determined by the collective effects due to the other units in the system. The main feature of collective behavior is that an individual unit's action is dominated by the influence of the ``others'' -- the unit behaves entirely differently from the way it would behave on its own. Such systems show interesting ordering phenomena as the units simultaneously change their behavior to a common pattern (see, e.g., \citet{VicsekNtr01}). For example, a group of feeding pigeons randomly oriented on the ground will order themselves into an orderly flying flock when leaving the scene after a big disturbance.\par

Understanding new phenomena (in our case, the transitions in systems of collectively moving units) is usually achieved by relating them to known ones: a more complex system is understood by analyzing its simpler variants. In the 1970s, there was a breakthrough in statistical physics in the form of the `renormalization group method' \citep{WilsonRenormGr75} which gave physicists a deep theoretical understanding of a general type of phase transition. The theory showed that the main features of transitions in equilibrium systems are insensitive to the details of the interactions between the objects in a system.\par

\subsection{The main difference between equilibrium and self-propelled systems}

The essential difference between collective phenomena in standard statistical physics and biology is that the ``collision rule'' is principally altering in the two kinds of systems: in the latter ones it does not preserve the momenta (the momentum of two self-propelled particles before and after their interaction is not the same), assuming that the we consider only the system made of the self-propelled particles exclusively (i.e., we do not consider the changes in the environment). Here the expression ``collision rule'' stands for specifying how the states (velocities) of two individual units change during their interaction. In particular, the momentum dissipated to the medium and within the medium itself in the realistic systems we consider cannot be neglected. In equilibrium systems, according to standard Newtonian mechanics the total momentum is preserved and that is how the well known Maxwellian velocity distribution is being built up from arbitrary initial conditions in a closed Galilean system. The mere condition of the units maintaining an approximate absolute velocity can be realized in an open system only and drives away the driven particles from any kind of equilibrium behavior.\par
Energy, sustaining self-propulsion, can be introduced into the system in various ways: uniformly in the bulk, for example in the case of or Janus particles, where the motion is ensured by a laser beam which causes self-thermophoresis \citep{JanusParticlesPRL}, at the boundaries (shaken rods), or at specific spatial points in space and time (fish).\par
Currents are bound to be generated and the overall momentum is gradually increasing if the initial state is random (in this case the initial momentum is very small because the moments of the oppositely moving particles cancel out). However, for this overall ordering to occur the random perturbations (acting against ordering) have to be small enough.\par
Most remarkably, however, in spite of this principal difference, a number of deep analogies can still be observed between equilibrium statistical physics systems and those made of self-propelled (living) units. In both only a few well defined macroscopic/collective states occur and the transitions between these states follow a similar scenario as well (discontinuity, algebraic divergences, etc).

\subsection{Goals to be achieved}

The approach of treating flocks, or even crowds, as systems of particles naturally leads to the idea of applying the successful methods of statistical physics, such as computer simulations or theories on scaling, to the detailed description of the collective behavior of organisms. Naturally, for better progress, observations/experiments and modeling have to be intimately related. Indeed, over the past few decades, an increasingly growing number of significant attempts have been made to both observe and describe flocking as well as modeling (simulation) the most conspicuous features of the observed natural systems ranging from molecules to groups of mammals.\par

It would be quite an achievement if we could establish a systematic chart of the types of collective motion, since many times understanding is achieved through classification. There are reasons and arguments for thinking that the same patterns of collective motion apply to the collection of molecules up to groups of humans. There must be some -- still to be discovered -- laws of such systems from which the above observation follows.


\section{Basics of the statistical mechanics of flocking}\label{subsec:ExperimDefsTechs}

Throughout this overview the notion of flocking is used as a synonym of any kind of \emph{coherent motion} of individual units. However, the notion of coherent motion needs some further elaboration since, as it turns out, it can be manifested in a number of specific ways. In any case, coherent or ordered motion is assumed to be a counterpart of disordered, random motion. In the various models of flocking it emerges through a kind of transition (from disorder to order) as a function of the relevant parameter(s) of the models. To demonstrate more clearly this aspect of collective motion the best approach is to adopt a few related definitions motivated mainly by statistical mechanics or statistical physics (the physics of many interacting molecules).\par

However, before we turn to the discussion of the statistical mechanics aspects of collective motion we give a somewhat more detailed list of features characterizing flocking in general. Thus we assume that a system exhibiting collective motion is made of units
\begin{itemize}
\item that are rather similar
\item moving with a nearly constant absolute velocity and are capable of changing their direction
\item interacting within a specific interaction range by changing their direction of motion, in a way involving an effective alignment
\item which are subject to a noise of a varying magnitude
\end{itemize}

\subsection{Principles and concepts}\label{sec:BasicDefs}

In a general sense, \textit{phase transition} is a process, during which a system, consisting of a huge number of interacting particles, undergoes a transition from one \emph{phase} to another, as a function of one or more external parameters. The most familiar \emph{phases} in which a physical system can be, are the solid, liquid and gaseous phases, and the best known example for a phase transition is the freezing of a fluid when the temperature drops. In this case the temperature is the external or ``control'' parameter.\par

Phase transitions are defined by the change of one or more specific system variables, called \emph{order parameters}. This name, order parameter, comes from the observation that phase transitions usually involve an abrupt change in a \emph{symmetry property} of the system. For example, in the solid state of matter, the atoms have a well-defined average position on the sites of an ordered crystal lattice, whereas positions in the liquid and gaseous phases are disordered and random. Accordingly, the order parameter refers to the degree of symmetry that characterizes a phase. Mathematically, this value is usually zero in one phase (in the \emph{disordered phase}) and non-zero in the other (which is the \emph{ordered phase}). In the case of collective motion the most naturally (but not necessarily) chosen \emph{order parameter} is the average normalized velocity $\varphi$,

\begin{equation}
\varphi=\frac{1}{N v_0}\left| \sum_{i=1}^N \vec{v}_i  \right|,
\label{eq:NormV}
\end{equation}
where $N$ is the total number of the units and $v_0$ is the average absolute velocity of the units in the system. If the motion is disordered, the velocities of the individual units point in random directions and average out to give a small magnitude vector, whereas for ordered motion the velocities all add up to a vector of absolute velocity close to $N v_0$ (thus the order parameter for large $N$ can vary from about zero to about 1).\par

If the order parameter changes discontinuously during the phase transition, we talk about a \emph{first order} transition. For example, water's volume changes like this (discontinuously) when it freezes to ice. In contrast, during \emph{second order} (or \emph{continuous}) phase transition the order parameter changes continuously, while its derivative, with respect to the control parameter, is discontinuous. Second order phase transitions are always accompanied by large fluctuations of some of the relevant quantities.\par

During a phase transition a spontaneous symmetry breaking takes place, i.e., the symmetry of the system changes as we pass the critical value of the control parameter (e.g., temperature, pressure, etc.). In the case of one of the simplest representations of a continuous phase transition (the Ising model), the spins are placed on a lattice, interact with their closest neighbors and can point up or down. For high temperatures the spins point in random directions (with a probability 1/2 either up or down), while for low temperatures (way below the critical point) most of them point in the same direction (which is either up or down), selected spontaneously. This is an example for the up and down symmetry (high temperatures) breaking during the transition (one of the directions becomes dominant). If there are infinitely many possible directions (continuous symmetry), during a transition a single preferred direction can still emerge spontaneously.\par

Phase transitions can occur in both equilibrium and non-equilibrium systems. In the context of collective motion -- although it is a truly non-equilibrium phenomenon -- there are reasons for the preference of the possible analogies with the equilibrium phase transitions rather than with those studied in the framework common in the interpretation of non-equilibrium phenomena. One important reason is that the investigations on the various universal behaviors that can be associated with
non-equilibrium systems have by now grown into a sub-discipline with its own language and formalism including specific processes the related research concentrates on.\par

In particular, most of the representative reviews on non-equilibrium phase transitions (see, e.g., \citet{OdorRMP} or \citet{HenkelKonyv}, Non-Equilibrium Phase Transitions: Volume 1: Absorbing Phase Transitions) are centered on such features as i) absorbing states of reaction-diffusion-type systems, ii) mapping to the universal behavior of interface growth models, iii) dynamical scaling in far-from-equilibrium relaxation behavior and ageing, iv) extension to non-equilibrium systems by using directed percolation as the main paradigm of absorbing phase transitions. On the other hand, there are some specific features of collective motion such as giant number fluctuations (GNF, see below) or various unusual transitions (e.g., to jamming) which obviously do not occur in systems at equilibrium, but we still have to consider them in this review. On balance, we find that the language and the spectrum of the various aspects of equilibrium phase transitions is surprisingly suitable for interpreting most of the observed phenomena in the context of collective motion; and this is an important point we intend to make.\par

A further important aspect of the phenomena taking place during collective motion is that in contrast with the standard assumptions of statistical mechanics, the number of units involved in the collective behavior typically ranges from a few dozens to a few thousands (in rare cases tens of thousands). On the other hand, for example, phase transitions in the framework of statistical mechanics are truly meaningful only for very large system sizes (consisting a number of units approaching infinity). Quantities like critical exponents cannot be properly interpreted for flocks of even moderate sizes. However, a simple transition from a disordered to an ordered state can take place even in cases when the number of units is in the range of a few dozens. Most of the real-life observations and the experiments involve this so called ``mesoscopic scale''. The states (e.g. rotation) and the transitions (e.g., from random to ordered motion) we later describe can be associated with the phenomena taking place in such mesoscopic systems.

\subsection{Definitions and expressions}

As for the definitions used in the statistical mechanics approach, phenomena associated with a continuous phase transition are often referred to as \textit{critical phenomena} because of their connection to a \textit{critical point} at which the phase transition occurs. (``Critical'', because here the system is extremely sensitive to small changes or perturbations.) Near to the critical point, the behavior of the quantities describing the system (e.g., pressure, density, heat capacity, etc.) are characterized by the so called \textit{critical exponents}. For example the (isothermal) compressibility $\kappa_T$ of a  liquid substance, near to its critical point, can be expressed by

\begin{equation}
\kappa_T \sim \left| T-T_c\right|^{-\gamma},
\label{eq:CrExpKT}
\end{equation}
where $T$ is the temperature, $T_c$ is the critical temperature (at which the phase transition occurs), $\sim$ denotes proportionality  and $\gamma$ is the critical exponent. In systems of self-propelled particles, noise ($\eta$) plays the role of temperature ($T$): an external parameter that endeavors to destroy order. Correspondingly, the fluctuation of the order parameter, $\sigma^2 = \left\langle \varphi^2 \right\rangle - \left\langle \varphi \right\rangle^2$, is described as
\begin{equation}
    \sigma \sim \left| 1- \eta/\eta_c \right|^{-\gamma},
\label{eq:CrExpSgm}
\end{equation}
where $\eta$ is the noise, $\eta_c$ is the critical noise that separates the ordered and disordered phases, and $\gamma$ is again the `susceptibility' critical exponent. By introducing $\chi=\sigma^2 L^2$, where $L$ is the linear size of the system, we get
\begin{equation}
\chi \sim (\eta-\eta_c)^{-\gamma}
\label{eq:ChiDlt}
\end{equation}
$\kappa_T$ in Eq. (\ref{eq:CrExpKT}) corresponds to $\chi$ in Eq. (\ref{eq:ChiDlt}).\par

An other descriptive expressing the change in the density between the liquid and the gas phases, $\rho_l - \rho_g$, obeys to

\begin{equation}
\rho_l - \rho_g \sim \left( T_c - T\right)^{\beta},
\label{eq:CrExpRhoBeta}
\end{equation}
where $\beta$ is the critical exponent. For systems of self-propelled particles, when $L \rightarrow \infty$, the corresponding equation is
\begin{equation}
\varphi \sim \left\{
    \begin{array}{ll}
        \left(  1- \eta/\eta_c \right)^{\beta}  & \mbox{for $\eta < \eta_c$}\\
        0 & \mbox{for $\eta > \eta_c$}
    \end{array}
\right.
\label{eq:CrSPPBeta}
\end{equation}

Regarding the relation between the order parameter $\varphi$ and the external bias field $h$ (``wind''), $\varphi$ scales as a function of $h$ according to the power law 
\begin{equation}
\varphi \sim h^{1/{\delta}}
\label{eq:CrExpDlt}
\end{equation}

for $\eta > \eta_c$, where $\delta$ is the relevant critical exponent.\par

Various similar expressions can be formulated involving further quantities as well as critical exponents. Interestingly, very different physical systems exhibiting seemingly different kind of phase transitions follow similar laws. For example the magnetization $M$ of a ferromagnetic material subject to a phase transition near to a critical temperature called Curie point, obeys $M \sim \left( T_c - T\right)^{\beta}$.\par

Another surprising observation is that these critical exponents are related to each other, that is, expressions like $\alpha+2\beta+\gamma=2$ or $\delta = 1+\frac{\gamma}{\beta}$ can be formulated, which hold independently of the physical system the critical exponents ($\alpha$, $\beta$, $\gamma$, $\delta$) belong to. Note that this is a far from trivial observation! For more details on this topic see \citep{IsiharaStatFiz, StatFizPath, StatFizCardy}, and for further analogies and differences between ferromagnetic models and systems of self-propelled particles see \citep{Csv97}.\par 

Since from a mathematical point of view, many of the results of statistical mechanics are exact only for infinitely large systems, structures are often described and analyzed in their \emph{``thermodynamic limits''}. This limit means that the number of particles constituting the system tends to infinity. Accordingly, finite structures do not have well defined phases but only in their thermodynamic limit, where the state-equations can develop singularities, and sharp phase transitions can exist between these well-defined phases.\par 
Another phenomenon accompanying phase transitions is the formation of clusters of units behaving (e.g., being directed or moving) in the same way. Units which can be reached through neighboring units belong to the same cluster, where neighboring stands for a predefined proximity criterion. Thus, the behavior of units in the same cluster is usually highly correlated. In general, \emph{correlation functions} represent a very useful tool to characterize the level of order in a system.

\subsection{Correlation functions}

Generally speaking, two series of data ($X$ and $Y$) are \textit{correlated} if there is some kind of relationship between their elements. A \textit{correlation function} measures the similarity between the data-sequences, or, in the continuous case, the similarity between two signals or functions. \textit{Auto}-correlation is the correlation of a signal with itself, typically as a function of time. This is often used to reveal repeating patterns, such as the presence of a periodic signal covered by noise. If the two signals compared are different, we consider \textit{cross}-correlation.\par

For example, let us consider two real-valued data-series $f_1$ and $f_2$, which differ only by a shift in the element-numbers, e.g., the 5th element in the first series is the same than the 12th in the second, the 6th corresponds to the 13th, etc. In this case the shift is $s=7$, that is, the first series has to be shifted with $7$ elements in order to be congruous with the second one. The corresponding (cross) correlation function will show a maximum at $7$. Formally, for discrete data-sequences $f_1$ and $f_2$, the correlation function is defined as:

\begin{equation}
c(s) = \sum_{n=-\infty}^{\infty} f_1^* \left[n\right] f_2 \left[n+s\right ],
\label{eq:DscrtCorrFuncDef}
\end{equation}

where `$*$' refers to the complex conjugate operation\footnote{A pair of complex numbers are said to be \emph{complex conjugates} if their real part is the same, but their imaginary parts are of opposite signs. For example, $2+3i$ and $2-3i$ are complex conjugates. It also follows, that the complex conjugate of a real number is itself.}.\par

Accordingly, in continuous case, when $f_1$ and $f_2$ are continuous functions (or ``signals''), the cross-correlation function will reveal how much one of the functions must be shifted (along the horizontal axis) to become congruous with the other one. Formally, 

\begin{equation}
c(\tau) = \int_{\tau=-\infty}^{\infty} f_1^*(t) f_2(t+\tau)dt
\label{eq:CntCorrFuncDef}
\end{equation}

Equation (\ref{eq:CntCorrFuncDef}) shifts $f_2$ along the horizontal axis (which is in this example the time-axis), and calculates the product at each time-step of the two functions. This value is maximal when $f_1$ and $f_2$ are congruous, because when lumps (positives areas) are aligned, they contribute to making the integral larger, and similarly, when the troughs (negative areas) align, they also make a positive contribution to the expression, since the product of two negative values is positive.\par

With this introduction we can now formulate some specific correlation functions that are often used in the field of collective motion.\par The \emph{velocity-velocity correlation function}, $c_{vv}$, is an auto-correlation function that shows how closely the velocity of a particle (unit, individual, etc.) at time $t$ is correlated with the velocity at a reference time. It is defined as follows:

\begin{equation}
c_{vv}(t) = \frac{1}{N}\sum_{i=1}^{N}\frac{\left \langle \vec{v}_i(t)\cdot \vec{v}_i(0) \right \rangle}{\left \langle \vec{v}_i(0) \cdot \vec{v}_i(0) \right \rangle},
\label{eq:Cvv}
\end{equation}
where $\vec{v}_i(0)$ is the starting (reference) time and $N$ is the number of particles within the system and $\left\langle \ldots \right\rangle$ denotes taking an average over a set of starting times. The way $c_{vv}(t)$ decays to zero shows how the velocities at later times become independent of the initial ones.\par

\begin{figure}
  \centerline{\includegraphics[angle=0,width=1\columnwidth]{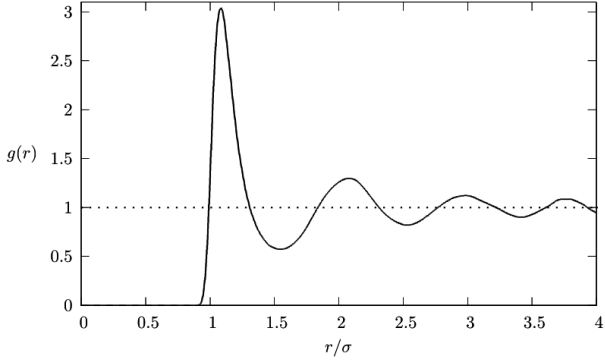}}
  \caption{\label{figpcf}Radial distribution function for the so called Lennard-Jones model. It describes how the unit density varies as a function of the distance from one particular element in the case of an interaction potential having both a shorter range, a strong repulsive and a longer range attractive part.}
\end{figure}

The \emph{pair correlation function}, $c_p(r)$, (or radial distribution function, $g(r)$), depicted on Fig. \ref{figpcf}, describes how the unit density varies as a function of the distance from one particular element. More precisely, if there is a unit at the origin, and if $n = N/V$ is the average number density ($N$ is the number of units in a system with volume $V$), then the local density at distance $r$ from the origin is $n g(r)$. It can be interpreted as a measure of local spatial ordering. Equation \ref{eq:pcf} gives the exact formula.
\begin{equation}
c_p(r)=\frac{V}{4 \pi r^2 N^2}\left < \sum_i \sum_{j \neq i} \delta(r-r_{ij})\right >
\label{eq:pcf}
\end{equation}

\begin{figure}
   \centerline{\includegraphics[angle=0,width=0.62\columnwidth]{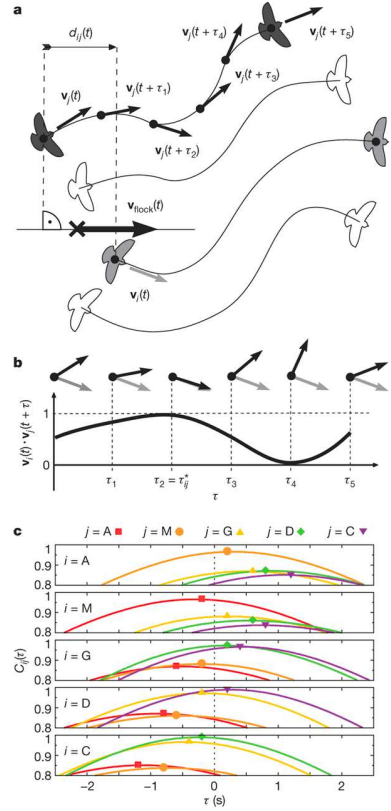}}
   \caption{\label{figDirCorrFunc}(Color online) Illustration for the directional correlation function, which is a tool for determining the
   leader-follower relationships within a flock or swarm. This example shows the case of a bird-flock. (a) Determining the projected distance $d_{ij}(t)$ of birds $i$ and $j$ onto the direction of motion of the whole flock at time step $t$. Bird $i$ is light gray on the draw, and bird $j$ is dark. The $\vec{v_j}(t)$ arrows indicate the direction of bird $j$ at each time step $t$. The center of mass of the flock is denoted with a cross, which moves with $\vec{v_{flock}}(t)$, the average velocity of the flock. The relative position of the birds $i$ and $j$ is projected onto $\vec{v_{flock}}(t)$. The directional correlation function for each $i \neq j$ pair is defined by Eq. (\ref{eq:DrctnCorr}). (b) Scalar products $\vec{v_i}(t) \cdot \vec{v_j}(t+\tau)$ of the normalized velocity vectors of bird $i$ at time $t$, and that of bird $j$ at time $t+\tau$. On this example bird $j$ follows bird $i$ with correlation time $\tau_{ij}^*$. (c) The directional correlation functions $c_{ij}(\tau)$ during the flock flight depicted on Fig. \ref{figBirdHrcy}. For better transparency only five of the items are shown, the data belonging to birds A, M, G, D and C. The solid symbols indicate the maximum value of the correlation functions, which have been used to determine the leader-follower network, depicted on \ref{figBirdHrcy} b. From \citet{HrarchGrDyn}.}
\end{figure}


The \emph{directional correlation function},
\begin{equation}
    c_{ij}(\tau)=\left\langle \vec{v_i}(t) \cdot \vec{v_j}(t+\tau) \right\rangle,
\label{eq:DrctnCorr}
\end{equation}
tells to what degree the velocity of the $i$th particle at time $t$ is correlated with that of particle $j$ at time $t+\tau$. $\left\langle \ldots \right\rangle $ denotes averaging over time, and $\vec{v_i}(t)$ is the normalized velocity of the $i$th SPP. (Note that $c_{ij}(\tau)=c_{ji}(-\tau)$.) The directional correlation \emph{delay} is primarily used to determine the leader-follower relationship within a flock of birds, fish, or more general, in a swarm of self-propelled particles (SPPs), as illustrated on Fig. \ref{figDirCorrFunc}  \citep{HrarchGrDyn}. The directional correlation delay \emph{for a pair} of SPPs ($i$ and $j$, where $i,j=1, 2, \ldots, N$ and $N$ is the number of SPPs within the flock) is calculated according to Eq. \ref {eq:DrctnCorr}. Then the maximum value of the $c_{ij}(\tau)$ correlation function is allocated, $\tau_{ij}^*$. Negative value means that the direction of motion of the $i$th SPP is falling behind that of the $j$th one, so this can be interpreted as $j$ is leading. The directional correlation function for an SPP with respect to the rest of a given flock or swarm, is $ c_{i}(\tau) = \left\langle \vec{v_i}(t) \cdot \vec{v_j}(t+\tau) \right\rangle_{i,j}$.\par 
Further useful correlation functions were suggested in \citet{Cavagna10} in order to characterize the response of a flock to external perturbation.\par

We end this Section with a brief discussion of two characteristic phenomena which occur in systems of self-propelled particles but have no direct analogy in processes occurring in equilibrium. Firstly, we point out that in a confined geometry any group of persistently moving units having a finite size have a chance to be trapped or jammed. The presence of walls or even a too narrow ``exit'' would inevitably lead to this phenomenon (see, e.g., \citep{PolarRodsKud, Panic}).\par

Another interesting feature is a very specific form of density changes called as giant number fluctuations or GNF.This expression stands for the following property of a system of self-propelled units: The fluctuation of the number of units in an increasing area of the system scale with the number of units (N) in this area linearly. In addition, these fluctuations relax anomalously slowly. This is in sharp contrast with a theoretical result valid for equilibrium systems according to which (along with the law of large numbers) the fluctuations in the number of units grow as a square root of $N$. This interesting feature was first pointed out in \citet{LongLivedFlucN} for a system of shaken elongated rods and subsequently commented upon by \citet{AransonEtAlSci} showing that inelastic collisions can lead to GNF even for spherical particles.\par

Since shaking introduces an average velocity and nematic or inelastic collisions involve a tendency for the particles to align, all these findings are expected to occur (and will be later discussed) in a variety of systems with collective motion.

\section{Observations and experiments}\label{sec:Experim}

It seems that collective motion (or flocking: these two notions will be used synonymously in this review although in principle there are some subtle differences in their meanings) is displayed by almost every living system consisting of at least dozens of units. \footnote{In such small systems, we associate with ``flocking'' a state of the group in which the units assume an approximately common direction (or orientation) developing through local communications among the entities.} One of the main points in this review is that the kinds of systems and the types of collective motion patterns have a greater variety than originally thought of. Below we give a -- naturally incomplete -- list of systems in which collective motion has been observed (with only some of the representative references included):
\begin{itemize}
	\item Non-living systems: nematic fluids, shaken metallic rods, nano swimmers, simple robots, boats, etc. \citep{KudrlDiff, LightPowMicroMtrs, SuematsuBoats, LongLivedFlucN}
	\item Macromolecules \citep{Filaments, ButtEtAl10}
	\item Bacteria colonies \citep{CmplxBactCln96, CzirMatVics, SokolovAKG, BacterialFluidMech}
	\item Amoeba \citep{KesslerSlimeMold, NaganoSlimeMold98, RotatingSlimeMold}
	\item Cells \citep{SzaboEtAl, FriedlGilm09, ArbCell2010, ChateSellSortingt}
	\item Insects \citep{CouzinScience, AntLaneCF}
	\item Fish \citep{HemelrijkK05, PVG02, BeccoFishExperim, QuorumSensing}
	\item Birds \citep{HeppnerPrsBook, BalleriniEtAl, JapiGeese, BHeppnerCikk}
	\item Mammals \citep{FischhoffZebra, MakakoSueur08, BaboonKing08}
    \item Humans \citep{FariaHumanLdr10, Pedestrian, Panic, MoussaidCrowd2011}
\end{itemize}

Although throughout the present review we primarily classified the observations and experiments based on the organizational complexity of the units constituting the systems, there are various other aspects as well, by which valid and meaningful categorizations can be made, such as:
\begin{itemize}
\item The patterns the units form (coherently moving clusters, mills, stripes, etc)
\item From an energetic viewpoint: 
\begin{itemize}
\item The way it is introduced to the system (uniformly, at the boundaries, at special points, etc.)
\item if the particles can preserve it (animals) or they dissipate it almost immediately (shaken rods)
\end{itemize}
\item The way the units interact with each other. This can be
\begin{itemize}
\item physical, chemical (chemotaxis), visual or medium-mediated
\item isotropic or anisotropic
\item polar or apolar
\item short or long-range (regarding its temporal characteristics)
\item through metric or topological distance
\end{itemize}
\end{itemize}

\subsection{Data collection techniques}

The main challenge during the implementation of observations and/or experiments on collective motion, is to keep a record on the individual trajectories of the group-members. This can easily turn out to be a difficult task, since (i) most of the colonies or groups being investigated consist of many members which (ii) usually look very much alike (iii) and are often moving fast -- just think of a flock of starlings. The applied technology is chosen by considering two factors: (i) the size of the moving units, and (ii) the size and dimension of the space in which the group can move. Both parameters range through many scales: bacteria and cells are subjects of such experiments as well as African buffalos or whales. Also, the area in which the observation is carried out can range from a Petri dish \citep{KellerSegelM} to the Georges Bank (a region separating the Gulf of Maine from the Atlantic Ocean, \citep{FishDensM}). Accordingly, there is a variety of technologies that have been applied during the last decades.\par

The method called ``Particle Image Velocimetry'', (PIV) \citep{PrtclImgVel} is used to visualize the motion of small particles moving in a well-confined area. 
Originally it is an optical technique used to produce the two dimensional instantaneous velocity vector field of fluids, by seeding the media with `tracer particles'. These particles are assumed to follow the flow dynamics accurately, and it is their motion that is then used to calculate velocity information. \citet{BacterialFluidMech} applied this method in order to evaluate the velocity filed of thousands of swimming bacteria. The task of tracing cells share many similarities with the challenge of tracing the motion of bacteria. \citet{CzirokCellMicroscopy} used computer controlled phase contrast video microscope system in order to follow the collective motion of cells. The trajectories were recorded for several days to determine the velocities of each cell of the types.\par
The movement of vertebrate flocks has been tracked mainly by camera-based techniques. Here, the observed animals are bigger, but the space in which they move is often unconfined. The simpler case is when the group to be observed moves only in two dimensions. In the '70-es \citet{SinclairBook} used aerial photos to investigate the individual's spatial positions within grazing African buffalo herds. Exactly because of the difficulties of analyzing three dimensional group motions, 
\citet{BeccoFishExperim} confined the motion of fish to two dimensions by putting them into a container which was ``basically'' two dimensional (in the sense that it was very shallow): $40 cm$ X $30 cm$ X $2 cm$. This arrangement was convenient to track fish with a single video recorder. A homogeneous light source was placed above the container and a CCD camera recorded the fish from below. (Obviously, the usage of the container confined the area of motion as well.)\par
In order to reconstruct the three-dimensional positions and orientations of the fish, \citet{Cullen65} used the so called ``shadow method''. With this technology, \citet{PartridgeEA80} investigated positioning behavior in fish groups of up to 30 individuals. \citet{ParrTurch} recorded the trajectories of fish in three dimensions with three orthogonally positioned video cameras.\par
\citet{MajorDill78} were the first to apply the stereo photography technique in order to record the three dimensional positions of birds within flocks of European starlings and dunlins. By using the same technique, recently \citet{BalleriniEtAl} reconstructed the three-dimensional positions of hundreds of starlings in airborne flocks with high precision. The stereo photography method allowed the detailed and accurate analysis of nearest neighbor distances in large flocks, but still did not make the trajectory reconstruction of the individual flock members possible.\par
As technology advances, newer and newer methods and ideas show up with the purpose of studying collective motion. Here we mention two of these: The recently developed OWARIS (``Acoustic Waveguide Remote Sensing'') exploits the wave propagation properties of the ocean environment \citep{FishObs}, and makes the instantaneous imaging and continuous monitoring of fish populations possible, covering thousands of square kilometers, that is, an area tens of thousands to millions of times greater than that of conventional methods. The other new method is based on the usage of the ``Global Positioning System'', commonly known as GPS. The idea is to put small GPS devices on moving animals by which the problem of trajectory-recording is basically solved. With this method the trajectory of flock members can be collected with high temporal resolution in their natural environment. The limits of this technique at this moment are on the one hand the growing cost of the research with the growing number of tracked flock members, and on the other hand the limited accuracy of the devices. \citet{BiroEtAl06} and \citet{HrarchGrDyn} analyzed GPS logged flight tracks of homing pigeon pairs in order to investigate hierarchical leadership relations inside the group and  \citet{DellsLipp} used this method to study the homing efficiency of a pigeon group consist of 6 birds.

\subsection{Physical, chemical and biomolecular systems}\label{subsec:ExperimPhysChem}

Along with the accumulating observations and experiments clarified the recognition, that flocking -- collective motion -- emerges not only in systems consisting of living beings, but also among interacting physical objects, based on mere physical interactions without communication. A very simple system has been described by \citet{LightPowMicroMtrs} who reported about simple autonomous micromotors, which are micrometer-sized silver chloride (AgCl) particles exhibiting collective motion in deionized water under UV illumination. The autonomous motion these particles exhibit under the above circumstances (deionized water and UV light) is due to their asymmetric photo-decomposition, and the spatial self-organization is due to the ions which are secreted by the AgCl particles as they move.\par

\begin{figure}
\centerline{\includegraphics[angle=0,width=0.58\columnwidth]{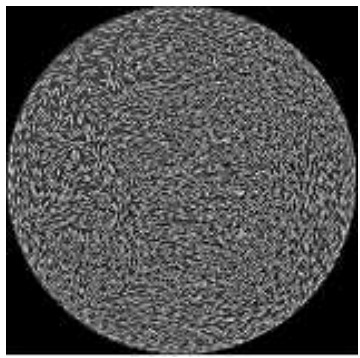}}
\caption{\label{figGrFlucN}A snapshot of the nematic order assumed by 2820 rods which are sinusoidally vibrated perpendicularly to the plane of the image. The large density fluctuations take several minutes to relax and to form elsewhere. From \citet{LongLivedFlucN}.}
\end{figure}

Various experiments on non-living self propelled particles (SPPs) possessing diverse features advocate that the shape and symmetry of the SPPs play an important role in their collective dynamics, and that large-scale inhomogeneity and coherent motion can appear in a system in which particles do not communicate except by contact.  Symmetric (or ``apolar'') rods on vibrating surfaces have been observed to form nematic order and under certain conditions found to exhibit persistent swirling as well \citep{NematicsSw1, GalanisSw2}. \citet{LongLivedFlucN} have also investigated symmetric macroscopic rods and have found giant number fluctuations lasting long, decaying only as a logarithmic function of time (see Fig. \ref{figGrFlucN}). This finding is in obtrusive contrast with the expected behaviour, since the the central limit theorem predicts number fluctuations proportional to the \emph{square root} of the particle number, in the homogeneous ordered phase of equilibrium systems,  away from the transition. \label{GNFAronson} By conducting complementary experiments with spherical particles, \citet{AransonEtAlSci} demonstrated that the giant number fluctuation phenomenon can arise either from dynamic inelastic clustering or from persistent density inhomogenity as well.

\begin{figure}
\centerline{\includegraphics[angle=0,width=0.5\columnwidth]{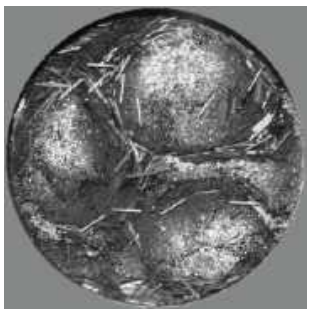}}
\caption{\label{figBNKR1} Periodically, vertically vibrated granular rods spontaneously form vortices which grow with time. From \citet{BlairNKudrolliRods}}
\end{figure}

\begin{figure}
\centerline{\includegraphics[angle=0,width=0.45\columnwidth]{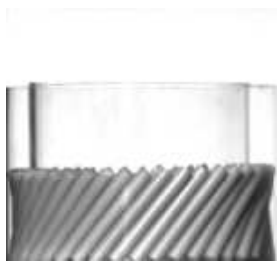}}
\caption{\label{figBNKR2}Experiments performed in an annulus with a single row of rods reveal that the rod motion is generated when these objects are inclined from the vertical, and is always in the direction of the inclination. From \citet{BlairNKudrolliRods}}
\end{figure}

Periodically, vertically vibrated granular rods form vortex patterns \citep{BlairNKudrolliRods}. Above a critical packing fraction, the ordered domains  -- consisting of nearly vertical rods -- spontaneously form and coexist with horizontal rods (see Fig. \ref{figBNKR1}). The vortices nucleate and grow as a function of time. Experiments performed in an annulus with a single row of rods revealed that the rod motion was generated when these objects were inclined from the vertical, and was always in the direction of the inclination (see Fig. \ref{figBNKR2}). The relationship between the covered area fraction and the diffusion properties in the case of self-propelled rods was also studied \citep{KudrlDiff}. 

\citet{PolarRodsKud} have made experiments with polar (non-symmetric) rods on a vibrating surface. Their rods had a symmetric shape, but a non-symmetrical mass distribution which caused them to move toward their lighter end. They have observed local ordering, aggregation at the side walls, and clustering behavior. 
Apolar, but \emph{round-shaped} disks have also been studied \citep{VibrPDisks}. By shaking a monolayer of these small objects with variable amplitude, large-scale collective motion and giant number fluctuations could be observed.\par

Another experiment studying inanimate objects used radio-controlled \emph{boats} moving in an annular pool, interacting through inelastic collisions only \citep{TarcaiEtAlHajos2011}. The team recorded various kinds of patterns, such as jamming, clustering, disordered and ordered motion, depending on the noise level. They also found that a few steerable boats -- acting as \emph{leaders} -- were able to determine the direction of the group. For this end, it was enough to manipulate $5-10$ \% of the boats.
In a somewhat similar experiments, the collective motion of \emph{camphor boats} were studied, interacting through the chemical field, swimming in an annular water channel. Here too, several patterns were observed, such as homogeneous state, cluster flow and congestion flow \citep{SuematsuBoats}.\par

\citet{TinsleyWaves} presented an experimental study on interacting particle-like \emph{waves} (see more about the design of wave propagation in \citet{IdWaves}) and suggested this method as an opportunity to investigate small groups of SPPs in laboratory environment. The stabilized wave-segments they have used were those appearing in the light-sensitive Belousov-Zhabotinsky reaction  \citep{BZreact70}. These constant-velocity chemical waves can be interpreted as self-propelled particles which are linked to each other via appropriate interaction potentials.\par

\begin{figure}[t]
  \centerline{\includegraphics[angle=0,width=0.9\columnwidth]{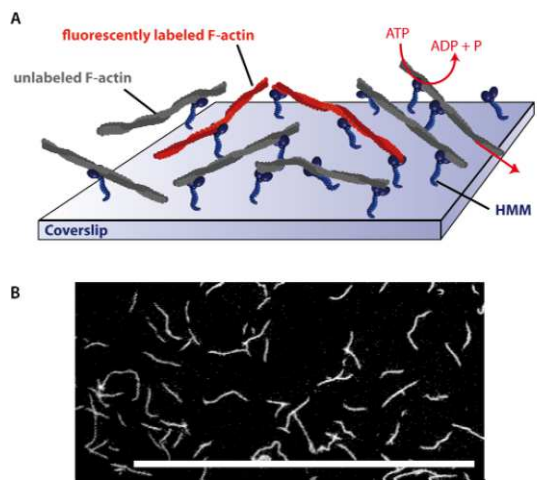}}
  \caption{\label{figFil1}(Color online) The setup of the experiment designed to investigate the effect of density on the collective motion of actin filaments. (a) Filament-motion is visualized by the usage of fluorescently labeled reporter filaments. The ratio of labeled to unlabeled molecules is around 1:200. (b) For low filament densities a disordered structure is found, where the bio-molecules perform persistent random walk without directional preference. Scale bar represents  50 $\mu m$. From \citet{Filaments}.}
\end{figure}

\begin{figure}[t]
  \centerline{\includegraphics[angle=0,width=0.64\columnwidth]{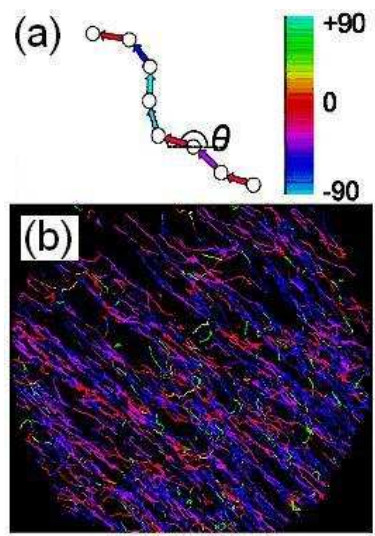}}
  \caption{\label{fig:ButtFila} (Color online) Motion of actin filaments in a motility assay. (a) The individual filaments were tracked automatically and the positions (denoted by circles) were used to estimate their (color coded) direction of motion. (b) The color coded (as above) paths plotted as a continuous track to highlight the trajectory of each filament. The tracks shown are from a 100-frame video sequence recorded at 25 frames/s. Adapted from \citet{ButtEtAl10}.}
\end{figure}

\begin{figure}[t]
  \centerline{\includegraphics[angle=0,width=0.95\columnwidth]{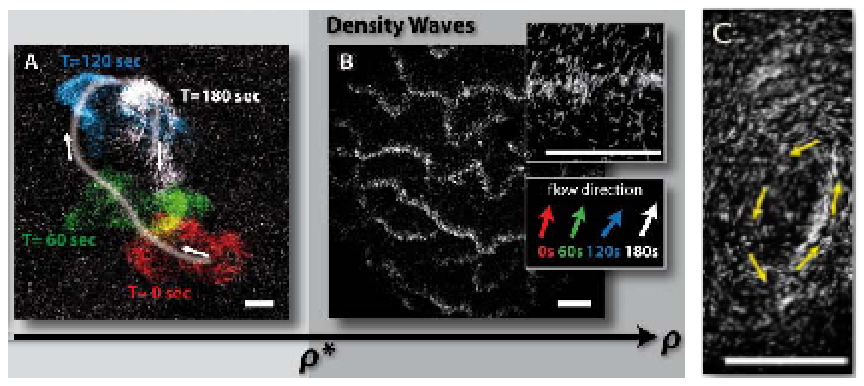}}
  \caption{\label{figFil23}(Color online) The typical behavior of the bio-molecule actin filament in the function of density. The disordered structure (a) becomes ordered (b) above a certain density $\rho^*$, which is around $0.8$ filaments per $\mu m^2$. In this high density regime, wave-like structures can be observed. Above $0.2$ filaments per $\mu m^2$, spirals or swirls can be observed as well (c), which are characterized by huge angular velocity gradients destabilizing the swirl. This limited stability is visible near to the central region of the pattern, where crushing events of the filament current are likely to develop. Scale bar is 50 $\mu m$. Adapted from \citet{Filaments}.}
\end{figure}

Along with the accumulation of the experimental results, the assumption that only a few parameters and factors play a crucial role in the emergence of the ubiquitous phenomena of collective motion has been increasingly supported. \emph{Particle--density} turns out to be one of these parameters, or more precisely, the density of the objects or living beings that exhibit collective motion.\par
The essential role of the density has been demonstrated in a set of elegant experiments on persistently moving biomolecules. In these investigations -- involving the smallest, experimentally realized self-propelled particles so far -- the so called in vitro motility assay is utilized to study the emergence of collective motion on a molecular level. In such an assay actin filaments and fluorescently labeled reporter filaments are propelled by immobilized molecular motors (myosin molecules) attached to a planar surface, as depicted on Fig. \ref{figFil1}. In a recent study, \citet{ButtEtAl10} were the first to observe the bulk alignment of the actin filaments sliding movement for high concentration values even though they were powered by randomly oriented myosin molecules (Fig. \ref{fig:ButtFila}). According to their observations, domains of oriented filaments formed spontaneously and were separated by distinct boundaries. The authors suggested that the self-alignment of actin filaments might make an important contribution to cell polarity and provide a mechanism by which cell migration direction might respond to chemical cues. At almost the same time, \citet{Filaments} undertook a very similar study, but in a somewhat different context of active matter and using extensive evaluation and computational techniques to characterize the phenomenon. They also found that the onset of collective motion was a result of increased filament density. In particular, for low filament densities a disordered phase has been discerned, in which individual filaments performed random walk without any directional preference. Above a certain density, which was around 0.2 filaments per $\mu m^2$, in an intermediate regime, the disordered phase became unstable and small clusters of coherently moving filaments emerged. Further concentration-increase caused growth in the cluster-sizes, but the bunches remained homogeneous. Then, above 0.8 filaments per $\mu m^2$ (signed with $\rho^*$ on Fig. \ref{figFil23}), persistent density fluctuations occurred, leading to the formation of wave-like structures. The authors also identified the weak and local alignment-interactions to be essential for the formation of the patterns and their dynamics. A simulation model was used to interpret the interplay between the underlying microscopic dynamics and the emergence of global patterns in a good agreement with the observations.


The collective behavior and pattern formation in granular, biological, and soft matter systems have been reviewed by \citet{AransonTsimringRMP}, and more recently in their book as well, \citep{AransonTsimringBook}, including both experiments and theoretical concepts.

\subsection{Bacterial colonies}\label{subsec:ExperimBaci}

Since microorganism colonies (such as bacteria colonies) are one of the simplest systems consisting of many interacting organisms, yet exhibiting a non-trivial macroscopic behavior, a number of studies have focused on the experimental and theoretical aspects of colony formation and on the related collective behavior \citep{ShapBaciBook, AltBaciBook, Vicsek01}.\par
The first investigations date back to the early 1970-es, when \citet{KellerSegelM} studied the motion of \emph{Escherichia coli} bands and developed a corresponding phenomenological theory as well. They used partial differential equations describing the consumption of the substrate and the change in bacterial density due to random motion and to chemotaxis. Since than, many observations have been made \citep{ChildressLS75, FujMat89, VCH90, BJEtAl94}, and it has become evident that the bacteria within the colonies growing on wet agar surfaces produce an exciting variety of collective motion patterns: among others, motions similar to super-diffusing particles, highly correlated turbulent as well as rotating states have been observed, and colony formations exhibiting various patterns including those reminiscent of fractals. A special category are those studies which contain not only an observation or a theoretical model, but a matching pair of them: detailed description of an observation together with a computational or mathematical model that accounts for the observations \citep{KellerSegelM, CmplxBactCln96, ReverseBaci, MutansBacik, CzirMatVics}.\par

\begin{figure}
  \centerline{\includegraphics[angle=0,width=0.872\columnwidth]{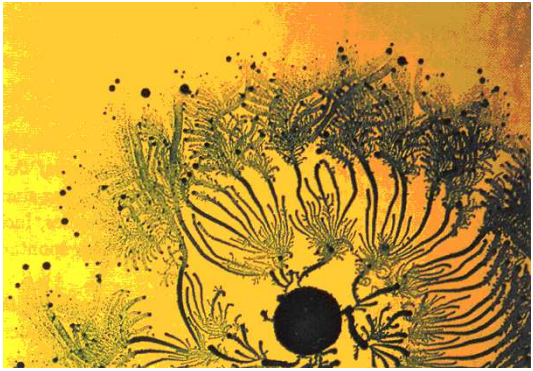}}
  \caption{\label{figBSVortex}(Color online) A typical colony of the \textit{vortex} morphotype of \textit{Bacillus subtilis}. The black discs contain thousands of bacteria circling together as the discs themselves glide outward on the surface during colony growth. Courtesy of E. Ben-Jacob.}
\end{figure}

\citet{CmplxBactCln96} were the first to interpret an experimentally observed complex behavior through a many-particle-type simulation, incorporating realistic rules. They have investigated the intricate colony formation and collective motion (formation of rotating dense aggregates, migration of bacteria in clusters, etc.) of a morphotype of \textit{Bacillus subtilis} using control parameters, such as the concentration of agar and peptone, which was the source of nutrient. Under standard (favorable) conditions bacterium colonies do not exhibit a high level of organization. However, under certain hostile environmental conditions (like limited nutrient source or hard agar surface) the complexity of the colony as a whole increases, characterized by the appearance of \emph{cell-differentiation} and \emph{long-range information transmission} \citep{ShapiroSciAm88}. For describing the observed hydrodynamics (vortex-formation, migration of clusters) of the bacteria in an intermediate level, \citet{CmplxBactCln96} proposed a simpler model of self-propelled particles, and more complex ones -- taking into account further biological details -- to capture the more elaborated collective behaviors, like the vortex and colony formation (see Sec. \ref{sec:BaciModels}). Figure \ref{figBSVortex} depicts a typical bacterium colony formed by the \textit{vortex} morphotype.
\par 

\begin{figure}
  \centerline{\includegraphics[angle=0,width=0.872\columnwidth]{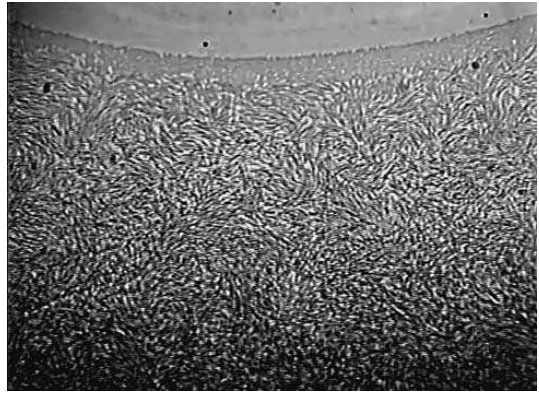}}
  \caption{\label{figBaciSubt}Swimming \textit{Bacillus subtilis} cells exhibiting collective dynamics. 
On a large scale, response to chemical gradients (oxygen, in this case) can initiate behavior that results in striking hydrodynamic flows. From \citet{BacterialFluidMech}.}
\end{figure}

\begin{figure}[t]
  \centerline{\includegraphics[angle=0,width=0.85\columnwidth]{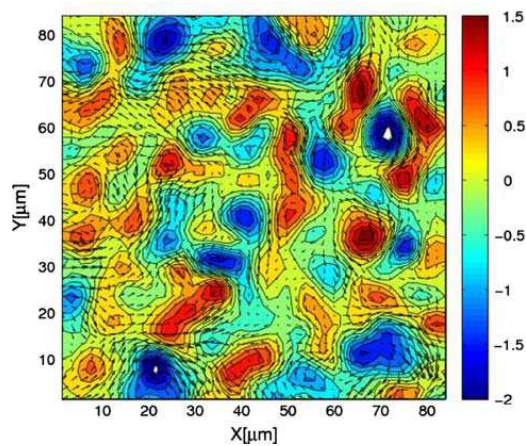}}
  \caption{\label{figBaciVrtFld}(Color) A snapshot of the vorticity field of the swimming bacteria \textit{Bacillus subtilis} (depicted on Fig. \ref{figBaciSubt}). The small arrows indicate the corresponding velocity field. The color bar indicates vorticity in $seconds^{-1}$. The turbulent motion of the suspension is well-observable. The regions of aligned motility are hundreds of times larger than the size of the bacteria, remaining coherent for the order of magnitude of a second. From \citet{BacterialFluidMech}.}
\end{figure}

This kind of bacterium, \textit{Bacillus subtilis}, when the cells are very concentrated (nearly close-packed), forms a special kind of collective \emph{phase} called ``Zooming BioNematics'' (ZBN) \citep{BacterialFluidMech}. This phase is characterized by large scale orientational coherence, analogous to the molecular alignment of nematic liquid crystals, in which the cells assemble together into co-directionally swimming clusters, which often move at speeds larger then the average speed of single bacteria. Figure \ref{figBaciSubt} shows a snapshot of swimming \textit{Bacillus subtilis} cells exhibiting collective dynamics, and Fig. \ref{figBaciVrtFld} depicts the corresponding vorticity field. 
By simultaneously measuring the positions, velocities and orientations of around one thousand bacteria, \citet{ZhangPNAS} demonstrated that under specific conditions, colonies of wild-type \textit{Bacillus subtilis} exhibit giant number fluctuations. More specifically, they found that the number of bacteria per unit area shows fluctuations far larger than those for populations in thermal equilibrium.\label{GNFBaci}\par

The collective behavior of motile aerobic bacteria (``aerobic'' are those bacteria which need the presence of oxygen for their survival), primarily in high cell-concentration, is governed by the interplay between buoyancy, oxygen consumption, mixing and hydrodynamic interactions. The pattern-formation of these bacteria is often governed by another physical mechanism as well, called \emph{bioconvection} \citep{Bioconv}, which appears on fluid medium having a surface open to the air. The authors argued that the patterns appear because bacteria, which are denser than the fluid they swim in, gather at the surface creating a heavy layer on the top of a lighter medium. When the density of the bacteria-cells exceed a certain threshold, this arrangement becomes unstable resulting in a large-scale cell circulation (or convection).\par

\citet{SklvEtAl09} have investigated the onset of large-scale collective motion of aerobic bacteria swimming in a thin fluid, a `film', which had adjustable thickness. They have demonstrated the existence of a clear transition between a quasi-two-dimensional collective motion state and a three-dimensional turbulent state that occurs at a certain fluid-thickness. 
In the turbulent state  -- which is qualitatively different from bioconvection -- an enhanced diffusivity of bacteria and oxygen can be observed, which -- supposedly -- serves the better survival of bacteria colonies under harsh conditions.

\begin{figure}
  \centerline{\includegraphics[angle=0,width=0.814\columnwidth]{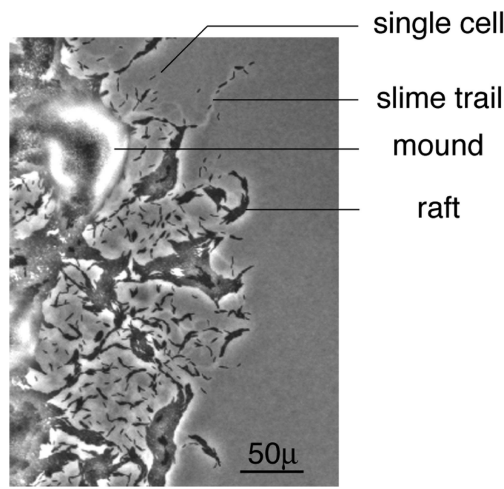}}
  \caption{\label{figReverseBaci}The edge of the expanding colony of bacteria \emph{M. xanthus}. Some individual cells and slime traits are labeled, along with some multicellular ``rafts'' and mounds. The colony is expanding in the radial direction, which is to the right in this image. From \citet{ReverseBaci}. }
\end{figure}

In another remarkable recent paper, while seeking to understand how certain bacteria colonies are able to spread so efficiently, \citet{ReverseBaci} reported on a completely unexpected phenomenon: they found that members of a certain kind of bacteria (\emph{Myxococcus xanthus}) regularly reverse their direction, heading back to the colony which they have just came from, which is -- seemingly -- only a waste of time and energy. Motivated by these observations, the authors constructed a detailed computational model that took into account both the behavior and the cell biology of the bacteria \emph{M. xanthus}. The most interesting result was revealed by the model, namely, that these reversals generate a more orderly swarm with more cells oriented in parallel, making the cells less likely to collide with each other. Without these turn-backs the cells would become disordered and as a whole, would move at a slower rate while finally coming to a standstill. The model predicts that the swarm expands at the greatest rate when cells reverse their direction approximately every eight minutes -- which is in exact match with the observations. Figure \ref{figReverseBaci} shows a snapshot of the expanding colony of bacteria \emph{M. xanthus}.\par

\begin{figure}
  \centerline{\includegraphics[angle=0,width=0.58\columnwidth]{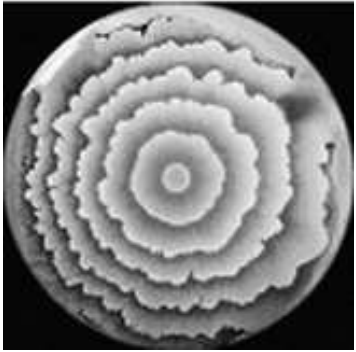}}
  \caption{\label{figBaciKorok}The colony of \emph{Bacillus subtilis} which -- similarly to \emph{P. mirabilis} -- grow in a peculiar concentric ring-like pattern, which is the result of a special swarming cycle reported by \citet{CzirMatVics} and \citet{YamazakiBaci}. From \citet{YamazakiBaci}.}
\end{figure}

Wild-type (``normal'') \emph{Myxococcus xanthus} has two different kinds of engines to move itself: a \emph{pilus} at its front end which \emph{pulls} the cell, and a slime secretion engine at its rear that \emph{pushes} the bacterium forward. \citet{MutansBacik} have investigated the coordinated motion and social interactions of this bacteria by using mutants: bacteria that were void of either one or the other type of engine. Based on their observations they have introduced a cell-based model to study the role of the two different kinds of engines and to show how the interactions between neighboring cells facilitate swarming.\par

\citet{CzirMatVics} reported on swarming cycles (exhibited by many bacterial species) resulting in a colony with concentric rings (see Fig. \ref{figBaciKorok}). (Although the phenomenon had been known for some time, in their study both the quantitative measurements and the theoretical interpretation have been the subject of research.) These zones develop as the bacteria (\emph{Proteus mirabilis} in their experiments) multiply and swarm following a periodically repeating scenario: when the bacteria cells are applied to the surface of a suitable hard agar medium, they grow as short, immotile ``vegetative'' rods. Then, after a certain time, cells start to differentiate at the colony margin into long motile ``swarmer'' cells which then migrate rapidly away from the colony until they stop and revert by a series of cell fissions into the vegetative cell form again. These cells then grow normally for a time, until the swarmer cell differentiation is initiated in the outermost zone again, and the process continues in periodic cycles resulting into the concentric ring-structure. For this process -- every step of which has been observed and described in detail -- a model has been developed as well, which is in excellent agreement with the observations. \citet{YamazakiBaci} investigated the above described periodic change between the motile and the immotile cell states experimentally, and concluded that the change between the two states was determined neither by biological nor by chemical factors, but by the local cell density.\par

Many papers deal with the effects of cell-density on the collective behavior of a bacterium colony. 
\citet{ZurosBaciC} artificially created regions in which a given type of bacteria-cells are strongly concentrated. In these regions the authors found striking collective effects with transient, reconstituting, high-speed jets straddled by vortexes, and suggested a corresponding modification for the Keller-Segel model which takes into account the hydrodynamic interactions as well. (The Keller-Segel model is probably the most prevalent model for chemical control of cell movements, which has been originally introduced by \citet{KellerSegelM}.) The relevance of the hydrodynamic effects was highlighted by \citet{SokolovAKG} as well, who presented experimental results on collective bacterial swimming in thin, two-dimensional fluid films by introducing a novel technique that made it possible to keep bacterium-cells in condensed populations exhibiting adjustable concentration.\par
\citet{BaciDyn} related the characteristic velocity, time and length scales of the collective motion of a given type of swarming bacteria colonies.\par
\begin{figure}[t]
  \centerline{\includegraphics[angle=0,width=1\columnwidth]{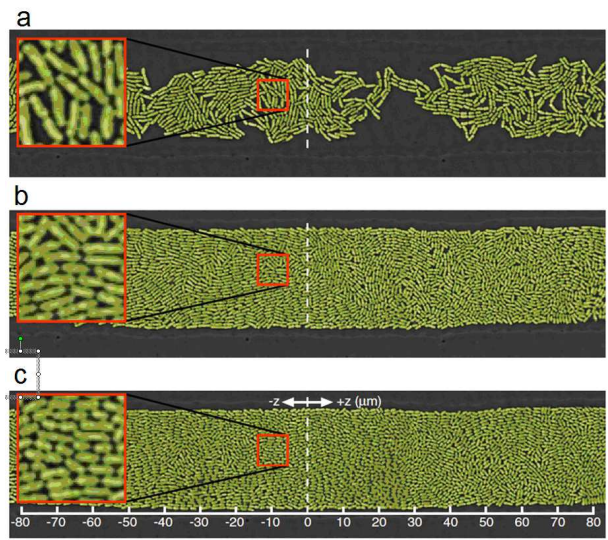}}
  \caption{\label{figBaciMechOrdr}(Color online) The growth and ordering of the bacteria \emph{E. coli} in a quasi 2D open microfluidic cavity. Originally, at the beginning of the test, the cells are distributed evenly and sparsely. The three snapshots are taken at (a) 60, (b) 90, and (c) 138 minutes from the beginning of the experiment, respectively. Growth and division in a dense colony leads to a dynamic transition from a disordered phase to a highly ordered one, characterized by orientational alignment of the rod-shaped cells. Adapted from \citet{BioMechBaciOrder}.}
\end{figure}

The effects of the biomechanical interactions (arising from the growth and division of the bacteria cells) on the colony formation -- although being ubiquitous -- have received little attention so far. \citet{BioMechBaciOrder} addressed this issue by observing and simulating the structure and dynamics of a growing two-dimensional colony of non-motile bacteria, \emph{Escherichia coli}. They found that growth and division in a dense colony led to a dynamic transition from a disordered phase to a highly ordered one, characterized by orientational alignment of the rod-shaped cells (see Fig. \ref{figBaciMechOrdr}). The authors highlighted, that this mechanism differed fundamentally from the one arranging the particles of liquid crystals, polymers or vibrated rods, since this latter one was due to the combination of fluctuation and steric exclusion.\par
Further studies also suggest that the local alignment among \emph{Escherichia coli} cells is accomplished by cell body collisions and/or short-range hydrodynamic interactions \citep{DarntonBaci10}. According to this experiment the directional correlation among the cells is anisotropic, and the speeds and the orientations are correlated over a short, several cell-length scale. The orientations of the bacteria were continually, randomly modified due to jostling by neighbors. Cells at the edge of the swarm were often observed to pause and swim back towards the swarm or along its edge.\par
In a recent paper \citet{DrescherBaci} also challenge the models that explain the motion of these bacteria based on long-range hydrodynamic effects. They argue that noise, due to intrinsic swimming stochasticity and orientational Brownian motion basically eliminates the hydrodynamic effects between two bacteria beyond a few microns, that is, short-range forces and noise dominate the interactions between swimming bacteria.\par

The web-page maintained by the Weibel lab,  \texttt{http://www.biochem.wisc.edu/faculty/weibel/} \texttt{lab/gallery/movies.aspx} contains movies of several types of swarming bacteria, among others, \emph{Escherichia coli}.\\
\par 
\citet{TokitaBaci09} studied the morphological diversity of the colonies of the same type of bacteria, \emph{Escherichia coli}, as a function of the agar and nutrient concentration. Various colony shapes were observed, classified into four fundamental types based on the main characteristics of the patterns. 

\citet{WuLibchaber} investigated a new situation, involving two types, active and passive particles moving in a fluid. The active particles were \emph{Escherichia coli} bacteria, and the passive units were micron-scale beads. The aim was to study the effects of bacterial motion on inactive particles on a quasi-two-dimensional geometry. They found large positional fluctuations for beads as large as $10 \mu m$ in diameter and measured mean-square displacements  indicating superdiffusion in short times and normal diffusion in long times.\par
Here we would like to note that \citet{BaciSwarmOfWeibel} published a very useful review on bacterial swarming from a biological point of view.

\subsection{Cells}\label{subsec:ExperimCells}

The basic observations and experiments regarding \emph{unicellular} organisms (also, cells) have been discussed in the previous section (\ref{subsec:ExperimBaci}) since many kinds of bacteria-stems proved to be good subject for various experiments. However, some interesting experiments regarding the collective motion of unicellular beings, have not been made on bacteria, but other kind of cells.\par

Here we only shortly mention some studies investigating the collective motion of the cells \emph{Dictyostelium discoideum} (commonly referred to as \emph{slime mold}). In order to describe the dynamics of these cells, \citet{RotatingSlimeMold} took into account their shape and plasticity. Based on the observations regarding how these amoeba cells aggregate into rotating ``pancake''-form structures, the authors built a model of the dynamics of self-propelled deformable objects (see also in Sec. \ref{sec:SysSpesMCells}). According to the experiments, these cells tended to form round structures which rotated around the center clockwise or counterclockwise (depending on some initial conditions) often persisting for tens of hours. Using the same genus, \emph{Dictyostelium discoideum}, \citet{SlimeMoldMcC} conducted experiments in order to elucidate the role of signal relay -- a process during which the individual cells amplify chemotactic signals by secreting additional attractants upon stimulation -- in the collective motion of these cells. They found that this process enhances the recruitment range, but does not effect the speed or directionality of the cells.\par

Next we discuss the collective motion of cells in highly structured, \emph{multi-cellular} organisms, in which cell migration plays a major role in both embryonic development (e.g. gastrulation, neural crest migration) and the normal physiological or patho-physiological responses of adults (e.g. wound healing, immune response or cancer metastasis). In these mechanisms cells have to be both motile and adhere to one other. In order to describe these features, \citet{SzaboUnnepPhysBiol} expanded the cellular Potts model by including active cell motility, and studied a corresponding computer simulations as well, which was compatible with the experimental findings.\par
\begin{figure}[!ht]
  \centerline{\includegraphics[angle=0,width=0.82\columnwidth]{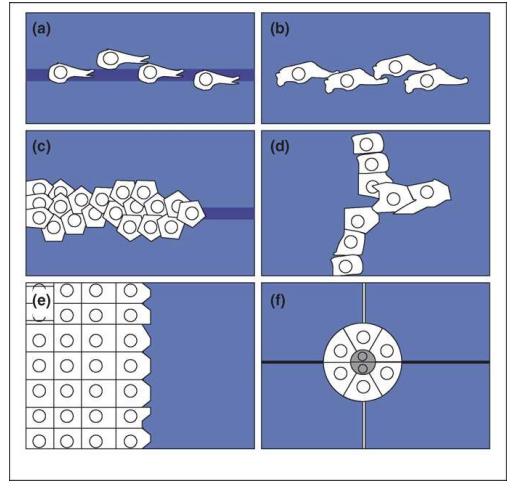}}
  \caption{\label{figRorthCellTbl}(Color online) The basic types of collective cell migrations with respect to the strength of the contact among the cells moving together. The schematic drawing of the cells are white with gray circles within them, which are the nuclei. The dark lines on (a) and (c) are migration-permissive tracks in the substrate. The movement is from left to right. (a) and (b) depict loosely associated cells which contact rarely (a), or more frequently, (b). Although these kind of motions are sometimes restricted by tracks, the cells mostly contact the substrate with a high degree of freedom. Neural-crest cells and germ-line cells belong to these categories. The cell-structure depicted in (c) has a well-defined front and back part. An example is the neuromast cells of the fish lateral line. (d) shows an example of tracheal or vascular-type branch outgrowth, during which the cells remain associated through the the central bud growing out from the existing epithelium cells. (e) shows an epithelial sheet moving to close a gap. These cells most probably have only a small degree of freedom. (f) A border-cell-cluster moves among giant nurse cells which are depicted by the surrounding squares. From \citet{CellRorth}.}
\end{figure}

 
In living organisms different strategies exist for cell movement, including both individual cell migration and the coordinated movement of groups of cells \citep{CellRorth}. Figure \ref{figRorthCellTbl} summarizes the basic types of collective cell migrations with respect to the strength of the contact among the cells moving together.\par

$i$) Groups can be associated loosely with occasional contact and much of the apparent cohesion might come from essentially solitary cells following the same tracks and cues. Examples are germ cells in many organisms; the rostral migratory stream supplying neurons to the olfactory bulb (RMS) and neural crest (NC) cells migrating from the developing neural tube to many distant locations in the embryos of mammals; sperm cells. Although the collective motion of these cells are often guided by chemical signals, in some cases they can also form patterns based on mere hydrodynamic effects \citep{SpermVertices}.\par

$ii$) Other migrating groups are more tightly associated and the cells normally never dissociate. Examples are the fish lateral line, structures performing branching and sprouting morphogenesis such as trachea or the vasculature and finally moving sheets of cells in morphogenesis or wound healing. These groups have an additional feature, in that the moving structure has an inherent polarity, a free 'front' and an attached 'back'.\par

$iii$) Drosophila border cells are a group or cluster of cells performing a directional movement during oogenesis. These migrating cells are associated tightly but the cluster is free, without an inherent 'back'. A particularly nice visualization of the collectively moving cells during the development of zebra fish (by three dimensional tracing of live-stained cell nuclei) very well demonstrates the relevance of collective motion during morphogenesis \citep{SchoetzBook}. 
\citet{GilmourGrZebraf08} investigate the mechanism that organize cells behind the leading edge. In particular, they study the role of the fibroblast growth factors (Fgfs) -- a signaling method known to regulate many types of developmental processes \citep{FgfHiv1, FgfHiv2} -- during the formation of sensory organs in zebrafish. The role of chemokine signaling in regulating the self-organizing migration of tissues during the morphogenesis of zebrafish is investigated by \citet{GilmZebraf06}. 

By using an interdisciplinary approach, \citet{CellHeisenbergGr} identified MCA (membrane-to-cortex attachement) as a key component in controlling directed cell migration during zebrafish gastrulation. They showed that reducing MCA in mesendoderm progenitor cells during gastrulation reduces the directionality of the cell migration as well. By investigating the same process, zebrafish gastrulation, \citet{ArbCell2010} found that individually moving mesendoderm cells are capable of normal directed migration on their own, but for moving coherently, that is, for participating in coordinated and directed migration as part of a group, cell-cell adhesion is required.\par

In order to distinguish individual, cell-autonomous displacements from convective displacements caused by large-scale morphogenetic tissue movements \citet{ZCell} have developed a technique. Using this methodology, they have separated the active and passive components of cell displacement directly, during the gastrulation process in a warm-blooded embryo. \citet{Sprouting} provided an excellent review on cell-movements and tissue-formation during vasculogenesis in warm-blooded vertebrates. As the key mechanism for this process, the authors identified the formation and rapid expansion of multicellular sprouts, by which the originally disconnected endothelial cell clusters join and form an interconnected network.\par

\begin{figure}[!ht]
  \centerline{\includegraphics[angle=0,width=1\columnwidth]{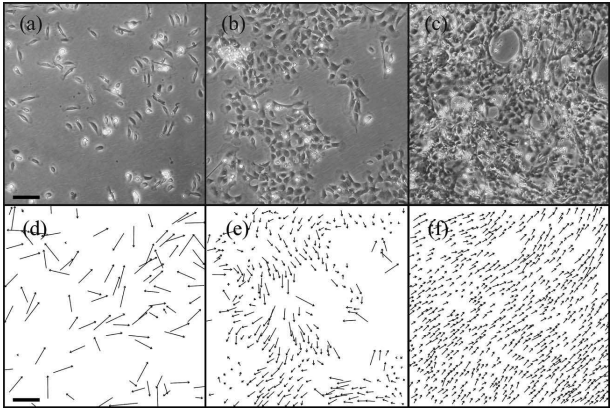}}
  \caption{\label{figCellSzabo}Phase contrast images showing the collective behavior of fish keratocites for three different densities. The normalized density, $\bar{\rho}$ is defined as $\bar{\rho}=\rho/\rho_{max}$, where $\rho_{max}$ is the maximal observed density, 25 cells/$100 \times 100 \mu m ^ 2$. (a) $\rho=1.8$ cells/$100 \times 100 \mu m ^ 2$ corresponding to $\bar{\rho}=0.072$ (b) $\rho=5.3$ cells/$100 \times 100 \mu m ^ 2$ which is $\bar{\rho}=0.212$, and (c) $\rho=14.7$ cells/$100 \times 100 \mu m ^ 2$, $\bar{\rho}=0.588$. The scale bar indicates $200 \mu m$. As cell density increases cell motility undergoes to collective ordering. The speed of coherently moving cells is smaller than that of solitary cells. (d)-(f) on the bottom panel depicts the corresponding velocities of the cells. From \citet{SzaboEtAl}. }
\end{figure}

\begin{figure}
  \centerline{\includegraphics[angle=0,width=0.82\columnwidth]{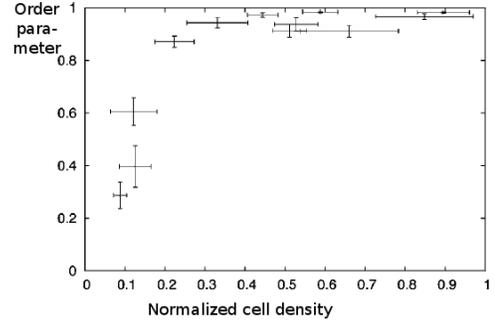}}
  \caption{\label{FigSzaboNormDens} Order parameter versus the normalized cell density. The order parameter is a measure describing the level of coherency of the motion (for more details see Eq. \ref{eq:NormV} in Seq. \ref{sec:BasicDefs}) and the normalized cell density is the measured density divided by the maximal observed cell density. The error bars show the standard error of the density and order parameter. From \citet{SzaboEtAl}.}
\end{figure}

A quantitative analysis of the experimentally obtained collective motion and the associated ordering transition of co-moving fish keratocites was carried out by \citet{SzaboEtAl}. They have determined the phase transition as a function of the cell density (Fig. \ref{FigSzaboNormDens}) and, motivated by their experimental results, have constructed the corresponding model as well (see Sec. \ref{sec:SysSpesMCells}). Figure \ref{figCellSzabo} shows the typical collective behavior of the keratocite cells for three different densities, and Fig. \ref{FigSzaboNormDens} depicts the phase transition (described by the order parameter) as a function of the normalized cell density. Other aspects of collective cell motion -- for example the relation between the viscosity of the substrate and the velocity of the cells -- were also studied \citep{MurrellCell2011}.\par

Further interesting examples for the collective motion of tissue cells are related to the following two processes:\par
\emph{Tissue repair and wound healing.} In tissue repair, collective migration is seen in vascular sprouts penetrating the wound or the horizontal migration of epithelial cell-sheets across 2d substrates upon self-renewal of keratinocytes migrating across the wound \citep{Friedl2004}. In epithelial tissue, the opening of a gap induces the proliferation and movement of the surrounding intact cells, which eventually closes the gap. \citet{CellKorny} studied the responses of artificially mechanically injured astrocytes (a characteristic star-shaped glial cell in the central nervous system) in vitro. In particular, the changes in the cell-motility, proliferation and morphology were analyzed. Their data suggested that the mechanical injury (basically a ``scratch'') was not sufficient to indicate changes in the motility of the astroglia cell, but did result in a local enhancement in the cell proliferation.\par

As discussed above, the widely accepted approach regarding the nature of the migration of groups of cells in organisms has assumed that ``leading cells'' situated at the front edge of the group guide the motion of all the cells. This is suggested by many studies \citep{CellContr1, Friedl2004, CellContr2}, among others by the ones carried out on zebrafish, a genus whose morphogenesis and organ formation is the subject of many experiments \citep{GilmourGrZebraf08, GilmZebraf06}.

\begin{figure}
  \centerline{\includegraphics[angle=0,width=1\columnwidth]{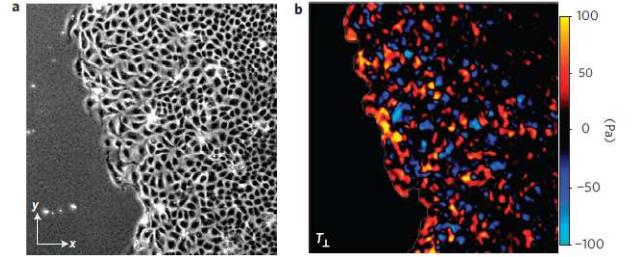}}
  \caption{\label{figCellForces}(Color) Traction forces generated by a sheet of collectively moving cells. (a) is the phase contrast image and (b) depicts the tractions normal to the edge. Adapted from \citet{CellMigrPhysF}.}
\end{figure}

However, in a recent paper \citet{CellMigrPhysF} argued that traction forces driving collective cell migration did not arise only (or primarily) in the leader cells which were at the front of the traveling cell sheet, but, as it can be seen on on Fig. \ref{figCellForces}, in many cell rows behind the leading front edge cells as well. Although, like some kind of ``tug of war'', the cell sheet as a whole moved in one direction, many cells within the cluster pulled the sheet in other directions.\par

\par

\emph{Cancer metastasis.} Two morphological and functional variants of collective migration have been described in tumors in vivo. The first results from protruding sheets and strands that maintain contact with the primary site, yet generate local invasion. The second shows detached cell clusters or cell files, histologically seen as 'nests', which detach from their origin and frequently extend along interstitial tissue gaps and paths of least resistance, as seen in epithelial cancer and melanoma \citep{Friedl2004}. Collective migration represents the predominant mode of tissue invasion in most epithelial cancers.\par
Furthermore, recently it has been argued that malignant tumor cells may be capable of developing collective patterns that resemble to evolved adaptive behaviors like collective decision-making or collective sensing of environmental conditions. \citet{CancerCellCB} presented a concept as to how these abilities could arise in tumors and why the emergence of such sophisticated swarm-like behavior would endow advantageous properties to the spatio-temporal expansion of tumors.\par
In a recently published review article \citet{FriedlGilm09} draw attention on the presumed common mechanistic themes underlying the different collective cell migration types by comparing them at the molecular and cellular level. This review paper summarizes the topic from a more biological point of view.

\subsection{Insects}\label{subsec:ExperimInsects}

Insects are one of the most diverse animal groups on Earth including more than a million described species which makes them represent more than half of all known living organisms \citep{BioDivInsct}. They can move about by walking, flying or occasionally swimming. Some of their species (like water striders) are even able to walk on the surface of water. Most of them live a solitary life, but some insects (such as certain ants, bees or termites) are social and are famous for their large and well-organized colonies. Some of these so called ``eusocial'' insects have evolved sophisticated communication system, such as the ``round dance'' and ``waggle dance'' of Western honey bee, \emph{Apis mellifera}. However, motion-patterns based on such highly developed communication are out of the scope of our review, but interested readers can find abundant literature on the topic, for example \citep{SocInsectBook2008}.\par
Many species of butterflies (e.g. \emph{Red Admiral}, \emph{Painted Lady}) and moths (\emph{Humming-bird Hawk-moth}, \emph{Silver-Y moth}) migrate twice a year between the two hemispheres: when it is autumn on the Northern hemisphere they form huge ``clouds'' and fly to south and come back only when spring arrives. Other insects being famous for exhibiting collective motion are \emph{ants}. Many of them create tracks between the nest and the food sources very efficiently, using pheromone trails. For example New World army ant \emph{Eciton burchelii} -- whose colonies may consist of million or more workers -- stage huge swarm raids with up to 200 thousand individuals forming trail systems that are in length up to 100 m or even more, and 20 m wide \citep{AntBookG, AntBlindF}. Based on the observations \citet{AntLaneCF} have investigated the formation of these elaborated traffic lanes, and created a corresponding model exploring the influences of turning rates and local perception on traffic flow. Furthermore, \citet{InsctPhaseTr} have investigated another fundamental question regarding these formations, namely what is the minimum number of workers that are required for this kind of self organization to occur. They have observed Pharaoh ants and they actually discovered that small groups forage in a disorganized way while larger ones are organized. Thus -- for the first time -- they have provided experimental evidence on a behavioral first-order phase-transition exhibiting hysteresis between organized and disorganized states.\par

\begin{figure}
  \centerline{\includegraphics[angle=0,width=0.64\columnwidth]{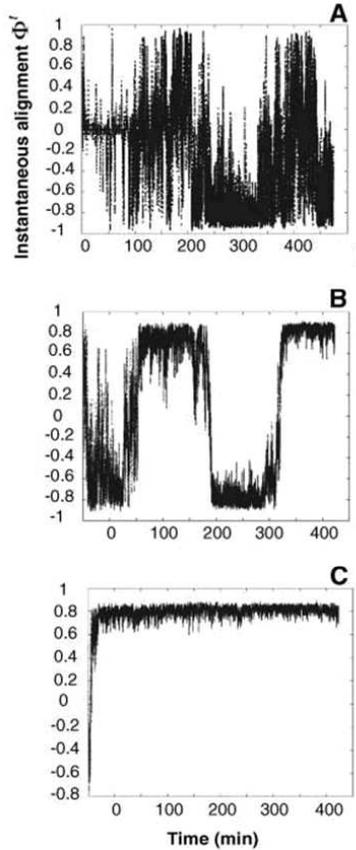}}
  \caption{\label{figLocustMerEredm}The alignment of the motion of locusts in the function of animal density. The alignment is defined as the normalized average of the orientation for all moving animals, which means that values close to -1 and 1 indicate strong alignment (all locusts move in the same direction) whereas values close to zero indicate uncoordinated motion. At low densities (a), the alignment among individuals - if it occurs - is sparse and sporadic, following a long initial period of disordered motion (5.3--17.2 locusts/$m^2$, equating to 2--7 moving locusts). (b) Intermediate densities (24.6--61.5 locusts/$m^2$, equating to 10--25 moving locusts) are  characterized by sharp and abrupt changes in direction, separating long periods of correlated motion. (c) At densities above 73.8 locusts/$m^2$ (equating to 30 or more moving locusts) the alignment of the motion is strong and persistent, individual locusts quickly adopt their motion to the others, and spontaneous changes in the direction do not occur. Adapted from \citep{CouzinScience}.}
\end{figure}

\begin{figure}
  \centerline{\includegraphics[angle=0,width=0.872\columnwidth]{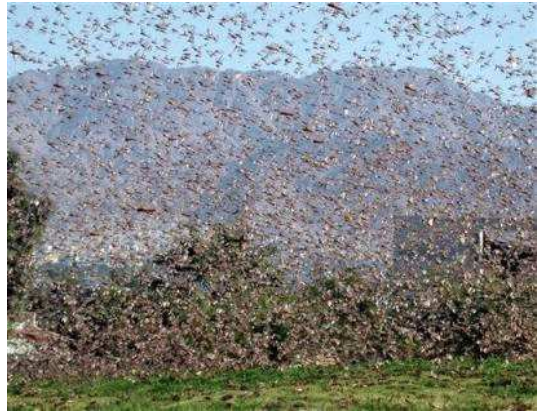}}
  \caption{\label{figLocustSwarm}Locust swarm. From Physorg.com}
\end{figure}

Traditionally, an aggregate is considered to be an evolutionarily advantageous state for its' members: it provides protection, information and choice of mates on the cost of limited resources and increased probability for various infections \citep{WSocioBi}. However, according to some recent studies, in the case of some insect-species the depletion of nutritional resources may easily lead to cannibalism among group-members. \citet{CannibalCrickets} reported how the local availability of protein and salt influenced the extent to which Mormon cricket bands marched, both through the direct effect of nutrient state on locomotion and indirectly through the threat of cannibalism by resource-deprived specimens. Similarly, \citet{CannibalLocusts} demonstrated that coordinated mass migration in juvenile desert locusts (see Fig. \ref{figLocustSwarm}) was influenced strongly by cannibalistic interactions: Individuals in marching bands tended to bite each other but also risk being bitten themselves. Surgical reduction of individuals' capacity to detect the approach of others from behind decreased their probability to start moving, dramatically reduced the mean proportion of moving individuals in the group and significantly increased cannibalism as well, but it did not influence the behavior of isolated locusts. They also showed that while abdominal biting and the sight of others approaching from behind triggered movement, the occlusion of the rear visual field inhibited individuals' propensity to march.\par 
In a field study \citet{BazaziSocLocusts} found that adult Mormon crickets were more likely to attack a stationary conspecific that was on the side-on than either head- or abdomen-on, from which it follows that an individual can reduce the risk of being attacked by aligning with its neighbors. The team also revealed a social effect on the cannibalistic behavior, namely that the more individuals were present around a stationary cricket, the higher the probability was for an encounter resulting in an attack.

Other characteristics of the collective motion of locusts have also been studied: \citet{YatesPnas09}  investigated the sudden coherent switches in direction, and \citet{CouzinScience} their behavior with respect to the effect of the animal-density on the transition between disordered and ordered states. The experimental results are depicted on Fig. \ref{figLocustMerEredm}, which shows the alignment of the motion in the function of locust density, for three different cases: (a) low density ($\bar{\rho}=0.26-0.86 \times 10^{-3}$ expressed in terms of normalized density), (b) intermediate density ($\bar{\rho}=1.23-3.1 \times 10^{-3}$) and (c) high density ($\bar{\rho}=3.7 \times 10^{-3}$). Normalized density is calculated as usually, $\bar{\rho}=\rho / \rho_{max}$, where $\rho_{max}$ is the maximal hopper density estimated to be 20000 hopper$/m^2$ after \citep{LocustBook, LocustGl}. As it can be seen, coordinated marching behavior strongly depends on the animal-density. With these experiments they also confirmed that the transition followed the theoretical predictions of the SPP-model \citep{VM}, and identified the critical density for the onset of coordinated marching as well.\par

Although zooplankton are not insects, here we briefly mention an interesting study in which Daphnia-swarms were artificially induced to carry out vortex motion by using optical stimulus. When the density of these tiny creatures is small, they exhibit circular motion around a vertical shaft of light, to which they are attracted. \citet{Daphnia} found that above a density-threshold a swarm-like motion emerges in which all Daphnia circle in the same -- randomly chosen -- direction. In order to reproduce the observed behavior, the authors developed a self-propelled agent based model based on random walks. 
They found that with two ingredients of the model the observed circular motion can be reproduced: (i) a short-range temporal correlation of the velocities (which is in the experiment the short-range alignment resulting from the water drag), and (ii) an attraction to a central point proportional to the agent's distance from it (which is in the experiment the attraction generated by the light beam.)

\subsection{Fish schools and shoals}\label{subsec:ExperimFish}

The largest groups of vertebrates exhibiting a rich set of collective motion patterns are certainly fish shoals and schools. Although these two terms cover very similar behaviors -- and thus are often mixed -- their meaning slightly differs: in a \emph{shoal} fish relate to each other in a looser manner than in a school, and they might include fish of various species as well \citep{PitcherFish83}. Shoals are more vulnerable to predator attack. In contrast, in a \emph{school} fish swim in a more tightly organized way considering their speed and direction, thus a school can be considered as a  special case of shoal \citep{FishBookH}. At the same time, from one second to the other a shoal can organize itself into a disciplined school and vice versa, according to the changes in the momentary activity: avoiding a predator, resting, feeding or traveling \citep{FishBookM, HoareF04}.\par
Schooling is a very basic feature of aquatic species and may have appeared in a very early stage of vertebrate evolution \citep{ShawFishOf}. Over 50 percent of bony fish species school and the same behavior has been reported in a number of cartilaginous fish species as well \citep{ShawFishOf, PorcosH}. Since the large-scale coherent motion of fish is also very important from a practical point of view (fishing industry), the observational and simulational aspects of fish schools have played a central role in the studies of coherent motion.\par

\begin{figure}[t]
  \centerline{\includegraphics[angle=0,width=1\columnwidth]{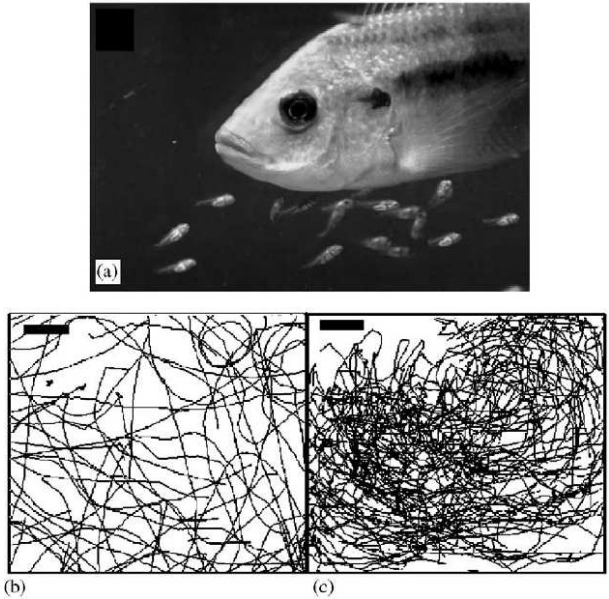}}
  \caption{\label{figFishBecco}\citet{BeccoFishExperim} recorded the trajectories of young Tilapia fish in a school. (a) The photo of a Tilapia together with her offsprings. (b) The trajectory of 20 fish (equating to 350 fish/$m^2$) recorded for 41 seconds. (c) Same as the previous one, but with 905 fish/$m^2$. The bar scale on sub--pictures (b) and (c) represents 1cm$\times$5cm. From \citet{BeccoFishExperim}. }
\end{figure}
\par

\citet{BeccoFishExperim} recorded the trajectories of young fish in a school (see Fig. \ref{figFishBecco}). Both individual and collective behavior were studied as a function of ``fish-density'', and a transition from disordered to correlated motion was found. Also by trajectory analysis, \citet{Katz2011} inferred the structure of the interactions among schooling golden shiners, \emph{Notemigonus crysoleucas}. They found that it is not an ``alignment rule'' but a speed regulation which is the key aspect during interaction, i.e., changes in speed effecting conspecifics both behind and in front of the fish are essential. They argue that alignment only \emph{modulates} the strength of speed regulation, rather than being an explicit force itself. Another important claim of the study is that the observed interactions can not be decomposed into the sum of two-body interactions, but rather three-body interactions are necessary in order to explain the observed dynamics.\par

\begin{figure}[t]
  \centerline{\includegraphics[angle=0,width=1\columnwidth]{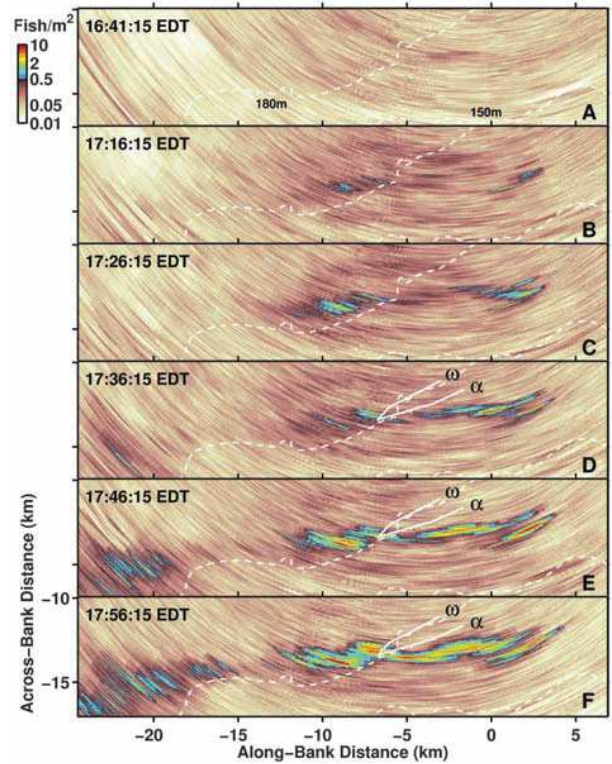}}
  \caption{\label{figFishRadar}(Color) OAWRS snapshots showing the formation of vast herring shoals, consisting of millions of Atlantic herring, on the northern flank of Georges Bank (situated between the USA and Canada) on 3 October 2006. Adapted from \citet{FishDensM}. }
\end{figure}

\begin{figure}
  \centerline{\includegraphics[angle=0,width=0.6\columnwidth]{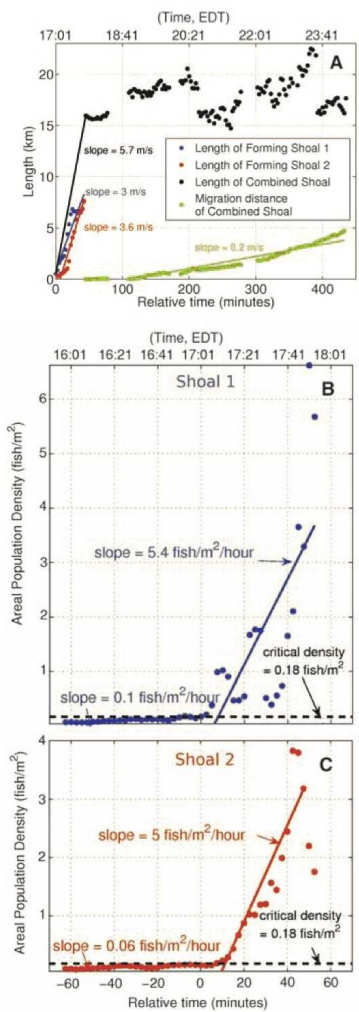}}
  \caption{\label{figFishRadarGrfkn}(Color online) The results obtained from evaluating the data recorded by the technique called ``Ocean Acoustic Waveguide Remote Sensing'' (OAWRS), see Fig. \ref{figFishRadar}. (a) The length of the three forming shoals (on the left, depicted by red, blue and black colors online) and the migration distance (bottom of the picture, green online) in the function of time. The solid lines are the best-fit slopes to the recorded data. (b) and (c) population density versus time for shoal 1 (blue data online) and shoal 2 (red data online). A slow growth in population density is followed by a rapid increase immediately after the critical fish-density is reached ($0.18 fish/m^2$ corresponding to approximately $\bar{\rho}=0.027$ normalized density. The normalized density is defined as $\bar{\rho}=\rho/\rho_{max}$, where $\rho_{max}$ is the maximal observed density, approximately 6.6 fish/$m^2$ according to sub-figure \emph{b}). Adapted from \citet{FishDensM}.}
\end{figure}

Using a novel technique called ``Ocean Acoustic Waveguide Remote Sensing'', OAWRS \citep{FishObs}, which enables instantaneous imaging and continuous monitoring of oceanic fish shoals over tens of thousands of square kilometers, \citet{FishDensM} observed vast herring populations during spawning (see Fig. \ref{figFishRadar}). The team observed a rapid transition from disordered to highly synchronized behavior at a critical density, followed by an organized group migration (see Fig. \ref{figFishRadarGrfkn}). Furthermore, in agreement with other studies (see Sec. \ref{subsec:ExprmLeadership}), they also found that a small set of leaders can significantly influence the actions of a much larger group.\par

One of the most fundamental question regarding gregarious animals -- thus fishes as well -- is how the \emph{common decision} is reached. If they are to stay together, they constantly have to face questions like: which direction to swim, where to stop and forage, how to guard against predators, etc. Is it governed by a leader or by some kind of consensus? \citep{ReebsFish, SumpterConsFish, BiroEtAl06} How does the size of the school influence decision making? \citep{GrunbaumFish}\par
Regarding the connotation of ``consensus \emph{decision},'' most scholars follow the definition proposed by \citet{ConradtRoper05}, who interpreted it as the process in which `the members of a group chose between two or more mutually exclusive actions with the aim of reaching a consensus', and ``\emph{leadership}'' was `the initiation of new directions of locomotion by one or more individuals, which were then readily followed by other group members' \citep{KrauseFish2000}.

In a recent experiment \citet{QuorumSensing} discovered that individual fish responded only when they saw a threshold number of conspecifics to perform a particular behavior (``quorum responses''). They experimentally investigated (and also modeled) the decision making process about movements of a school in the case of a specific kind of fish. \citet{ReebsFish} trained twelve golden shiners to expect food around midday in one of the brightly lit corners of their tank and investigated whether these informed individuals were able to lead their shoal-mates to the site of the food source later or not. He found that a minority of informed individuals (even one) can lead a shoal to the food-site. He also observed that the shoals never split up and were always led by the same fish.\par
Some of the experiments suggesting these results utilized ``replica fish'' or fish robots in order to study the decision-making behavior \citep{SumpterConsFish, RoboFish}.

\subsection{Bird flocks}\label{subsec:ExperimBirds}

Flocking of birds have been the subject of speculation and investigation for many years. Some of the nearly paradoxical aspects of the extremely highly-coordinated motion patterns were pointed out already in the mid 1980-es \citep{Potts84}. In this paper Potts discussed how the flock movements were initiated and coordinated, through a frame-by-frame analysis of high-speed film of sandpiper flocks. He argued that any individual can initiate a flock movement, which then propagates through the flock in a wave-like form radiating out from the initiation site.\par
\begin{figure}[h!]
  \centerline{\includegraphics[angle=0,width=0.82\columnwidth]{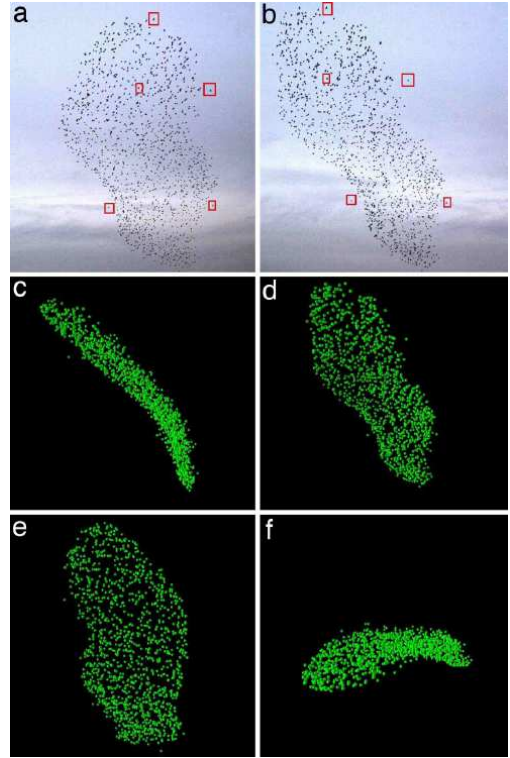}}  
  \caption{\label{figBallFlock}(Color online) A typical starling flock and its 3D reconstruction. (a) and (b) is the photograph of one of the analyzed flocks. The pictures were made at the same moment by two different cameras, 25 meters apart. For reconstructing the flocks in 3D, each bird's image on the left had to be matched to its corresponding image on the right. The small red squares indicate five of these matched pairs. (c-f) The 3D reconstructions of the analyzed flock from four different perspectives. (d) The reconstructed flock from the same view-point as (b). From \citet{BalleriniEtAl}. }
\end{figure}
\begin{figure}
  \centerline{\includegraphics[angle=0,width=0.95\columnwidth]{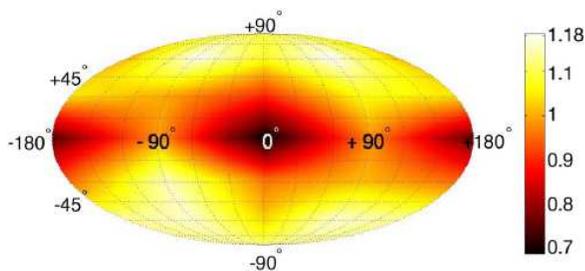}}
  \caption{\label{figBallStruct}(Color online) The average angular density of the birds' nearest neighbors. The map shows a striking lack of nearest neighbors along the direction of motion, thus the structure is strongly anisotropic. (A possible explanation for this phenomenon might lie in the anatomical structure of this genera's visual apparatus.) According to the authors, the observed anisotropy is the effect of the interaction among the individuals. Adapted from \citet{BalleriniEtAl}.}
\end{figure}

Research has investigated various features of the group flight of birds, including positional effects on vigilance (mainly anti-predatory) \citep{Elgar89, Beauchamp03}, flock size, positional effects and intra-specific aggression in European starlings \citep{KeysD90}, landing mechanisms \citep{KunalekLanding}. Skeins of wild geese are famous for their characteristic V-shaped formations, which was spatio-temporally analyzed by \citet{JapiGeese}. By performing field measures, he observed long-term fluctuations with single-sided propagation through the string, and proposed a corresponding model as well. \citet{AnimalGrIn3D} published a remarkable collection of papers about the state of the art of the research on animal congregations in three dimensions. 
The most recent and impressive experimental observational study was carried out within the framework of a EU FP6 NEST project (Starflag, 2005-07). In this project the team measured the 3D positions of individual birds (European Starlings, \textit{Sturnus vulgaris}) within flocks containing up to 2,600 individuals, using stereometric and computer vision techniques (see Fig. \ref{figBallFlock}). They characterized the structure of the flock by the spatial distribution of the nearest neighbors of each bird. Given a reference bird, they measured the angular orientation of its nearest neighbor with respect to the flock's direction of motion, and repeated this process for all individuals within a flock as reference bird. Figure \ref{figBallStruct} depicts the average angular position of the nearest neighbors. The important main observation of the research team in Rome was that starlings in huge flocks interact with their 6-7 closest neighbors (``topological approach'') instead of those being within a given distance (``metrical approach''). Thus, they argued, the effect of density was quantitatively different in these (and probably most) flocks from that one would expect from models assuming a spatially limited interaction range \citep{BalleriniEtAl}. A topological flock model based on the above findings was recently considered by \citet{BodeTplgclMdl2011}.\par 

On the other hand, other experiments -- concerning various other species -- rendered the opposite view also probable, namely that the range of interaction did not change with density \citep{CouzinScience}. This question is still the subject of investigations, and it may easily lead to the conclusion that this mechanism differs from species to species.\par

\citet{Cavagna10} obtained high resolution spatial data of thousands of starlings using stereo imaging in order to calculate the response of a large flock to external perturbation. They were attempting to understand the origin of collective response, namely the way the group as a whole reacts to its environment. The authors argued that collective response in animal groups may be achieved through scale-free behavioral correlations. This suggestion was based on measuring to what extent the velocity fluctuations of different birds are correlated to each other. They found that behavioral correlations decay as a power law with a surprisingly small exponent, thereby providing each animal with an effective perception range much larger than the direct inter-individual interaction range. Further simulations are needed to clarify the origin of the experimental findings.\par

\begin{figure}
  \centerline{\includegraphics[angle=0,width=0.756\columnwidth]{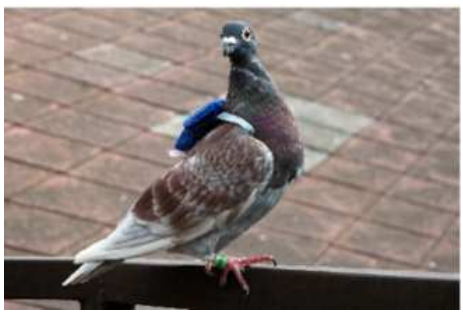}}
  \caption{\label{figPgnGPS}(Color online) A pre-trained homing pigeon with a small GPS device on his back, a recent technology to obtain information about the position of the individual birds. From \citet{HrarchGrDyn}.}
\end{figure}

A very recent direction (made possible by technological advances) is to obtain information about the position of individual birds during the observations using ultra light GPS devices (see Fig. \ref{figPgnGPS}). Although the present technology is still not suitable for large scale, high-precision studies (only a couple of birds per experiment and a resolution of the order of meters have been achieved yet), this method has already called to forth important results \citep{REtAlBirdGPS, AkosNagyV}.\par

Applying GPS data-loggers in six highly pre-trained pigeons, the efficiency of a flock was investigated by \citet{DellsLipp}. They found that the homing performance of the birds flying as a flock was significantly better than that of the birds released individually.
Employing high-precision GPS tracking of pairs of pigeons \citet{BiroEtAl06} found that if conflict between two birds' directional preferences was small, individuals averaged their routes, whereas if conflict arose over a critical threshold, the pair split or one of the birds became the leader.\par

\begin{figure}
  \centerline{\includegraphics[angle=0,width=0.7\columnwidth]{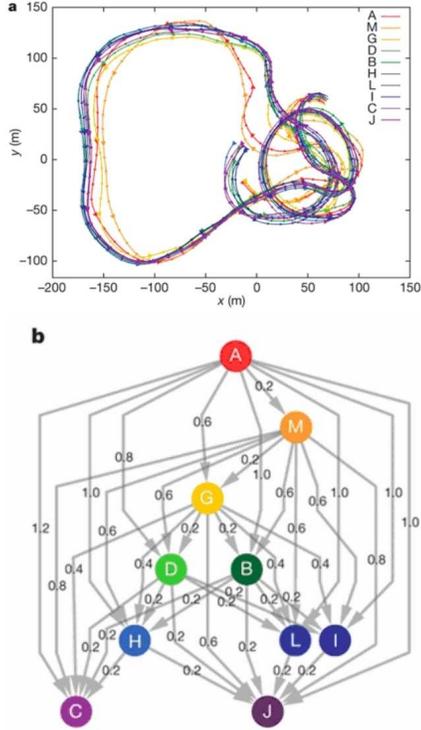}}
  \caption{\label{figBirdHrcy}(Color online) The route of a flight and the corresponding leadership-network of a pigeon flock. (a) A two-minute segment of the trajectory of ten pigeons recorded by small GPS devices (as depicted on Fig. \ref{figPgnGPS}). The different letters (and colors online) refer to the different individuals. The small dots on the lines indicate 1 second, the triangles indicate 5, and they point in the direction of the flight. (b) The leadership-network for the flight depicted on sub-figure (a). Each node (letter) represent a bird, among which the directed edges point from the leader to the follower. The numbers on the edges indicate the time delay (in seconds) in the two birds' motion. For those bird-pairs which are not connected directly with each other with an edge, directionality could not be resolved by means of the applied threshold. Adapted from \citet{HrarchGrDyn}.}
\end{figure}

Using a similar method, track-logs obtained from high-precision lightweight GPS devices, \citet{HrarchGrDyn} found a well-defined hierarchy among pigeons belonging to the same flock by analyzing data concerning leading roles in pairwise interactions (see Fig. \ref{figBirdHrcy}). They showed that the average spatial position of a pigeon within the flock strongly correlates with its place in the hierarchy.\par

\begin{figure}[t]
  \centerline{\includegraphics[angle=0,width=0.8\columnwidth]{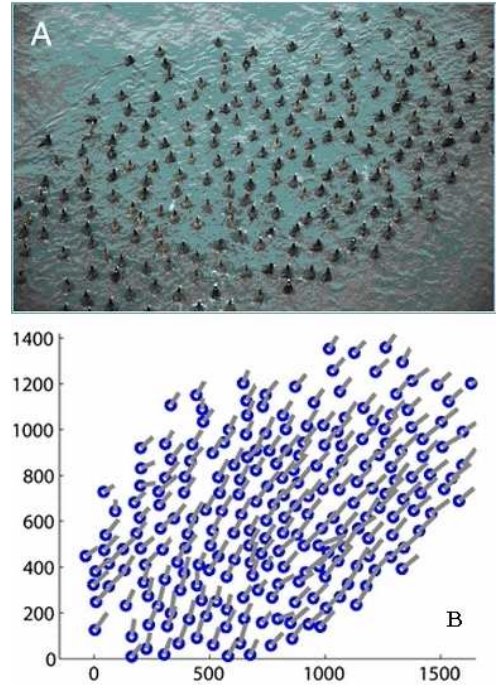}}
  \caption{\label{Lukeman}(Color online) A flock of surf scoter (M. perspicillata) swimming on the water surface (A). The actual coordinates and velocities after correction for perspective and drift currents effects. After \citet{Lukeman10}.}
\end{figure}

One of the long standing questions about the collective behavior of organisms is the measurement and interpretation of their positions relative to each other during flocking. Precise data of this sort would make the reconstruction of the rules of interaction between the individual organisms possible. In a very recent paper \citet{Lukeman10} carried out an investigation with this specific goal. They analysed a high-quality dataset of flocking surf scoters, forming well spaced groups of hundreds of individuals on the water surface (Fig. \ref{Lukeman}). \citet{Lukeman10} were able to fit the data to zonal interaction models (see, e.g., \citet{CouzinCollMem02}) and characterize which individual interaction forces suffice to explain observed spatial patterns. The main finding is that important features of observed flocking surf scoters can be accounted for by zonal models with specific, well-defined rules of interaction.

\subsection{Leadership in groups of mammals and crowds}\label{subsec:ExprmLeadership}

Many insect, fish and bird species live in large groups in which members are considered to be identical (from the viewpoint of collective motion), unable to recognize each other on an individual level (although not all, \citep{HrarchGrDyn}). Such groups might reach a consensus either without leader (by quorum response, mean value, etc.) or with a leader. However, even in this latter case, leadership is temporal since it is based on temporal differences, such as pertinent information of food location or differences in some inner states (hunger, spawning inducement, etc). As an important difference, most mammals do have the capacity for individual recognition enabling the emergence of hierarchical group structures. Although the assumption that the dominant individuals are at the same time the ones that lead the herd seems quite plausible, in fact, recent biological studies reveal that in many cases there is \emph{no} direct relationship between dominance and leadership. Most probably it is an interaction among kinship, dominance, inner state and some outer conditions. \par
Zebras, like many other mammals, need significantly more water and energy during the lactation period than they need otherwise. \citet{FischhoffZebra} investigated the effect of two factors, \emph{identity} and \emph{inner state}, on leadership in herds of zebras, \emph{Equus burchellii}. ``Identity'' covers both dominance and kinship, while the inner state was interpreted as the reproductive state (if the individual is in its lactation period or not). Zebra harems consist of tightly knit individuals in which females were observed to have habitual roles (``personal differences'') in the initiation of group movements. The authors also found that lactating females initiate movements more often than non-lactating ones, thus lactation, as inner state, plays an important role in leadership. Others find more direct relationship to hierarchy. \citet{SarovaCow} recorded the motion of a herd of 15 beef cows, \emph{Bos taurus}, for a three-week period using GPS devices. They found that short-distance travels and foraging movements are not lead by particular individual, instead, they are rather influenced in a \emph{graded} manner, i.e., the higher an individual was in the group hierarchy, the bigger influence it exerted on the motion of the herd. According to the observations, Rhesus macaques (\emph{Macaca mulatta}) preferred to join related or high-ranking individuals too, whereas Tonkean macaques (\emph{Macaca tonkeana}) exhibited no specific order at departure \citep{MakakoSueur08}. In a recent review article \citet{DMRulesPetit10} interpreted the process of collective decision making (regarding group movements) as a combination of two kinds of rules: `individual-based' and `self-organized'. The first one covers the differences of the [mostly inner] states of the animals, that is, differences in social status, physiology, energetic state, etc. The second one, self-organization, corresponds to the interactions, simple responses among individuals.\par
Regarding the case when leadership emerges solely from differences in the inner states of the group members (those ones lead who have pertinent information), \citet{CouzinNtrCollDcsn} suggested a simple model to show how a few informed individuals can lead a whole group. In this model (which is detailed in Sec. \ref{sec:LdrshpModels}) group members do not signal and do not know which of them (if any) has information regarding the desired direction. This model predicts that even if the portion of the informed individuals within the group is very small, the group as a whole can achieve great accuracy in its movement. In fact, the larger the group size, the smaller the portion of informed members are needed to lead the group. \citet{ConsInHum08} tested these predictions on \emph{human} groups in which the experimental subjects were na\"{i}ve and they did not use verbal communication or any other active signaling. The experiments indeed supported the predictions. Other experiments investigated the relationship between the spatial position of informed individuals and the speed and accuracy of the group motion \citep{LeaderSh09}. The results proved valid in larger crowds as well (100 and 200 people) which can have important implications on plans aiming to guide human groups for example in case of emergency.\par
\begin{figure}[t!]
  \centerline{\includegraphics[angle=0,width=0.65\columnwidth]{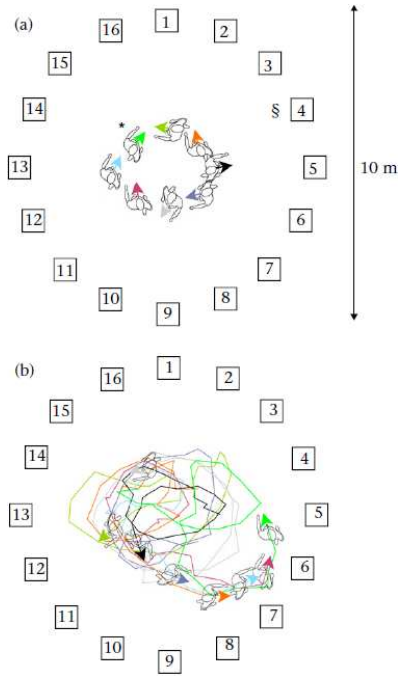}}
  \caption{\label{figHumanLdrFaria}(Color online) Orientations (arrows) and walking trajectories (lines) of the eight participants during the second treatment. The colors identify the participants. The leader -- marked by a '*' on subpicture (a) -- was instructed to reach the randomly selected target which is marked by a '\S'. (a) depicts the situation at the beginning of the trial and (b) 25 s later. Adapted from \citet{FariaHumanLdr10}.}
\end{figure}

\citet{FariaHumanLdr10} studied the effect of the knowledge regarding the presence and identity of a leader in small human groups, and also investigated those inadvertent social cues by which group members might identify leaders. With this object they conducted 3 treatments: in the \emph{first}, participants did not know that there was a leader, in the \emph{second} treatment they were instructed to follow the leader but they did not know who it was, while in the \emph{third} they knew who the leader was. The experiments took place in a circular area with $10 m$ diameter labeled by numbers from 1 to 16. These marks were spaced equally around the perimeter, as shown in Fig. \ref{figHumanLdrFaria}. In all the trials, participants were instructed (i) not to talk or to make any gesture, (ii) to walk continuously, and (iii) to remain together as a group. Further instructions were provided on a piece of paper: a (randomly chosen) person was asked to move to a (randomly chosen) target but stay with the group. She/he was the ``informed individual'', the ``leader''. The rest of the group was uninformed whose instructions differed from treatment to treatment: In the first one, they were only asked to stay with the group. In the second treatment they were told to follow the leader, but they did not know who it was. In the third one, they were asked to follow the leader whose identity was provided (by the color if his/her sash). Although the accuracy of the group movement significantly differed from treatment to treatment, the leader always succeeded to guide the group to the target. The least accurate group motions were measured during the first treatment, while the second and third ones resulted group motions whose accuracy were close to the possible maximum value. Three main factors were exposed as inadvertent social cues that might help uninformed group members to identify the leader(s): (i) time to start walking -- informed individuals usually started walking sooner, (ii) distance from the group center -- leaders were farther from the center than others, and (iii) proportion of time spent following -- informed people spent significantly less time following.

Many authors study animal groups from the viewpoint of a cost-benefit interpretation. They highlight that for an individual, living in a group brings more benefit than disadvantage, which is after all the ultimate reason for group formation. However, when reaching a consensus, if individuals differ in state and experience -- which is reasonable to assume -- then some individuals will have to pay bigger ``consensus costs'' than others (which is the coast that an individual pays by foregoing its optimal behavior to defer to the common decision \citep{BaboonKing08}). Theoretical models estimate ``democratic decisions'' less costly (in terms of average consensus cost) than ``despotic decisions'' \citep{ConradtRoperNtr03} which estimation is supported by a number of observations as well \citep{ConradtRoper05}. However, many animal groups (including primates and humans) often follow despotic decisions. Field experiments (for example on wild baboons \citep{BaboonKing08}) highlight the role of social relationships and leader incentives in such cases. From a more theoretical viewpoint, \citet{ConradtRoper10} discussed the cost/benefit ratio during group movements, separately for timing and spatial decisions.\par
In his recent book on collective animal behavior, \citet{SumpterBook} dedicated a whole chapter to decision making.


\subsection{Lessons from the observations}\label{subsec:ExperimLessions}

The main, commonly assumed advantages of flocking are:
\begin{enumerate}
	\item Defense against predators
	\item More efficient exploration for resources or hunting
	\item Improved decision making in larger groups (e.g., where to land)
\end{enumerate}
In general, it can be argued that with the increasing size of a group the process of decision making is likely to become more efficient \citep{SelfOrgBk, GrpDecSrv09, KrauseRuxton}. In addition, based on the numerous observations the following hypotheses can be made about the nature of the patterns of motion arising:
\begin{enumerate}
	\item Motion and a tendency to adopt the direction of motion of the neighbors is the main reason for ordered motion.
	\item Apparently the same, or very similar behaviors occur in systems of very different origin. This suggests the possibility of the existence of universal classes of collective motion patterns.
	\item Boundary conditions may significantly affect the  essential features of flocking.
	\item Collective decision making is usually made in a globally highly disordered, locally moderately ordered state (associated with a relatively slowly, but consistently decaying velocity correlation function) in which large scale mixing of the local information is enhanced.
	
\end{enumerate}


\section{Basic models}\label{sec:Models}

\subsection{Simplest self-propelled particles (SPP) models}\label{sec:OrigSPPMdl}
 
\begin{figure}
\centerline{\includegraphics[angle=0,width=1\columnwidth]{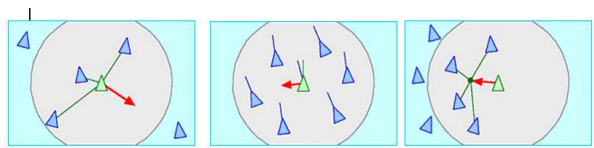}}
\caption{\label{figBoids} (Color online) The three basic steering behaviors determining the motion of the objects (called ``boids''). (a) \emph{Separation}, in order to avoid crowding local flock-mates. (Each boid reacts only to flock-mates within a certain neighborhood around itself, they are the ``local flock-mates''.) (b) \emph{Alignment}: objects steer towards the average heading direction of their local flock-mates. (c) \emph{Cohesion}: objects move toward the average position of their neighboring boids. From \texttt{http://www.red3d.com/cwr/boids/}.}
\end{figure}

Modeling of flocks has simultaneously been considered by the, initially somewhat divergent communities of computer graphics specialists, biologists and physicists. Perhaps the first widely-known flocking simulation was published by \citet{Reynolds87}, who was primarily motivated by the visual appearance of a few dozen coherently flying objects, among them imaginary birds and spaceships. His bird-like objects, which he called ``boids'', moved along trajectories determined by differential equations taking into account three types of interactions: avoidance of collisions, heading in the direction of the neighbors and finally, trying to stay close to the center of mass of the flock, as illustrated on Fig. \ref{figBoids}. The model was deterministic and had a number of relatively easily adjustable parameters. The website \texttt{http://www.red3d.com/cwr/boids/}, created and maintained until 2001 by Reynolds, is a unique source of links to all sorts of information (programs, demos, articles, visualizations, essays, etc.) related to group motion.\par

Reynolds's model shares features with an earlier simulation carried out by \citet{AokiEarlyMod}, who used the following rules (similar to those assumed by Reynolds) in order to simulate the collective motion of fish: (a) avoidance, (b) parallel orientation movements and (c) approach. The speed and direction of the individuals were considered to be stochastic, but the direction of the units was related to the location and heading of the neighbors (the velocity component, for the sake of simplicity, was considered to be independent of other individuals). In this pioneering paper of the field it was already stated that collective motion can occur without a leader and the individuals having information regarding the movement of the entire school.\par

In order to establish a quantitative interpretation of the behavior of huge flocks in the presence of perturbations, a statistical physics type of approach to flocking was introduced in 1995 by \citet{VM}, which nowadays is widely referred to as ``Vicsek Model'' (VM) e.g., \citep{albano1, BaglAlb09, VMKul, ChEurPhys08, VmChns1, JadbabaieIeee, GinRods09}. In the present paper we will refer to this approach as the ``SVM'', corresponding to Standard Vicsek Model as suggested in \citep{aldana3, SVMBertin}. In this model the perturbations, which are considered to be a natural consequence of the many stochastic and deterministic factors affecting the motion of the flocking organisms, are taken into account by adding a random angle to the average direction (Eq. \ref{eq:UjThi}). In this cellular-automaton-like approach of self-propelled particles (SPP-s) the units move with a fixed absolute velocity $v_0$ and assume the average direction of others within a given distance $R$. Thus, the equations of motion for the velocity ($\vec{v_i}$) and position ($\vec{x_i}$) of particle $i$ having neighbors labeled with $j$ are
 \begin{equation}
 \vec{v_i}(t+1)=v_0 \frac{ \left\langle \vec{v_j}(t) \right\rangle_R }{\left|   \left\langle \vec{v_j}(t) \right\rangle_R   \right|   }+perturbation
 \label{eq:UjVi}
 \end{equation}
 
 \begin{equation}
 \vec{x_i}(t+1)=\vec{x_i}(t)+\vec{v_i}(t+1)
 \label{eq:UjXi}
 \end{equation}
Here $\left\langle \ldots \right\rangle_R$ denotes averaging (or summation) of the velocities within a circle of radius $R$ surrounding particle $i$. The expression $\frac{ \left\langle \vec{v_j}(t) \right\rangle_R }{\left|   \left\langle \vec{v_j}(t) \right\rangle  \right|_R}$ provides a unit vector pointing in the average direction of motion (characterized by its angle $\vartheta_i(t)$ ) within this circle. It should be pointed out that the processes accounted for by such an alignment rule can be of very different origin (stickiness, hydrodynamics, pre-programmed, information processing, etc). 
 Perturbations can be taken into account in various ways. In the standard version they are represented by adding a random angle to the angle corresponding to the average direction of motion in the neighborhood of particle $i$. The angle of the direction of motion $\vartheta_i(t+1)$ at time $t+1$, is obtained from $\vartheta_i(t)=arc tan \left[\frac{<v_{j,x}>_R}{<v_{j,y}>_R} \right]$, as
 
 \begin{equation}
 \vartheta_i(t+1)=\vartheta_i(t)+\Delta_i(t),
 \label{eq:UjThi}
 \end{equation}
where $v_{j,x}$ and $v_{j,y}$ are the $x$ and $y$ coordinates of the velocity of the $j$th particle in the neighborhood of particle $i$, and the perturbations are represented by $\Delta_i(t)$, which is a random number taken from a uniform distribution in the interval $[-\eta \pi, \eta \pi]$ (i.e., the final direction of particle $i$ is obtained after rotating the average direction of the neighbors with a random angle). The only parameters of the model are the density $\rho$ (number of particles in a volume $R^d$, where $d$ is the dimension), the velocity $v_0$ and the level of perturbations $\eta < 1$. For order parameter $\varphi$, the normalized average velocity is suitable, $\varphi \equiv \frac{1}{N v_0}\left| \sum_{i=1}^N \vec{v}_i  \right|$, as defined by Eq. (\ref{eq:NormV}).

\begin{figure}
\centerline{\includegraphics[angle=0,width=0.64\columnwidth]{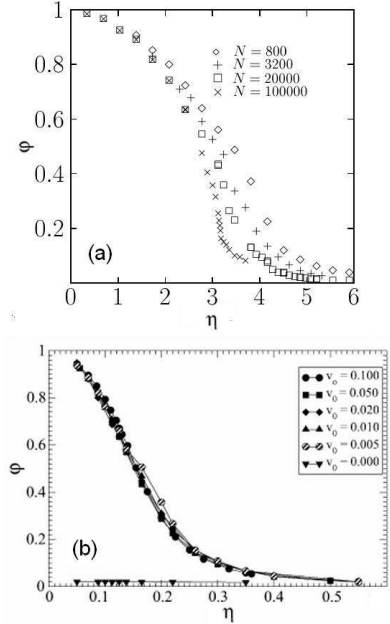}}
\caption{\label{figOrdParamFgv} Order parameter ($\varphi$) versus noise ($\eta$) in the SVM. (a) The different kind of points belong to different system sizes. (b) The different curves belong to different $v_0$ velocities, with which each particles move. As it can be seen, the concrete value of $v_0$ does not effect the nature of the transition (except when $v_0=0$, that is, when the units do not move at all). (a) is adapted from \citet{Csv97} and (b) is from \citet{albano2}.}
\end{figure}

This extremely simple model allows the simulation of many thousands of flocking particles and displays a second order type phase transition from disordered to an ordered (particles moving in parallel) state as the level of perturbations is decreased (see Fig. \ref{figOrdParamFgv}). At the point of the transition features of both order and disorder are simultaneously present leading to flocks of all sizes (and an algebraically decaying velocity correlation function).\par

\citet{CollMotElShimoyama96} proposed a mathematical model (neglecting noise) from which they obtained a categorization of the different types of collective motion patterns and determined the corresponding phase diagrams as well.\par

As we shall see in the upcoming sections, by varying some parameters, initial conditions and settings, the simulations exhibit a rich variety of collective motion patterns,  such as `marching groups', mills, rotating chains, bands, etc. (see Sec. \ref{SecVariants} and \ref{sec:ContMediaAndMF}). In some cases (i.e. by applying some certain parameter and initial condition settings) these ordered phases can exhibit some remarkable features as well, such as giant number fluctuations (GNF, see for example page \pageref{ChateGNF}, \pageref{RamGNF}) or band formation. According to the studies, noise, density, the type of interaction (attractive or repulsive, polar or apolar, the range of the interaction) and the boundary conditions (in case of finite size models) all proved to play an important role in the formation of certain patterns.

\subsubsection{The order of the phase transition}\label{sec:PhaseTrOrder}

In the paper introducing the original variant (OVM) of the SVM \citep{VM}, a second order phase transition from disordered to ordered motion was shown to exist. In particular, in the thermodynamic limit, the model was argued to exhibit a kinetic phase transition analogous to the continuous ones in equilibrium systems, that is, 
\begin{eqnarray}
	\varphi \sim \left[\eta_c(\rho)-\eta \right]^\beta  \nonumber\\
	\varphi \sim \left[\rho - \rho_c(\eta) \right]^\delta
\label{VMCrVals}	
\end{eqnarray}
which defines the behavior of the order parameter at criticality, in the case of a standard second order transition. 
$\beta$ and $\delta$ are critical exponents, $\eta$ is the noise (in the form of random perturbations), $\rho$ is the particle density, and $\eta_c(\rho)$ and $\rho_c(\eta)$ are the critical noise and critical density, respectively, for $L\rightarrow \infty$. ($L$ is the linear size of the system.)\par

However, the continuous nature of this transition has been questioned \citep{Gr04} resulting in a number of studies investigating this fundamental aspect of collective motion. Chat\'e and coauthors \citep{ChGinGrR08}, in their extensive follow-up study, presented numerical results indicating that there exists a ``crossover'' system size, which they call $L^{*}$, beyond which the discontinuous character of the transition appears independent of the magnitude of the velocity. They demonstrate that this discontinuous character is the ``true'' asymptotic behavior in the infinite-size limit. Importantly, \citet{ChGinGrR08} showed and presented results in favor of their picture that $L^{*}$ diverges in various limits: both the low and high density limits, as well as in the small velocity limit. In particular, an extrapolation of their estimates towards the small velocity regimes considered in prior works gives values of $L^{*}$ so large that do not make the corresponding simulations feasible.\par
Studies aiming to reveal the nature of the above phase transition (whether it is first or second order) find that the noise (more precisely, the way it is introduced into the system), and the velocity with which the particles move, play a key role. Accordingly, while simulations show that for relatively large velocities  $(v_0 > 0.5)$ the transition is discontinuous, \citet{albano2} demonstrated that for smaller velocities, even in the limit when the velocity goes to zero (except when it is exactly equal to zero), the transition to ordering is continuous (is independent of the actual value of the velocity, as it can be seen in Fig. \ref{figOrdParamFgv} b). Very recently \citet{Ihle10} and \citet{MisBaskranMarchetti10} were able to see band-like structures in their solutions obtained from a continuum theory approach. Bands usually signal first order phase transition, however, \citet{Ihle10} found them for a large velocity case, while \citet{MisBaskranMarchetti10} assumed throughout their calculation that the transition from disorder to order was continuous.\par
\citet{AldanaPhsTrans07} demonstrated that the type of the phase transition depends on the way in which the noise is introduced into the system. They analyzed two network models that capture some of the main aspects characterizing the interactions in systems of self-propelled particles. In the so called  ``vectorial noise model'' the perturbation (in the form of a random vector) is first added to the average of the velocities and the final direction is determined only after this \citep{Gr04}. When the average velocity is small (disordered motion) this seemingly subtle difference in the definition of the final direction leads to a qualitatively different ordering mechanism (sudden -- first order-type -- transition to the ordered state).\par
Correspondingly, \citet{Aldana09} analyzed the order-disorder phase transitions driven by two different kinds of noises: ``intrinsic'' (the original form, perturbing the final angle) and ``extrinsic'' (the vectorial one, perturbing the direction of the individual particles before averaging). Intrinsic is related to the decision mechanism through which the particles update their positions, while extrinsic affects the signal that the particles receive from the environment. The first one calls continuous phase transitions forth, whereas the second type produces discontinuous phase transitions \citep{AldanaLongPahsetransit}. Finally, \citet{NagyDaruka07} showed that vectorial noise results in a behavior which can be associated with an instability.\par

\subsubsection{Finite size scaling}

So far, the most complete study regarding the scaling behavior of systems of self-propelled particles exhibiting simple alignment plus perturbation, has been carried out by \citet{albano1}. They performed extensive simulations of the SVM, and analyzed them both by a finite-size scaling method (a method used to determine the values of the critical exponents and of the critical point by observing how the measured quantities vary for different lattice sizes), and by a dynamic scaling approach. They observed the transition to be continuous. In addition they demonstrated the existence of a complete set of critical exponents for the two dimensional case (including those corresponding to finite size scaling\footnote{Numerical simulations carried out on systems having finite size $L$ in at least one space dimension exhibit so called \emph{finite size effects}, most importantly rounding and shifting effects during second-order phase transitions. These artifacts are particularly emphasized near the critical points, but they can be accounted for by means of the so called \emph{finite-size scaling}. See more on this topic in \citep{StatFizCardy, StatFizBrankov}.}) and numerically determined their values as well. In particular, within the framework of finite-size scaling theory, the scaling ansatz for the order parameter $\varphi$ of the SVM has been rewritten as

\begin{equation}
	\varphi(\eta, L)=L^{-\beta/\nu}\tilde{\varphi}(\eta-\eta_c)L^{1/\nu},
\label{eq:FSSOrdPrm}
\end{equation}
where $L$ is the finite size of the system, $\tilde{\varphi}$ is a suitable scaling function, and finally, $\beta$ and $\nu$ are two of the critical exponents in question: $\beta$ is the one belonging to the order parameter, and $\nu$ is the correlation length critical exponent.\par

Similarly, the fluctuation of the order parameter, $\chi=\sigma^2L^2$, takes the form
\begin{equation}
	\chi(\eta, L)=L^{\gamma/\nu}\tilde{\chi}((\eta-\eta_c)L^{1/\nu}),
\label{eq:FSSSusc}
\end{equation}
where $\tilde{\chi}$ is a suitable scaling function, $\gamma$ is the susceptibility critical exponent, and $\sigma^2\equiv \left\langle \varphi^2 \right\rangle - \left\langle \varphi \right\rangle^2$ is the variance of the order parameter. In the thermodynamic limit, $\chi$ obeys $\chi \sim (\eta-\eta_c)^{-\gamma}$. (See also Eq. (\ref{eq:ChiDlt}))\par
Equations (\ref{eq:FSSOrdPrm}) and (\ref{eq:FSSSusc}) are convenient to determine the critical exponents within the framework of finite size scaling theory. As a crucial result, the authors found that the exponents they calculated satisfy the so-called \emph{hyperscaling relationship}
 \begin{equation}
 		d\nu-2\beta=\gamma
 \label{eq:HypRel}
 \end{equation}
which is, in general, valid for standard (equilibrium) critical phenomena. $d$ denotes the dimension, $d=2$.
\par

The nature of ``intermittency'' -- intermittent bursts during which the order is temporarily lost in such systems -- has also been a subject of investigations recently \citep{HuepeIntrmtncy}.

\subsection{Variants of the original SPP model}\label{SecVariants}

Several variants of the above-introduced, simplest SPP model have been proposed over the years. One of the main directions comprises those studies that investigate systems in which the particles (units) do not follow any kind of explicit alignment rule, only collisions occur between them in the presence of some kind of interaction potential. We shall overview this approach in Sec. \ref{sec:MdlWithoutAlgnt}. Models assuming some kind of alignment rule for the units, will be dealt with in Sec. \ref{sec:MdlWithAlgnt}.

\subsubsection{Models without explicit alignment rule}\label{sec:MdlWithoutAlgnt}

As mentioned in Sec. \ref{sec:OrigSPPMdl}, in the most simple SPP models, an alignment term is assumed. However, according to very recent studies (see Sec. \ref{subsec:ExperimPhysChem}), the motion of particles may become ordered even if no explicit alignment rule is applied, but alignment is introduced into the collision in an indirect way by the local interaction rules. The simplest (most minimal) model of ordered motion emerging in a system of self-propelled particles looks like this: The particles are trying to maintain a given absolute velocity and the only interaction between them is a repulsive linear force $(\vec{F})$ within a short distance (i.e., they do not ``calculate'' the average of the velocity of their neighbors, and the only interaction is through a pair-wise central force). The corresponding equations are:

\begin{equation}
\label{eq:ViDeriv}
	\frac{d\vec{v}_i}{dt}=\vec{v}_i\left(\frac{v_0}{|\vec{v}_i|}-1\right)+\vec{F}_i + \vec{\xi}_i
\end{equation}

where $\vec{\xi}_i$ is noise (random perturbations, typically white noise)

\begin{equation}
	\label{eq:FiVec}
	\vec{F}_i = \sum_{i \neq j} \vec{F}_{ij} + \vec{F}_i \mbox{  (wall)}
\end{equation}

\begin{equation}
	\label{eq:rijVec}
	\vec{r}_{ij} = \vec{x}_i - \vec{x}_j
\end{equation}

\begin{equation}
\label{eq:FijEsetek}
\vec{F}_{ij} = \left\{
	\begin{array}{ll}
		C \vec{r}_{ij} (\frac{r_0}{|\vec{r}_{ij}|}-1)& \mbox{,if $|\vec{r}_{ij}| \leq r_0$, and}\\
		0 & \mbox{,otherwise}
	\end{array}	
\right.
\end{equation}

\begin{figure}
\centerline{\includegraphics[angle=0,width=.75\columnwidth]{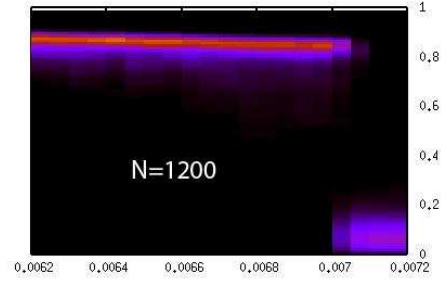}}
\caption{\label{figDerzsi} (Color online) Probability density distribution of the order parameter versus noise for 1200 particles. The first order nature of the transition is indicated by the behavior of the order parameter, depicted on the vertical axis, which abruptly falls, in this case at noise level 0.007. From \texttt{http://hal.elte.hu/\~{ }vicsek/SPP-minimal/}.}
\end{figure}

Simulations of the above minimal model result in a first order transition from disordered to coherent collective motion \citep{DerzsiMinimalHp}, as it can be seen in Fig. \ref{figDerzsi}.\par
\begin{figure}
\centerline{\includegraphics[angle=0,width=1\columnwidth]{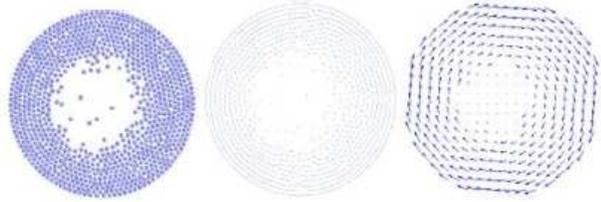}}
\caption{\label{figGrsmn} (Color online) Vortex formation in a reflective round boundary. Reflecting boundaries cause particles to move parallel to them. Both clockwise and counter clockwise vortices can form according to the randomly chosen initial direction. (a) A snapshot of the particles ($N=900$). (b) Their movement within a short period of time. (c) Coarse graining average velocity. From \citet{GrossmanEtAl08}.}
\end{figure}

Analogous results were recently obtained for another simple model assuming only a specific form of inelastic collisions between the particles \citep{GrossmanEtAl08}. In their numerical experiments, self-propelled isotropic particles move and collide on a two-dimensional frictionless flat surface. Imposing reflecting boundary-conditions produce a number of collective phenomena: ordered migration, the formation of vortices (see Fig. \ref{figGrsmn}) and random chaotic-like motion of subgroups. Changing the particle density and the physical boundary of the system -- for example from a circular to an elliptical shape -- again results in different types of collective motion; for certain densities and boundary-types the system exhibits nontrivial spatio-temporal behavior of compact subgroups of units. The reason why coherent collective motion appears in such a system is that each of these inelastic collisions between isotropic particles induce alignment, resulting in an increased overall velocity correlation (it can be shown that the collisions do not preserve the momentum, but lead to at least a slight increase each time). Such numerical experiments are fundamental in clarifying the question regarding the minimal requirements for a system to exhibit collective motion, based solely on physical interactions.\par

\begin{figure}
\centerline{\includegraphics[angle=0,width=1\columnwidth]{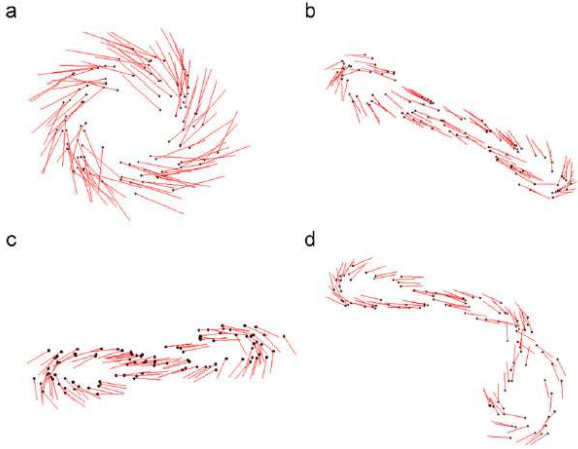}}
\caption{\label{StrombomPatterns} (Color online) (a) A ``mill'', (b) a ``rotating chain'' without intersection, (c) with one junction, and (d) with two self-intersections. From \citet{Strombom2011}.}
\end{figure}

\citet{Strombom2011} also considered an SPP model in which only one kind of social interaction rule was taken into account: attraction. By using simulations he found a variety of patterns, such as swarms (a set of particles with low and varying alignment), undirected mills (a group in which the particles move in a circular path around a common center) and moving aligned groups (in which the units move in a highly aligned manner). Importantly, these structures were stable only in the presence of noise. Introducing a blind angle (which is the region behind each unit in which other particles are ``invisible'', and which, accordingly,  incorporates some sort of alignment into the system) had a fundamental effect on the emergent patterns: undirected mills become directed, and ``rotating chains'' appeared (see Fig. \ref{StrombomPatterns}). In these chains the units move on a closed curve having zero (Fig. \ref{StrombomPatterns}b), one (Fig. \ref{StrombomPatterns}c) or two (Fig. \ref{StrombomPatterns}d) junctions. These formations are called ``rotating'', because the chains with zero or two intersections often rotate around a slowly moving axis. (Some corresponding videos can be seen on the author's webpage, \texttt{https://sites.google.com/site/danielstrmbm/} \texttt{research}.)

The first work in which the relevance of the simultaneous presence of volume exclusion and self-propulsion for an effective alignment of the particles was published by \citet{PeruaniDB06}. They stressed the importance of the particle shape by showing that self-propelled objects moving in a dissipative medium and interacting by inelastic collision, can self-organize into large coherently moving clusters. Their simulations have direct relevance to the experiments on shaken rods \citep{KudrlDiff} and on the collective motion patterns by mixobacteria \citep{ReverseBaci}. Furthermore, \citet{PeruaniDB06} showed that self-propelled rods exhibit non-equilibrium phase transition between a monodisperse phase to an aggregation phase that depends on the aspect ratio and density of the self-propelled rods. To see all this, no specific boundary conditions had to be applied due to the elongated shape of the particles.\par
\begin{figure}
\centerline{\includegraphics[angle=0,width=.95\columnwidth]{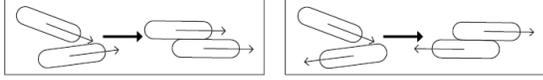}}
\caption{\label{figNematicAlgnmt}Nematic collision means that, if the included angle of the two velocity vectors belonging to the colliding rod-like units was smaller than $180^{\circ}$ before they impinge on each other, they would continue their motion in the same direction, in parallel, after the collision. If this angle was bigger than $180^{\circ}$, then they would continue their travel in parallel, but in the opposite direction. From \citet{GinRods09}.}
\end{figure}
\begin{figure}
\includegraphics[angle=0,width=1\columnwidth]{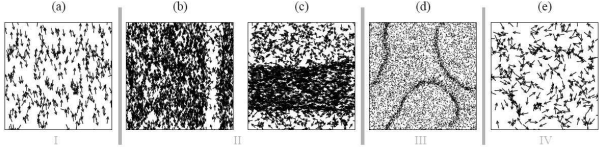}
\caption{\label{figGinelliPhases} Steady-state snapshots for the four different phases observed by \citet{GinRods09}. The linear size is $L=2048$. Arrows show the orientation of particles (except in (d)). The phases belong to different noise values: (a) $\eta=0.08$, (b) $\eta=0.1$, (c) $\eta=0.13$, (d) $\eta=1.168$ and (e) $\eta=0.2$. The order-disorder transition occurs between phases II and III. Both of these phases are characterized by spontaneous segregations into bands, but in phase III these bands are thinner and are more unstable. From \citet{GinRods09}.}
\end{figure}

In a similar spirit, \citet{GinRods09} investigated in more detail the properties of a collection of elongated, asymmetric (``polar'') units moving in two dimensions with constant speed, interacting only by ``nematic collisions'', in the presence of noise. Nematic collision, illustrated on Fig. \ref{figNematicAlgnmt}, means the following: if the included angle of the two velocity vectors belonging to the colliding rod-like units was smaller than $180^{\circ}$ before they impinge on each other, they would continue their motion in the same direction, in parallel, after the collision. If this angle was bigger than $180^{\circ}$, then they would continue their travel in parallel, but in the opposite direction. Four phases were observed, depending on the strength of the noise (labeled I to IV by increasing noise-values, see Fig. \ref{figGinelliPhases}). Phase I is spatially homogeneous and ordered, from which phase II differs in low-density disordered regions, which appear in the steady state. The order-disorder transition occurs between phases II and III. Both of these phases (II and III) are characterized by spontaneous segregations into bands, but in phase III these bands are thinner and are more unstable, constantly bending, breaking, reforming and merging, displaying a persistent space-time chaos. Phase IV is spatially homogeneous with global and local disorder on small length and timescales.\par

In the large scale experiments of \citet{Filaments} propagating density waves were observed  and produced by the related simulations as well (see also Sec. \ref{subsec:ExperimPhysChem}). Such waves are also generated by the model of \citet{VM} in the large velocity limit (see \citet{NagyDaruka07}), where large means that the jumps made by the particles between two updates are compatible or larger than the interaction radius. For these parameter values the trajectories of two particles can cross each other without an interaction taking place. And, indeed, this is what happens in a large number of cases in the motility assay. Sometimes the filaments align, some other times they simply cross each others' trajectories. Furthermore, if in the simulational model the parameters are chosen in such a way that crossing cannot occur (this limit corresponds to the low velocity case), the waves do not show up any more (Schaller, private communication).\par

A swarm of identical self-propelled particles interacting via a harmonic attractive pair potential in two dimensions in the presence of noise was also considered. By numerical simulations \citet{Erdmn05} found that, if the noise is increased above a certain limit, a transition occurs during which the translational motion breaks down and instead of it, rotational motion takes shape.

\subsubsection{Models with alignment rule}\label{sec:MdlWithAlgnt}

Units in every system exhibiting any kind of collective motion (or more generally, collective behavior) \emph{interact} with each other. In the original SVM, this interaction occurs in the so called ``metric'' way, that is, each unit interacts only with those particles which are closer than a pre-defined distance, called ``range of interaction''. An alternative to this approach is the ``topological'' representation, in which each particle communicates with its $n$ closest neighbour (a typical value for $n$ is around 6-7.) These approaches are closely related, since by varying the range of interaction (or if it is set to be unit, as in most cases, then by varying the particle density) the number of the nearest neighbours, with whom a unit communicates, can be -- at least in average -- adjusted. The important difference here is that since in the metric approach the density can be prescribed, thus, the number of the particles falling in the range of interaction might change as well. There is a reoccurring subtle point here. \citet{GinelliChate2010} compared the two approaches and pointed out the main difference, because in the topological distance model they obtain a second order phase transition to order, while they claim that in the SVM model the nature of the transition is of first order. We have discussed this point in Sec. \ref{sec:PhaseTrOrder} and argued that the resolution for the controversy lies in the very specific feature of the SVM, i.e., in the metric model for low velocities the transition is continuous (like in the topological model), while for large velocities it is of first order (see \citep{NagyDaruka07}).\par

Compared to the original SVM, an important additional feature has been introduced by \citet{Gr03} who added adhesion between the particles to avoid ``evaporation'' of isolated clusters in simulations with open boundary conditions. Adding this new feature has changed the universality class (order of transition) and the observed ordering was discontinuous as a function of perturbations.\par
\begin{figure}
\centerline{\includegraphics[angle=0,width=.6\columnwidth]{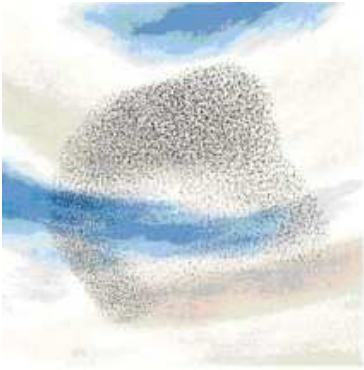}}
\caption{\label{figChAdh}(Color online) The SVM augmented with cohesive interactions among the particles. A snapshot of a flock consisting of 16,384 particles, moving `cohesively'. Adapted from \citet{ChEurPhys08}.}
\end{figure}

The most common way to introduce cohesion to a system without resorting to global interactions, is to 
complement the interaction rules defining the units' behavior with some kind of pairwise attraction-repulsion mechanism. In this spirit, \citet{ChEurPhys08} have added a new term to Eq. (\ref{eq:UjVi}), which determines a pairwise attraction-repulsion force between the particles (See Fig. \ref{figChAdh}).\par
Another generalization has been considered by \citet{SzaboCplng}. By extending the factors influencing the ordering, the model assumes that the velocity of the particles depends both on the velocity and the acceleration of neighboring particles. (Recall, that in the original model it depends solely on the velocity). Changing the value of a weight parameter determining the relative influence of the velocity and acceleration terms, the system undergoes a kinetic phase transition. Below a critical value the system exhibits disordered motion, while above the critical value the dynamics resembles that of the original SPP model.\\

\begin{figure}
\centerline{\includegraphics[angle=0,width=.5\columnwidth]{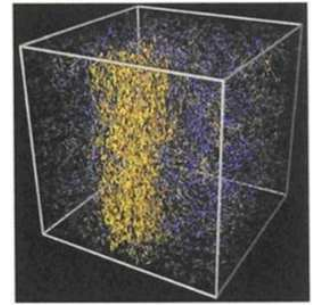}}
\caption{\label{figChElong}(Color online) A snapshot of the simulations with point-like particles, but subject to a nematic-type interaction, performed by \citet{ChEurPhys08}. The behavior of the system is qualitatively different from those with isotropic particles and exhibits characteristic density and velocity fluctuations. Color code refers to the local denseness from blue (low density) to yellow (high density). Adapted from \citet{ChEurPhys08}.}
\end{figure}

One might interpret the particles of the SVM as \emph{polar} units, since they carry a velocity vector. Accordingly, \citet{ChateNmtcs} consider a bipolar version of the SVM, in which after the angle corresponding to the local average velocity is determined, the particles can `decide' whether they move along this direction or in a direction opposite to it. Such a model arises from the consideration of the self-propelled motion of elongated particles preferably moving along their main axis. The authors find a distinctively different disorder-order transition involving giant density fluctuations (GNF),\label{ChateGNF} compared to the previously considered cases. According to \citet{ChEurPhys08}, the expression $\left\langle \vec{v_j}(t) \right\rangle_{S_i}$ appearing in the interaction rule of the SVM (Eq. (\ref{eq:UjVi})) -- expression which is in close relation to the \emph{local} order parameter around particle $i$ in its neighborhood $S$ -- can be replaced by the eigenvector of the largest eigenvalue belonging to the nematic tensor calculated on the same neighborhood. Denoting the angle defining the direction of $\vec{v}_j$ by $\theta_j$, this eigenvalue, which is also directly related to the local order parameter, for uniaxial nematics in two space dimensions is calculated as $| \left\langle exp (2i\theta_j(t))\right\rangle_{S_i} |$. $j$ denotes those particles that are within the neighborhood $S$ of particle $i$, $S_i$, at time-step $t$. 
Since each particle $i$ chooses the direction defined by $\theta_i$ or the opposite direction $\theta_i + \pi$ with the same probability $1/2$, Eq. (\ref{eq:UjXi}) gets the form

\begin{equation*}
\vec{x_i}(t+1)=\vec{x_i}(t) \pm \vec{v_i}(t+1)
\end{equation*}\par

A snapshot of the resulting collective motion pattern can be seen in Fig. \ref{figChElong}.\par

By using a novel set of diagnostic tools related to the particles' spatial distribution \citet{aldana3} compared three simple models qualitatively reproducing the emergent behavior of various animal swarms. The most important aim of introducing the above measures is to unveil previously unreported qualitative differences and characteristics (which were unclear) among the various models in question. Comparing only the standard order parameters (measuring the degree of alignment), the authors find very similar order-disorder phase transitions in the investigated models, as a function of the noise. They demonstrated that the distribution of cluster sizes is typically exponential at high noise-values, approaches a power-law distribution at reduced noise levels, and interestingly, that this trend is sometimes reversed near to the critical noise value, suggesting a non-trivial critical behavior.\par

\citet{SmithM09} used a Lagrangian individual based model with open boundary conditions to show that the Morse and the Lennard-Jones potentials (coupled with an alignment potential) are also capable to describe many aspects of flocking behavior.

\begin{figure}
\centerline{\includegraphics[angle=0,width=\columnwidth]{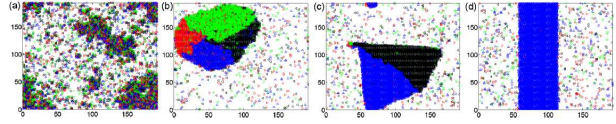}}
\caption{\label{figPrnPhases}(Color online) As the particles in the model exhibit an increasing tendency to align, different pattern arise: (a) orientational disorder, while particles self-segregate. (b) traffic jam, (c) glider, and (d) band. The colors code the four possible orientation: red is for `right', `left' is depicted with black color, green marks `down', and `up' is coded with blue. From \citet{PeruaniPRL2011}.}
\end{figure}

With the accumulation of experimental data and modeling results within this field, it is becoming more and more clear that very simple local interaction rules can produce a huge variety of patterns within the same system in a way that the type of the emerging pattern depends only on a few parameters. 
Recently \citet{PeruaniPRL2011} recorded various kinds of self-organized spatial patterns by using a simple model: a two-dimensional lattice with volume exclusion. In a \emph{lattice}, ``volume exclusion'' means that a node could be occupied by at most one particle, and also the rotational symmetry is broken which otherwise characterizes all models assuming continuum spatial dimensions. The units had the tendency to align `ferromagnetically'. As the susceptibility of the particles to align to their neighbors increased, the system went through some distinct phases (see Fig. \ref{figPrnPhases}): first, for weak alignment strength the units self-segregated into disordered aggregates (Fig. \ref{figPrnPhases}a), which then, by strengthening the alignment, turns into locally ordered, high density regions, which the authors call ``traffic jams'' (Fig. \ref{figPrnPhases}b). By further enhancing the susceptibility of alignment, triangular high density aggregates emerge (called ``gliders'') that migrate in a well-defined direction (Fig. \ref{figPrnPhases}c). Finally, these structures self-organize into highly-ordered, elongated high density regions: bands (Fig. \ref{figPrnPhases}d).

\subsection{Continuous media and mean-field approaches}\label{sec:ContMediaAndMF}

Self-propelled particles, during their motion, consume energy and dissipate it in the media they move in, meanwhile performing rich collective behavior at large scales. Recent studies devoted to deriving hydrodynamic equations for specific microscopic models have led to new ideas and approaches within this field.\par
The continuous media approaches to collective motion have been carried out, on one hand, in the context of giving a macroscopic description of SPP systems, however, at the same time, also to interpret the so called  ``active matter'' systems associated mainly with  applications from physics, such as active nematics and active suspensions.\par
The first theory describing the full nonlinear higher dimensional dynamics was presented in \citep{TonerTu95, TonerTu98}. Tu and Toner followed the historical precedent \citep{ForsterNSt} of the Navier-Stokes equation by deriving the continuum, long-wavelength description \emph{not} by explicitly coarse graining the microscopic dynamics, but, rather, by writing down the most general continuum equations of motion for the velocity field $\vec{v}$ and density $\rho$ consistent with the symmetries and conservation laws of the problem. This approach allows to introduce a few phenomenological parameters (like the viscosity in the Navier-Stokes equation), whose numerical values will depend on the detailed microscopic behavior of the particles. The terms in the equations describing the large-scale behavior, however, should depend only on symmetries and conservation laws, and \emph{not} on the microscopic rules.\par
The only symmetry of the system is rotation invariance: since the particles lack a compass, all direction of space are equivalent to other directions. Thus, the ``hydrodynamic'' equation of motion cannot have built into it any special direction picked \emph{a priori}; all directions must be spontaneously selected. Note that the model does \emph{not} have Galilean invariance: changing the velocities of all the particles by some constant boost $\vec{v}_b$ does \emph{not} leave the model
invariant.\par
To reduce the complexity of the equations of motion still further, a spatial-temporal gradient expansion can be performed keeping only the lowest order terms in gradients and time derivatives of $\vec{v}$ and $\rho$.  This is motivated and justified by the aim to consider \emph{only} the long distance, long time properties of the system. The resulting equations are
\begin{multline}
\partial_t \vec{v}+\lambda_1(\vec{v}\nabla)\vec{v}+\lambda_2(\nabla \vec{v})\vec{v}+\lambda_3 \nabla (|\vec{v}|^2)=\\
\alpha\vec{v}-\beta|\vec{v}|^2\vec{v} -\nabla P+D_L\nabla(\nabla\vec{v}) + D_1\nabla^2\vec{v} + D_2(\vec{v}\nabla)^2\vec{v}+\vec{\xi}
\label{eq:TTHosszu}
\end{multline}
and
\begin{equation}
\partial_t \rho + \nabla(\rho\vec{v})=0.
\label{eq:TTMasCons}
\end{equation}

In Eq. (\ref{eq:TTHosszu}), the terms $\alpha, \beta>0$ give $v$ a nonzero magnitude, $D_{L,1,2}$ are diffusion constants and $\vec{\xi}$ is an uncorrelated Gaussian random noise. The $\lambda$ terms on the left hand side of the equation are the analogs of the usual convective derivative of the coarse-grained velocity field $\vec{v}$ in the Navier-Stokes equation. Here the absence of Galilean invariance allows all \emph{three} combinations of one spatial gradient and two velocities that transform like vectors; if Galilean invariance \emph{did} hold, it would force $\lambda_2=\lambda_3=0$ and $\lambda_1=1$. However, Galilean invariance does \emph{not} hold, and so all three coefficients can be non-zero phenomenological parameters whose values are determined by the microscopic rules.
Eq. (\ref{eq:TTMasCons}) reflects the conservation of mass (birds). The pressure $P$ depends on the local density only, as given by the expansion
\begin{equation}
P=P(\rho)=\sum_{n=1}^\infty\sigma_n(\rho-\rho_0)^n
\label{eq:TTPrsr}
\end{equation}
where $\rho_0$ is the mean of the local number density and $\sigma_n$ is a coefficient in the pressure expansion.\par
It is possible to treat the whole problem analytically using dynamical renormalization group and show the existence of an ordered phase in 2D, and extract exponents characterizing the velocity-velocity and density-density correlation functions \citep{TonerTu98}. The most dramatic result is that an intrinsically non-equilibrium and
nonlinear feature, namely, convection, suppresses fluctuations of the velocity $\vec{v}$ at long wavelengths, making them much smaller than the analogous fluctuations found in ferromagnets, for all spatial dimensions $d<4$.
In other words, the existence of the convective term makes the dynamics ``non-potential'' and further stabilizes the ordered phase. Heuristically, this term accounts for the stabilization effect resulting from the feature that the actual neighbours of each unit continually change due to the local differences in the direction. Thus, particles (birds) which initially were not neighbours, and thus did not interact with each other, at a later time-step might be within each other's interaction range.\par
Further predictions of the above model were tested by numerically studying a discrete model very similar to the SVM. Compared to the original model, an extra interaction term was introduced, in order to prevent cluster formation:
\begin{equation}
\vec{g}_{ij} = g_0 \left( \vec{r_i}-\vec{r_j} \right) \left( \left( \frac{l_0}{r_{ij}} \right)^3 - \left( \frac{l_0}{r_{ij}} \right)^2 \right)
\label{eq:ClustFormPrev}
\end{equation}
This expression sets the average distance between boids in the flock to be $l_0$ by creating an attraction force between particles $i$ and $j$ if the $r_{ij}$ distance between them is bigger than than $l_0$, and making them repel each other if $r_{ij} < l_0$. $g_0$ is a model parameter, defining the strength of the above mentioned attraction-repulsive force. The linear size $L$ of the simulated system was $L=400$ with $N=320,000$ boids moving in it. The range of interaction was set to be unit (1), $g_0=0.6$, the velocity $v_0=1.0$ and $l_0=0.707$. In finite size systems, the average direction of the flock $<\vec{v}>$ slowly changes in time, due to the noise. In contrast, analytic results assume infinite systems size in which the direction $<\vec{v}>$ is constant. In order to handle this disagreement, the boundary condition in one direction (say the $x$) was set to be periodic, while in the other (the $y$) a reflective boundary condition was applied. Hence, the symmetry broken velocity was forced to lie along the $x$ direction. As an interesting phenomenon, in the direction $y$ (the one perpendicular to $<\vec{v}>$) the individual boids exhibit an anomalous diffusion. Furthermore, as an even more surprising result, in the flock's moving direction ($x$), the fluctuations of the velocity and the that of the density were propagating with different velocity.\par
\citet{SVMBertin2, SVMBertin} made a very important step towards a fundamental theory of collective motion by deriving the hydrodynamic equations for the density and velocity fields of a gas of self-propelled particles with binary interactions from the corresponding microscopic rules. They gave explicit expressions for the transport coefficients as a function of the microscopic parameters. Comparison with numerical simulations on a standard model of self-propelled particles (SVM, see  Sec. \ref{sec:OrigSPPMdl}) resulted in an agreement as well as in a demonstration of the robustness of the phase diagram they obtained. \citet{Ihle10} showed how to explicitly coarse grain the microscopic dynamics of the SVM to obtain expressions for all transport coefficients as a function of the three main parameters, noise, density and velocity. 

\begin{figure}
\centerline{\includegraphics[angle=0,width=0.75\columnwidth]{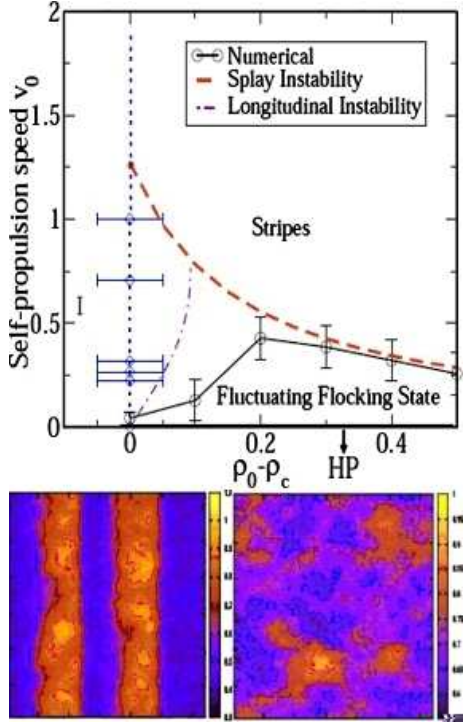}}
\caption{\label{fig:MishraEtAl} (Color online) (A) Phase diagram of the solutions of the Eqs. (\ref{eq:TTHosszu}) and (\ref{eq:TTMasCons}) in the $(v_0, \rho_0)$ plane. At $v_0=0$ the system exhibits a continuous mean-field transition at $\rho_0=\rho_c$ from an isotropic (I) to a homogeneous polarized (HP) state. For  $\rho_0>\rho_c$ there is a critical $v_c(\rho_0)$ separating a polarized moving state with large anomalous fluctuations, named the fluctuating flocking state, at low self-propulsion speed from a high-speed phase of traveling stripes. The circles denote the values of $v_c(\rho_0)$ obtained numerically. The dashed-dotted line (purple online) is the longitudinal instability boundary  $v_{c1}^L(\rho_0)$ obtained in the calculations. The dashed line is the splay instability boundary  $v_c^S(\rho_0)$. (B) shows a snapshot of the density profile in the striped phase. The stripes travel horizontally. (C) shows a snapshot of the density profile in the coarsening transient leading to the fluctuating flocking state at  $v_0<v_c$. Density values grow from dark to light. After \citet{MisBaskranMarchetti10}.}
\end{figure}

Over the last 4 years, the hydrodynamic equations became increasingly precise by including higher order terms and more precise coefficients. Very recently  \citet{MisBaskranMarchetti10} solved the equations describing the collective motion of self-propelled polar rods moving on an inert substrate. From their theoretical considerations and numerical analysis, the authors obtained a remarkable phase diagram for this system (Fig. \ref{fig:MishraEtAl}). They showed that the same physics that leads to global ordering destabilizes the homogeneous ordered state above a critical value of self-propulsion speed and allows the nonlinear equations to admit a propagating front solution that yields the striped phase identified numerically. The two phases they observed, namely, the striped phase and the fluctuating flocking phase, have been identified earlier in the context of numerical studies of the SVM model, thus, the approach of \citet{MisBaskranMarchetti10} identified the origin of these phenomena in the model independent framework of the dynamics of conserved quantities and broken symmetry variable.\par

It is important to point out that the above mentioned equations \citep{SVMBertin, SVMBertin2, MisBaskranMarchetti10} obtained as a result of detailed derivations based on microscopic dynamics have an analogous structure and contain the same major terms as the ones (Eqs. \ref{eq:TTHosszu} and \ref{eq:TTMasCons}) proposed by Toner and Tu inspired by general considerations.\par


A further important approach involving continuum mechanics is based on considering the hydrodynamic properties of systems consisting of microscopic swimmers (see also Sec. \ref{sec:EffectsOfMedium}). By developing a kinetic theory, \citet{SaintillanShelleyPrl08}a and \citet{SaintillanShelleyPhysOfFluids08}b studied the collective dynamics and pattern formation in suspensions of self-propelled particles. They investigated the stability both of aligned and isotropic suspensions, and -- by generalizing the predictions of \citet{SimhaHydrDyn}  -- they showed that aligned suspensions of self-propelled particles are always unstable to fluctuations. Furthermore, they showed that in the case of initially isotropic suspensions an instability for the particle stress takes place for pushers -- particles propelled from the rear -- but not for pullers.\par
\begin{figure}
\centerline{\includegraphics[angle=0,width=0.75\columnwidth]{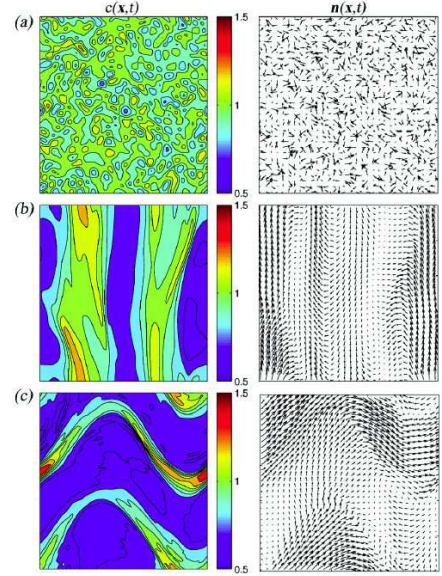}}
\caption{\label{figSaintillanShelleyPrl}(Color online) Snapshots of the simulations performed to study the long-time dynamics and pattern formation of suspensions of pushers. The left column shows the concentration field $c$ and the right column depicts the mean director field $\vec{n}$ at three different times: (a) $t=0$, (b) $t=60$ and (c) $t=85$. From \citet{SaintillanShelleyPrl08}a.}
\end{figure}

Figure \ref{figSaintillanShelleyPrl} shows three snapshots of the simulations they performed in order to study the long-time dynamics and pattern formation of suspensions of pushers. The left column shows the concentration field $c$, and the right column depicts the mean director field $\vec{n}$ at various times. The instability develops at $t=60$, when short-scale fluctuations disappear and a smooth director field appears with correlated orientations over the size of the box. The dense regions that can be observed on Fig. \ref{figSaintillanShelleyPrl} (b) typically form bands, and as time passes by, they become unstable and fold onto themselves, the bands break up and reorganizes in the transverse direction. These dynamics repeat quasi-periodically.\par

Starting with a simple physical model of interacting active particles (swimmers) in a fluid, \citet{BaskrnMarchtt09} derived a continuum description of the large-scale behavior of such active suspensions. They differentiated ``shakers'' from ``movers''. Both of them are active, but a mover, in contrast with a shaker, is self propelled. Shakers are also active, but they do not move themselves. Furthermore ``pushers'' are propelled from the rear (like most bacteria), while ``pullers'' are propelled by flagella at the head of the organism.\par

\begin{figure}
\centerline{\includegraphics[angle=0,width=1\columnwidth]{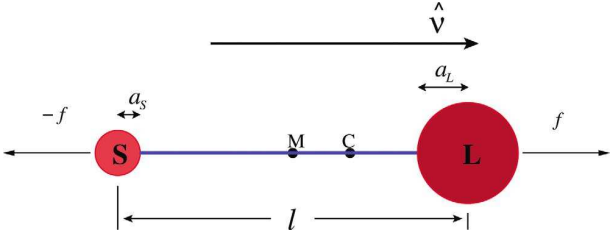}}
   \caption{\label{figDumbell}(Color online) The simplified physical model of the active self-propelled particles in the paper of \citet{BaskrnMarchtt09} are basically asymmetric rigid dumbbells. Two different size of spheres ($S$ and $L$) are connected with an infinitely rigid rod having a length  $l$. The radii of the smaller and larger spheres are $a_S$ and $a_L$ respectively. The geometrical midpoint of the swimmer is depicted by $M$, while the hydrodynamic center is marked with $C$, at which the propulsion is centered. The orientation of these asymmetric particles are characterized by a unit vector $\hat{\nu}$. $|f|$ denotes the force they exert on the fluid they swim in. With this notation, pullers correspond to $f<0$ and pushers to $f>0$. From \citet{BaskrnMarchtt09}.}
\end{figure}

The simplified physical model of a swimmer is basically an asymmetric rigid dumbbell, as depicted in Fig. \ref{figDumbell}. Each of these units has a length  $l$, and their orientation is characterized by a unit vector $\hat{\nu}$, directed along its axis from the small sphere (having radius $a_S$) to the large sphere (having radius $a_L$). They exert a force dipole of strength $|f|$ on the fluid they swim in, which has a viscosity $\tilde{\eta}$. The velocity of the particles are $\vec{v}_{SP}=\nu_0 \hat{\nu}$.

The dynamics of a swimming particle $\alpha$ is given by
\begin{eqnarray}
	\partial_t \vec{r}_{L\alpha} = \vec{u}(\vec{r}_{L\alpha}),  \nonumber \\
	\partial_t \vec{r}_{S\alpha} = \vec{u}(\vec{r}_{S\alpha}),
\label{eq:SwmDyn}
\end{eqnarray}
where $\vec{r}_{S\alpha}$ and $\vec{r}_{L\alpha}$ denote the position of the small and large ``heads'' of swimmer $\alpha$, respectively, with respects to a fixed pole. $\vec{u}(\vec{r})$ is the flow velocity of the fluid at point $\vec{r}$ which is determined by the solution of the Stokes equation, that is,

\begin{equation}
  \tilde{\eta}\nabla^2\vec{u}(\vec{r}) = \nabla p - \vec{F}_{active}+\vec{F}_{noise},
\label{eq:SokesSltn}
\end{equation}
where $\vec{F}_{noise}$ describes the effect of the fluid-fluctuations, and $\vec{F}_{active}$ is the active force 
exerted by swimmer $\alpha$ on the fluid.\par
Closed formulas can be obtained for the translational and rotational motion, and for the hydrodynamical force and the torque between two swimming particles as well.
\citet{marchettit} analyzed a simple model that captured two crucial properties of self-propelled systems: the orientable shape of the particles and the self propulsion. Using the tools of non-equilibrium statistical mechanics they derived a modified Smoluchowski equation for SPP and used it to identify the microscopic origin of several observed or observable large scale phenomena.\par
\citet{peruani1} suggested a mean-field theory for self-propelled particles which accounted for ferromagnetic (F) and liquid-crystal (LC) alignment. The approach predicted a continuous phase transition with the order parameter scaling with an exponent of 1/2 in both cases. The critical noise amplitude below which orientational order emerges found to be smaller for LC-alignment than for F-alignment. 
\citet{CsahokCz97} presented a hydrodynamic approach to describe the motion of migrating bacteria as a special class of SPP systems.\par

As a novel application of the hydrodynamic equations, \citet{ElectronFl} used this approach in order to describe a recently observed electromagnetic phenomenon. According to the observations (\citep{ManiEtAl2002, ZudovEtAl2002}) high-mobility two-dimensional electron systems subject to a perpendicular magnetic field exhibit zero-resistance states, when driven with microwave radiation. By studying the transition from normal state (with non-zero resistance) to the zero-resistance phase, the authors find analogy with ``flocking'' systems. In particular, the two frames become identical in the limit of zero magnetic field and short range electron-electron interactions.\par
The equations of motion for the current and density fluctuations are constructed based on symmetries and conservation laws. Two cases were analyzed: a model valid for small length scales (on which the density does not vary noticeably) characterized by an imposed symmetry under a global uniform shift of the density, and a description argued to be appropriate for describing the system on long length scales. In the first case, the type of the phase transition is predicted to be continuous in case of short-range interactions and first order otherwise (long-range), while the second model predicts first order phase transition in both cases.\par
This is a quickly growing field of its own has recently been reviewed by \citet{HydrodymOf09}.


\subsection{Exact results}

By exact results here we mean results obtained with a minimum or completely missing amount of any kind of assumptions or approximations concerning the behavior of the moving units (beyond the rules/definitions they obey). Originally most results in this area were obtained only for systems in which the noise (an otherwise essential aspect of flocking) was completely neglected. Thus, one could consider the related systems as fully deterministic. However, it has recently been shown (see later) that the theorems we review below are in most cases valid for systems with a low level of perturbations as well.

\subsubsection{The Cucker-Smale model}
An exact formulation of the convergence to consensus in a population of autonomous agents was achieved by \citet{CuckerSmaleIeee}a and \citet{CuckerSmaleJJM}b based on their model (CS). Following their train of thought, let us consider birds, denoted by $i=1, \ldots, k$, moving in 3 dimensional (Euclidean) space, $\Re^3$, endeavoring to reach a common direction -- which is in this case the topic of ``consensus''. The position of the $i$th bird is given by $x_i (\in \Re^3)$. (Of course, $x_i=x_i(t)$.) Let us define the adjacency matrix $A=(a_{ij})$, where the element $a_{ij}$ measures the ability of birds $i$ and $j$ to communicate with each other, or one could say, the \emph{influence} they exert on each other. The elements of $A$ should take values from the interval $(0...1]$, and the closer unit $i$ is to unit $j$, the bigger $a_{ij}$ should be (since they influence each other stronger). $\beta$ is a ``tuning parameter'', effecting the strength of the influence. An appropriate expression for the above requirements is
\begin{equation}
a_{ij}=\frac{1}{(1+\left\| x_i - x_j\right\|^2)^{\beta}},
\label{eq:cmai}
\end{equation}
where $\beta \geq 0$ (not to be confused with the critical exponent introduced in Sec. \ref{sec:BasicDefs}). The main advantage of this form of the distance dependence of the interaction is that it is a smooth function allowing analytical treatment. Importantly, this adjacency matrix $A$ changes with time, since the positions of the birds change with time.\par
For the more manifest usage of graphs the authors introduce the \emph{Laplacian matrix} of $A$ as well, $L=D-A$, where $D$ is a $k\times k$ diagonal matrix whose $i$th diagonal element is defined as $d_i=\sum_{j=1}^k a_{ij}$. The Laplacian matrix -- a form by which a graph can be represented in matrix-form -- is often used to find various properties of a graph. In particular, as we will see, the \emph{eigenvalues} of $L$ bear important information. \par
Denoting the velocity of bird $i$ at time $t$ by $v_i(t)(\in \Re^3)$, 
\begin{equation}
v_i(t+h)-v_i(t)=h \sum_{j=1}^k a_{ij} (v_j - v_i)
\label{eq:CSvi}
\end{equation}
Recall, that the $a_{ij}$ value measures the strength of the communication between birds $i$ and $j$, thus the right hand side of Eq. (\ref{eq:CSvi}) signifies a \emph{local averaging} around bird $i$.\par
The \emph{equations of flocking} are obtained by letting $h$ tend to zero:
\begin{eqnarray}
x'=v  \nonumber \\
v'=-Lv
\label{eq:CSflocking}
\end{eqnarray}
on $(\Re^3)^k\times(\Re^3)^k$, where $L$ gives the local averaging. (Note that the matrices $A$ and $L$ are acting on $(\Re^3)^k$ by mapping $(v_1, \ldots, v_k)$ to $(a_{i1}v_1+ \ldots +a_{ik}v_k)_{i\leq k}$.)\\
\par
After the above preparations, one can ask that \emph{under what conditions does a system (described by the above equations) exhibit flocking behavior?} Or in other words, when do the solutions of $v_i(t)$ converge to a common $v^* (\in \Re^3)$?\par
\emph{One of the most important results of \citet{CuckerSmaleIeee}a is that the emergence of the flocking behavior depends on $\beta$; namely if $\beta$ is small enough ($\beta < 1/2$) then flocking always emerges.} Formally,\par
\textbf{Theorem:} For the equations of flocking (Eqs. \ref{eq:CSflocking}) there exists a unique solution for all $t \in \Re$. \par
If $\beta < 1/2$ then the velocities $v_i(t)$ tend to a common limit $v^* (\in \Re^3)$ as $t \rightarrow \infty$, where $v^*$ is independent of $i$, and the vectors $x_i - x_j$ tend to a limit-vector $\hat{x}_{ij}$ for all $i,j \leq k$, as $t \rightarrow \infty$, that is, the relative positions remain bounded.\par
If $\beta \geq 1/2$ dispersal, the split-up of the flock is possible. But, provided that some certain initial conditions are satisfied, flocking will occur.\\
\par

To obtain more general results for the conditions of flocking, one can investigate the eigenvalues of the corresponding Laplacian matrix $L$. Let $G$ denote a graph and $A$ be the corresponding adjacency matrix defined as usually, that is,
\begin{equation}
a_{ij} = \left\{
	\begin{array}{ll}
		1 & \mbox{if $i$ and $j$ are connected,}\\
		0 & \mbox{if not}
	\end{array}	
\right.
\label{eq:AdjcMtx}
\end{equation}
Let $D$ be a diagonal matrix with the same dimensions as $A$, defined by $d(i,i)=\sum_j a(i,j)$. Then the general form of the Laplacian matrix of $G$, is given as $L=L(G)=D-A$. The eigenvalues of $L$ can be expressed by
\begin{equation}
0=\lambda_1 \leq \lambda_2 \leq \lambda_3 \leq \ldots
\end{equation}
$\lambda_1$, the first eigenvalue is always zero. The second one in ascending order, $\lambda_2$, is the so-called \emph{Fiedler-number}, $F$, which is zero if the graph $G$ is separated (in this case the flock disintegrates to two or more smaller flocks), and non-zero if and only if $G$ is connected. This number is a crucial descriptive measure of the conditions needed to be satisfied for the emergence of flocking.\par
Importantly, in the case of flocking, $F=F(t)$ (it is a function of time), because the elements of $G$ depend on the $x_i$ positions of the individual birds. 
By using the Fiedler-number, we can say that \emph{one obtains flocking, if and only if
\begin{equation}
0 < const \leq F=F(x(t)).
\end{equation}
Otherwise the flock disperses.}\\
The above definitions can be extended to weighted, general matrices as well.\\
\par
In addition, \citet{CuckerDong10} extended the model by adding to it a repelling force between particles. They showed that, for this modified model, convergence to flocking is established along the same lines while, in addition, avoidance of collisions (i.e., the respect of a minimal distance between particles) is ensured.\par
The main differences between the systems described by Cucker and Smale and by the SVM are, from the one hand, the definition of the range of interaction, and from the other hand, the existence (or absence) of noise. The SVM comprises noise, while the original Cucker-Smale model does not. Regarding the range of interaction, in the present model it is a long-range effect decaying with the distance according to $\beta$ (see Eq. (\ref{eq:cmai})), while in the SVM it has the same intensity for all the neighboring units around a given particle, but only within a well-defined range (see Eq. (\ref{eq:UjVi})).\par 
Very recently perturbations have also been considered in the CS model. This has been done with various forms of noise by \citet{CuckerMord08, Shang09}, modeled with stochastic differential equations by \citet{HaEtAl09}, and taking into account random failures between agent's connections by \citet{CckrSmaleFlck09}.\par
Other works developing the Cucker-Smale (CS) model in several directions include an extension to fluid-like swarms \citep{HaLiu09, HaTadmor08, CSCarrillo}, collision avoiding flocking \citep{CuckerDong10}, the inclusion of agents with a preferred velocity direction \citep{CuckerHuepe08}, and its proposal as a control law for the spacecrafts of the Darwin mission of the European Space Agency \citep{CSPerea09}.

\subsubsection{Network and control theoretical aspects}

Networks have recently been proposed to represent a useful approach to the interpretation of the intricate underlying structure of connections among the elements of complex systems. A number of important features of such networks have been uncovered \citep{BarabasiRMP, SmllWrld}. It has been shown that in many complex systems ranging from the set of protein interactions to the collaboration of scientists the distribution of the number of connections is described by a power law as opposed to a previously supposed Poissonian. Most of the networks in life and technology are dynamically changing and are highly structured. In particular, such networks are typically made of modules that are relatively more densely connected parts within the entire network (e.g., interacting flocks) \citep{NewmanDet, NewmanMtx, PallaNature05, SocNtwAB}. The evolution of these modules plays a central role in the behavior of the system as a whole \citep{PallaNature07}.\par

In these terms, a dynamically changing network can be associated with a flock of collectively moving organisms (or robots, agents, units, dynamic systems, etc.). In such a network two units are connected if they interact. Obviously, if two units are closer in space have a better chance to influence the motion of each other, but their interaction can also be disabled by environment or internal disturbances. Since the units are moving and the environment is also changing, the network of momentarily interacting units is evolving in time in a complex way. Using the conventional terminology of control theory, this kind of topology (that is, when certain number of edges are added or removed from the graph from time to time), is called ``switching topology''.\par

\citet{JadbabaieIeee} investigated a theoretical explanation for a fundamental aspect of the SVM, namely, that by applying the nearest neighbor rule, all particles tend to align into the same direction despite the absence of centralized coordination and despite the fact that each agent's set of nearest neighbors changes in time. 
By addressing the question of global ordering in models analogous to Eqs. (\ref{eq:UjVi}) and (\ref{eq:UjThi}) they presented some rigorous conditions for the graph of interactions needed for arriving at a consensus.\par
Several further control theory inspired papers discussed both the question of convergence of the simplest SPP models as well as the close relation of flocking to such alternative problems as consensus finding, synchronization and ``gossip algorithms''\citep{Blondel05, BoydGossip}. 

\citet{RenBeard} considered the problem of consensus finding under the conditions of limited and unreliable information exchange for both discrete and continuous update schemes. They found that in systems with dynamic interaction-topologies consensus can be reached asymptotically, if the union of the directed communication network across some time intervals has spanning trees frequently enough, as the system evolves. Similarly, \citet{Xiao06} found the existence of spanning trees to be crucial in the directed graphs representing the interaction topologies, in systems in which the topology, weighting factors and time delays are time-invariant. They studied the consensus-problem for dynamic networks with bounded time-varying communication delays under discrete-time updating scheme, based on the properties of non-negative matrices.\par

An efficient algorithm controlling a flock of unmanned aerial vehicles (UAVs) is considered by \citet{BenAsherUAV}. The units are organized into a minimal set of rooted spanning trees (preserving the geographical distances) which can be used for both distributed computing and for communication as well, in addition to computation and propagation of the task assignment commands. The proposed protocol continually attempts to keep the number of trees minimal by fusing separate adjacent trees into single ones: as soon as radio connection between two nodes belonging to separate trees occurs, the corresponding networks fuse. This arrangement overcomes the typical deficiencies of a centralized solution. The motion of the certain UAVs is coordinated by Reynolds's algorithm \citep{Reynolds87} (see also Sec. \ref {sec:OrigSPPMdl}).\par

\citet{TannerCT1} proposed a control law for flocking in free-space. Dynamically changing topology of the interacting units has also been considered \citep{TannerCT2}. 
\citet{LindheCT} suggested a flocking algorithm providing stable and collision-free flocking in environments with complex obstacles. \citet{HollandCT} proposed a flocking scheme for unmanned ground vehicles similar to Reynolds' algorithm based on avoidance, flock centering and alignment behaviors, where the units receive the range, bearing and velocity information from the base station based on pattern recognition techniques. Very recently, many further papers have appeared both on the original flocking problem as well as on interesting variants including  the role of ``leaders'', delays in communication, convergence time, etc.\par

One of the most general theoretical frameworks for design and analysis of distributed flocking algorithms was discussed by \citet{OlfatiFl06}. Three algorithms were investigated in detail: two for free-flocking (one fragmented and one not) and one for constrained flocking. The basic driving rules and principles and their relation to specific underlying network structures were discussed.\par

Formally, from a control theoretical view-point, the problem looks as next: given a set of agents, who want to reach a \emph{consensus}, which, in this terminology, means a common value (an ``agreement'') regarding a certain quantity that depends on the state of the agents. (For example, this 'certain quantity' can be the direction of motion.) The interaction rule that defines the information exchange between a unit and its neighbors is called the consensus algorithm (or ``\textit{protocol}'').\par
This system can be represented by a graph $G=(V,E)$, in which the agents are the nodes $V=\left\lbrace 1,2, \ldots, n \right\rbrace$. Two nodes are connected with an edge $e \in E$ if, and only if, they communicate with each other. In this case they are \textit{neighbors}. Accordingly, the neighbors of node $i$ are $N_i={j \in V: (i,j) \in E}$. If the state of the $i$th agent (regarding the quantity of interest) is denoted by $x_i$, then the agreement is 
\begin{equation}
 x_1 = x_2 = x_3 = \ldots = x_n . 
\label{eq:agrmt}
\end{equation}
Within this framework, \emph{reaching a consensus} means to converge asymptotically to an agreement (defined by Eq. \ref{eq:agrmt}) via local communication.\par
Let  $A=(a_{ij})$ denote the adjacency matrix, which defines the communication pattern among the agents: if $i$ and $j$ interact with each other, then $a_{ij} > 0$, otherwise $a_{ij} = 0$. Notably, in the case of flocks $A=A(t)$ and $G=G(t)$, that is, they vary with time. Such graphs -- called \emph{dynamic graphs} -- are useful tools for describing the (time-dependent) topology of flocks and mobile sensor networks \citep{OlfatiFl06}.\par
Assuming a simple protocol, the state of agent $i$ can change according to
\begin{equation}
\dot{x}_i(t) = \sum_{j \in N_i} a_{ij}\left( x_j(t)-x_i(t)  \right) 
\label{eq:xDeriv}
\end{equation}
which linear system always converges to a collective decision, that is, it defines a \emph{distributed consensus algorithm} \citep{OlfatiM04}.\par
In the case of undirected graphs (when $a_{ij}=a_{ji}$ for all $i,j \in V$) the sum of the sate-values does not change, that is, $\sum_i \dot{x}_i = 0$. Applying this condition for $t=0$ and $t=\infty$,
\begin{equation}
\alpha = \frac{1}{n} \sum_i {x}_i(0),
\label{eq:EqvSoltn}
\end{equation}
that is, the collective decision ($\alpha$) is the average of the initial state of the nodes.\par
In fact, regarding the protocol defined by Eq. (\ref{eq:xDeriv}), a more strict statement can also be formulated \citep{OlfatiOf07}:\par

\textbf{Lemma:} Let $G$ be a connected undirected graph. Then, the algorithm defined by Eq. (\ref{eq:xDeriv}) asymptotically solves an average-consensus problem for all initial states.\par

\subsection{Relation to collective robotics}

The collective robotics literature is on the one hand about mathematical questions concerning the control theoretical aspects of coherently moving devices whereas, on the other hand, it represents important efforts to eventually produce and describe the collective patterns of behavior of a collection (ranging from 5 to a few dozen) of robots moving on a plane surface. Experimental attempts to produce flocking of aerial devices have been very limited.\par

In one of the earliest attempts towards obtaining flocking in a group of actual robots, \citet{Mataric94} combined a set of ``basic behaviors''; namely safe-wandering, aggregation, dispersion and homing. In this study, the robots were able to sense the obstacles in the environment, localize themselves with respect to a set of stationary beacons and broadcast the position information to the other robots in the group. \citet{KellyR} used a group of ten robots, which were able to sense the obstacles around them through ultrasound sensors, and the relative range and bearing of neighboring robots through the use of a custom-made active infra-red (IR) system. The proximity sensors on most mobile robots (such as ultrasound and IR-based systems) can sense only the range to the closest point of a neighboring robot and multiple range-readings can be returned from a close neighboring robot. Furthermore, as \citet{SelfOrgRobs} pointed out, the sensing of bearing, velocity and orientation of neighboring robots is still difficult with off-the-shelf sensors available on robots. Hence, there exist a major gap between the studies that propose flocking behaviors and robotics.\par

An interesting experiment on flocking in 3D was carried out by \citet{WelsbyR} using motorized balloon-like objects. The slow coherent wondering of 3 of the robots was observed. Model (toy) helicopters were also proposed to observe flocking in three dimensions \citep{NardiR}.\par

Several major efforts have been documented about the collective exploration of swarms of robots. A variety of algorithms have been published about the optimal strategy to locate a given object or uncover the details of an area (in which the robots could move) having a complex shape. Recent papers have demonstrated that, using an appropriate algorithm, such tasks can be achieved effectively. The largest collection of swarming robots has now over 100 miniature devices (\texttt{http://www.swarmrobot.org/}).\par

\begin{figure}
  \centerline{\includegraphics[angle=0,width=0.68\columnwidth]{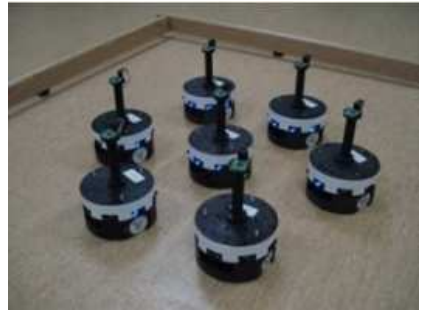}}
  \caption{\label{figKobots}(Color online) A photo of seven mobile robots (called ``Kobots'') moving in a swarm. According to \citet{SelfOrgRobs}, the main factor defining the size of the swarm (the number of Kobots flocking together) is the range of communication, and it is highly independent from both the noise (encumbering the sensing systems) and from the number of neighbors each robot had. Adapted from \citet{SelfOrgRobs}.}
\end{figure}

\citet{SelfOrgRobs} examined the spatial self organization properties of robot swarms using mobile units (called ``Kobots'', see Fig. \ref{figKobots}). Every unit were equipped with a digital compass, an infrared-based short range sensing system (capable of measuring the distance from obstacles and detecting other robots) and an other appliance sensing the relative direction of the neighboring units. The group investigated the behavior of the flock in the function of: (1) the amount and nature of the noise encumbering the sensing systems (2) the number of neighbors each unit had, and (3) the range of the communication. They found that the main factor defining the size of the swarm (number of units that can flock together) is the range of communication, and that this size is highly robust against the other two parameters. The motion of such robot swarms can be influenced by externally guiding some of their members towards a desired direction \citep{htr}.\par

Only a few examples are known about trying to combine robots and animals into a single system and monitor the joint behavior. In a beautiful paper \citet{CsotanyRobi} investigated whether the behavior of a population of cockroaches can be influenced by micro-robots imitating cockroaches (these micro-robots had about the same size and had the same odor than the cockroaches). It turned out that it was possible to increase the number of cockroaches hiding under a given ``shelter'' if the mini-robots were moving there upon switching on the light.\par

In a more theoretical work, \citet{JapiRobtSwrm} investigated the formations of motile elements (robots) as a function of various control parameters. Their kinetic model -- inspired by living creatures, such as birds, fishes, etc. -- is defined by Eqs. (\ref{eq:JpDin1}) and (\ref{eq:JpDin2}). 

\begin{equation}
	m \frac{d\vec{v}_i}{dt}=-\gamma \vec{v}_i + a \vec{n}_i + \sum_{j \neq i} \alpha_{ij}\vec{f}_{ij} + \vec{g}_i
\label{eq:JpDin1}
\end{equation}

\begin{equation}
	\tau \frac{d \theta_i}{dt}=sin(\phi_i - \theta_i)+\sum_{j \neq i}J_{ij}sin(\theta_j-\theta_i),
\label{eq:JpDin2}
\end{equation}
where $\vec{r}_i$ is the position, $\vec{v}_i$ is the velocity, and $\vec{n}_i$ is the \emph{heading unit vector} of the $i$th element of a swarm consisting of $N$ units ($i \in {1, 2, \ldots, N}$), respectively. The velocity $\vec{v}_i$ is relative to the medium (air, fluid, etc.) in which the motion occurs. The last quantity, $\vec{n}_i$ is parallel to the axis of unit $i$, but not necessarily parallel to its velocity $\vec{v}_i$. For example, bigger birds often glide, during which the heading direction $\vec{n}_i$ and the velocity $\vec{v}_i$ encloses an angle, which is assumed to disappear within a relaxation time $\tau$. In other words, $\tau$ is the time needed to $\vec{n}_i$ and $\vec{v}_i$ to relax to parallel.
$\theta_i$ and $\phi_i$ are the angles between the $x$ axis and the vectors $\vec{n}_i$ and $\vec{v}_i$, respectively. $m$ is the mass of the elements (of \emph{all} the elements -- apart from the initial conditions, every unit is identical in this model). $a$ is the motile force acting in the direction of $\vec{n}_i$, and $\gamma$ is a quantity proportional to the relaxation time in velocity.
The term $\alpha_{ij}$ denotes a ``direction sensibility factor'' which is introduced to account for the possible anisotropy of the interaction. For example, if the robots gather information about the motion of their mates through vision (that is, with camera), than the interaction is strong towards the visual field that is covered by the camera, and zero elsewhere. It is defined as

\begin{equation}
	\alpha_{ij}=1+d \cos \Phi,
\label{eq:JpAlpha}
\end{equation}
where $\Phi$ is the angle enclosed by $\vec{n}_i$ and $\vec{r}_j - \vec{r}_i$, and $d$ is the sensitivity control parameter, $0 \leq d \leq 1$.\par

$J_{ij}$ is introduced to account for the observation that animals tend to align with each other \citep{HunterHal} through an interaction which is supposed to decrease in the linear function of distance between individuals $i$ and $j$:
\begin{equation}
	J_{ij}=k \left( \frac{\left|\vec{r}_j - \vec{r}_i \right|}{r_c} \right)^{-1},
\label{eq:JpJij}
\end{equation}
where $k$ is the control parameter and $r_c$ is the preferred distance between neighbors. The term $g_i$ is a force directed towards the center of the group, and finally, $f_{ij}$ denotes a mutual attractive/repulsive force between elements $i$ and $j$, in analogy with the intermolecular forces, as suggested by \citet{Breder54}.

\begin{multline}
	\vec{f}_{ij} = -c \left\{  \left( \frac{\left| \vec{r}_j - \vec{r}_i \right|}{r_c} \right)^{-3} - \left( \frac{\left| \vec{r}_j - \vec{r}_i \right|}{r_c} \right)^{-2}  \right\} \\
	\cdot \left( \frac{\vec{r}_j - \vec{r}_i}{r_c} \right)\cdot e^{-\frac{\left| \vec{r}_j - \vec{r}_i \right|}{r_c}},
\label{eq:Jpf}
\end{multline}
where $c$ is the control parameter defining the magnitude of the interaction.\par

\begin{figure}
\centerline{\includegraphics[angle=0,width=1\columnwidth]{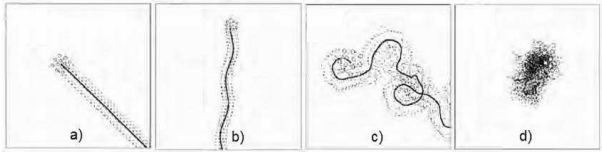}}
\caption{\label{figJpRbt}Trajectories with various control parameters, obtained from numerical simulations. The solid line shows the center of mass. (a) marching, (b) oscillation, (c) wandering, and (d) swarming. Adapted from \citet{JapiRobtSwrm}.}
\end{figure}

Using numerical simulations and experiments with small mobile robots (called ``Khepera'', a popular device for such experiments), the authors observed various formations, depending on the control parameters (see Fig. \ref{figJpRbt}). They classified the observed collective motions into four categories: (1) \emph{``Marching''}. Obtained when the value of the anisotropy of mutual attraction is kept small. This state exhibits only small velocity-fluctuations and it is stable against disturbance. (Fig. \ref{figJpRbt} \emph{a}) (2) The category called \emph{``Oscillation''} includes motions exhibiting regular oscillations, such as the wavy motion of the swarm, along its linear trajectory, depicted on Fig. \ref{figJpRbt} \emph{b}. The stability of this state is weaker than that of the marching, and these two phases (1 and 2) may coexist for some parameters. (3) \emph{``Wandering''}. When $d \neq 0$ (see Eq. (\ref{eq:JpAlpha})), the mutual positions of the units abruptly vary, according to stochastic changes in the direction of motion. Such a phase is often exhibited by -- for example -- small non-migratory birds. (Fig. \ref{figJpRbt} \emph{c}) (4) \emph{``Swarming''}. Irregularly moving units within a persistent cluster. The mobility of the entire cluster is small. For example mosquitoes travel in such swarms. (See Fig. \ref{figJpRbt} \emph{d}).

In general, the point of organizing robots into a swarm is to accomplish tasks (preferably without centralized control), that are too challenging for an individual agent. The fields of the possible applications are extremely wide, including practical applications (such as the localization of hazardous emission sources in unknown large-scale areas \citep{CuiRbt}, the surveillance in hostile or dangerous places \citep{MThesis}, the optimization of telecommunication networks \citep{TelComOpt}) as well as theoretical topics (like discrete optimization \citep{Dopt99} or providing new heuristics for the Traveling Salesman Problem \citep{Dopt97}).\par
Furthermore, within these robot swarms, the appearance of the most variable forms of collective behavior (like co-operative, altruistic, selfish, etc.) can be studied as well through various genetic algorithms, conditions and tasks. Many homepages maintained by research groups working on this field contain further information for those who are interested (for example Laboratory of Autonomous Robotics and Artificial Life: \texttt{http://laral.istc.cnr.it/}, Laboratory of Intelligent Systems: \texttt{http://lis.epfl.ch/}, Distributed Robotics Lab of MIT: \texttt{http://} \texttt{groups.csail.mit.edu/drl/wiki/index.php/}, or the web-page of the Swarm-bots Project: \texttt{http://www.swarm-bots.org/} and \texttt{http://www.swarm-robotics.org/index.php/}), to mention a few. For a more engineering viewpoint of the topic, see also \citep{SwrmIntBook}.\par

\section{Modeling actual systems}\label{sec:SystmSpecModels}  

\subsection{Systems involving physical and chemical interactions}

\subsubsection{The effects of the medium}\label{sec:EffectsOfMedium}

In the case of microorganisms swimming in a medium, the hydrodynamic effects are often significant enough to generate collective motion passively, that is, various coherent structures (e.g., clusters, vortices, etc.) arise merely as a result of hydrodynamic interactions. One of the first general methods for computing the hydrodynamic interactions among an infinite suspension of particles under some well-defined conditions was presented by \citet{BradyFldMech, BradyStDyn}. Their method was accurate and computationally efficient forming the basis of the Stokesian-dynamics simulation method, a technique used in order to yield approximate expressions for the velocities of hydrodynamically interacting particles.\par
\citet{SimhaHydrDyn, HatwalneHydroDyn} constructed hydrodynamic equations for suspensions of SPPs suitable for making testable predictions for systems consisting of bacteria, cells with motors or artificial machines moving in a fluid.\par

Based on similar studies, in particular experiments on living cells moving on a solid surface (\citep{GrulerEtAl99, KemkemerEtAl2000}) and studies on vertically vibrated layers of rods (\citep{NeicuCondMat03}), further continuous equations were formulated by \citet{RamaswamyEPL}. They considered systems of active nematogenic particles without total momentum conservation (the momentum was assumed to being damped by friction with the substrate). The two most important predictions implied by their results are: (i) Giant number fluctuations (GNF):\label{RamGNF} the standard deviation in the number $N$ of particles is enormous: it scales as $N$ in the entire nematic phase for two dimensional systems, which is in deep contrast with $\sqrt N$ (in the limit of $N \to \infty$), characterizing equilibrium systems being not at the point of continuous phase transition. (ii) for $d \geq 2$ spatial dimensions, the autocorrelation of the particle velocity of a tagged element decays with time $t$ as $t^{-d/2}$, despite the absence of a hydrodynamic velocity field. Importantly, the above results imply that the nematic phases of rod-like powders can not be described by equilibrium statistical mechanics.\par

Another approach, the so called slender-body theory was used by \citet{SaintiHydroDyn} in order to numerically study the dynamics and orientational order of self-propelled slender rods. This method was used to obtain an approximation to the field surrounding a slender object and to get an estimation for the net effect of the field on the body \citep{SlenderBodyTh1, SlenderBodyTh2}. They found local nematic ordering over short length scales as well having a significant impact on the mean swimming speed. 
\citet{SaintillanGrMElectr} investigated the role of hydrodynamic interactions in case of metal rod-like particles in the presence of an externally applied electric field, both by simulations and experiments. In both cases the particles were observed to experience repeated pairing interactions in which they come together axially, approaching one other with their ends, slide past each other until their centers approach, and then push apart. 
\citet{SanFm} showed that polar self-propelled particles were prone to exhibit various types of instabilities through the interplay of polarity, activity and the existence of a free surface, by using a thin-film hydrodynamic model.\par

\begin{figure}
\centerline{\includegraphics[angle=0,width=1\columnwidth]{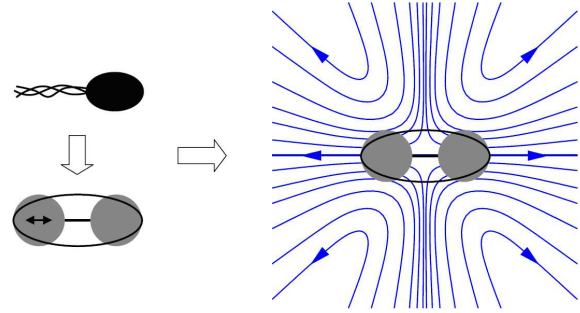}}
\caption{\label{figUnderhPP}(Color online) The scheme of a pusher and the fluid disturbance it causes. Each SPP is represented by two spheres connected by a rod. The propulsion is provided by a ``phantom flagellum'' (``phantom'', because it is not treated explicitly, only through the effect it exerts to the swimming body and to the fluid.) The force exerted by this flagellum acts at the center of the first sphere. A puller produces the same streamlines (dark gray curves) but the arrows point in the opposite direction. From \citet{UnderhillHOG}.}
\end{figure}

The motion of the fluid generated by the particles swimming in it seems to depend strongly on the way these organisms propel themselves \citep{HydrodymOf09}. \citet{UnderhillHOG} simulated \emph{pushers} (organisms propelled from the rear, like most bacteria) and \emph{puller}s (creatures that are propelled at the head of the organism) separately to capture the differences in the effects of the forces these creatures exert on the fluid while swimming in it. Figure \ref{figUnderhPP} shows the scheme of their self-propelled swimmers. Each of them consists of two beads connected by a rod. They propel themselves by a ``phantom flagellum''. (``Phantom'', because its physical appearance is not taken care of, only its effect on the swimmer and on the fluid.) It exerts an $F_f$ force on bead 1, and $-F_f$ force on the fluid. Pushers and pullers are distinguished by the \emph{direction} of $F_f$: if it points from bead 1 to bead 2, then it is a pusher, and if it points in the opposite direction, then it is a puller. The motion of the particles is calculated by solving the force balance for each bead, as given by Eq. (\ref{eq:UFBal}):

\begin{equation}
 F_f+F_{h1}+F_{c1}+F_{e1}=0,
\label{eq:UFBal}
\end{equation}
where $F_{c1}$ is the force exerted by the rod (connecting the two beads), $F_{h1}$ is the hydrodynamic drag force and  $F_{e1}$ is the excluded volume force on the bead. The force balance defining the motion of bead 2 is the same as Eq. (\ref{eq:UFBal}), but without the $F_f$ flagellum force.

Using this model, \citet{UnderhillHOG} observed qualitative differences between the effects of pushers and pullers, exerted on the fluid they move in: SPPs that are pushed from the behind show greater enhancement than particles that are pulled from the front. This model -- supported by \citet{MNFm} as well -- describes the far-field behavior of interacting swimming particles.\par

The notion of ``squirmer'' has also been introduced in order to apprehend the most important features of swimming microorganisms (with respect to their motion in a fluid) \citep{Lighthill52}. These are neutrally buoyant squirming spheres with a tangential surface velocity and with anisotropic structures, that is, their center of mass and geometric center do not necessarily coincide.

\begin{figure}
\centerline{\includegraphics[angle=0,width=.9\columnwidth]{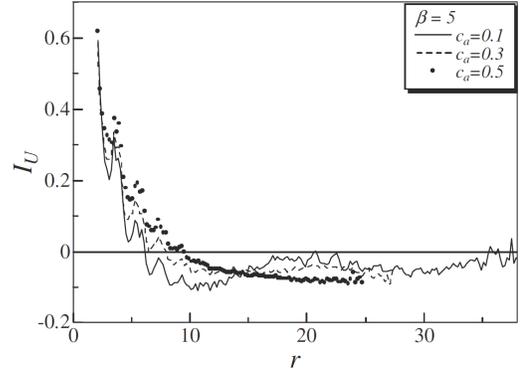}}
\caption{\label{figIshiIu}The velocity correlation function among the particles ($I_U$) as a function of the distance among the units $r$ for three different sphere-sizes. In the region $r<6$, $I_U$ is positive, denoting that nearby particles tend to swim together in similar direction. In the region $r>10$, the correlation turns into anti-correlation, since $I_U$ is negative, meaning that squirmers at least 10 radii apart tend to swim in opposite direction. From \citet{IshikHydroDyn}.}
\end{figure}

\citet{IshikHydroDyn} simulated the motion of such squirmers in a monolayer, that is, in two dimensions. In order to do so, they included not only the far-field fluid dynamics, but the near-field components too, which gave the novelty of their approach. These simulations demonstrated that various types of processes resulting in coherent structures (such as aggregation, band formation or mesoscale spatio-temporal motion) can be generated by pure hydrodynamic interactions. Accordingly, Fig. \ref{figIshiIu} shows the velocity correlation function among the particles $I_U=I_U(r)$, as a function of the distance $r$ separating the squirmers. $c_a$ is the areal fraction of the particles in the monolayer, thus it refers to their sizes: bigger $c_a$ denotes larger sphere. However, these simulations did not show vigorous coherent structures in fully three-dimensional cases, that is, when the particles were not restricted to move on a monolayer \citep{Ishik2}.

\citet{SubrKoch} examine the stability properties of a bacteria suspension, in which the the bacteria execute a ``run-and-tumble'' motion, that is, after swimming in a given direction, the runs are interrupted by tumbles, leading to an abrupt motion. Due to the features of the force field produced by the bacteria, instability is predicted to occur in suspensions of pushers. They argue that for speeds smaller than a critical value, the destabilizing stress remains minor and a dilute suspension of such swimmers responds to long-wavelength perturbations in a similar way than suspensions of passive rigid particles would respond.\par

An interesting example of collective motion is the synchronized beating of flagella on the surface of unicellular, or simple multicellular organisms. It was shown that such a coherent motion of flagella leads to a highly increased exchange rate of the nutrients around such organisms \citep{ShortEtAl06}.\par


\subsubsection{The role of adhesion}\label{sec:SysSpesMCells}

The problem regarding the mechanisms determining tissue movements dates back to the beginning of the 20th century. In 1907 Wilson discovered that sponge cells which have been previously squeezed through a mesh of fine bolting-cloth are able to reunite again reconstituting themselves into a functioning sponge \citep{Wilson1907}. Early studies mainly envisioned cell sorting as a resultant of inhomogeneities (for example of pressure) in the immediate environment. Since then many theoretical and experimental studies have been dedicated to this question supporting the idea that the movements are due to intrinsic properties of the individual tissues themselves (landmarked by, among many others \citet{St2}, \citet{St3}, \citet{St4}, \citet{St5}, \citet{St6}, \citet{St7}).\par
To explain the phenomenon of cell sorting, \citet{SteinbergCellStrng} developed the hypothesis that the local rearrangement behavior (characterizing cells during the process of sorting out and tissue reconstruction) follows directly from their \emph{motility} and quantitative \emph{differences in adhesiveness}. (This theory is often referred to as ``differential adhesion hypothesis'', DAH). Based on the basic ideas of DAH, \citet{ChateSellSortingt} introduced a simple self-propelled particle model to study cell sorting (see Sec. \ref{sec:CellSorting}).\par
Regarding the collective motion and phase transition observed in migrating keratocyte cells (cells taken from the scales of goldfish), \citet{SzaboEtAl} constructed a model describing their experimental observations (see also Sec. \ref{subsec:ExperimCells}). Using long-term video-microscopy they observed kinetic phase transition from disordered to ordered state, taking place as the cell density exceeds a relatively well-defined critical value. Short-range attractive-repulsive inter-cellular forces are suggested to account for the organization of the motile keratocyte cells into coherent groups.\par
Instead of applying an explicit averaging rule (which would not be realistic), the model-cells (self-propelled particles) adjust their direction toward the direction of the net-force acting on them (see Eq. (\ref{eq:Sz2})). In this two-dimensional flocking model, $N$ SPPs move with a constant speed $v_0$ and mobility $\mu$ in the direction of the unit vector $\vec{n}_i(t)$ while the $i$ and $j$ particles experiences the inter-cellular force $ \vec{F}(\vec{r}_i\vec{r}_j)$. The motion of cell $i (\in {1, \ldots, N})$ in the position $\vec{r}_i(t)$  is described by
\begin{equation}
\frac{d\vec{r}_i(t)}{dt}=v_0 \vec{n}_i(t) + \mu \sum_{j=1}^{N} \vec{F}(\vec{r}_i\vec{r}_j).
\label{eq:Sz2}
\end{equation}

The direction $\vec{n}_i(t)$ can be described by $\theta^{n}_i(t)$ as well, which attempts to relax to $\vec{v}_i(t)=d\vec{r}_i(t)/dt$ within a relaxation time $\tau$. Denoting the noise by $\xi$ and the unit vector orthogonal to the plane of motion by $\vec{e}_z$,

\begin{equation}
\frac{d\theta^{n}_i(t)}{dt} = \frac{1}{\tau} arcsin\left[ \left( \vec{n}_i(t)\frac{\vec{v}_i(t)}{|\vec{v}_i(t)|}  \right) \cdot \vec{e}_z \right]+\xi
\label{eq:Sz3}
\end{equation}

\begin{figure}
\centerline{\includegraphics[angle=0,width=1\columnwidth]{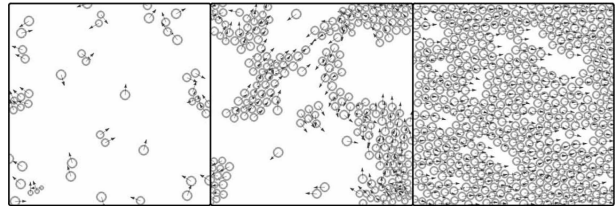}}
\caption{\label{figSzaboEtAlMdl}Computer simulations obtained by solving Eqs. (\ref{eq:Sz2}) and (\ref{eq:Sz3}) for different particle densities. In agreement with the observations, the model exhibit a continuous phase transition from disordered to ordered phase. From \citet{SzaboEtAl}.}
\end{figure}

Figure \ref{figSzaboEtAlMdl} shows the typical simulation results obtained by solving Eqs. (\ref{eq:Sz2}) and (\ref{eq:Sz3}) with periodic boundary conditions. The model -- in good agreement with the observations -- exhibit a continuous phase transition from disordered to ordered phase. (For the corresponding observations see also Fig. \ref{figCellSzabo} in Sec. \ref{subsec:ExperimCells}.)\par

Some authors put much emphasis on the actual \emph{shape} and \emph{plasticity} of the cells as well, since these properties also play an important role in the emergent behavior of the system \citep{GG92, GG93, GGH97, GGH01}. Following this line, \citet{RotatingSlimeMold} suggested a model consisting of self-propelled deformable objects to explain their experimental results on the dynamics of \emph{Dictyostelium discoideum} (see also in Sec. \ref{subsec:ExperimCells}). Their model reproduces the observed self-organized vortex states (the ``pancake''-structures), as the resultants of the coupling between the self-generated propulsive force and the cell's configuration, and of the cohesive energy between the cells.\par

A number of recent interdisciplinary studies focus on the detailed mechanisms by which organisms -- from bacteria to vertebrates -- generate sophisticated multicellular patterns (for example organs) during ontogenesis. We mention a representative example by \citet{Sprouting} who investigated the formation and regulation of multicellular sprouting during vasculogenesis. Based on \emph{in vivo} and \emph{in vitro} observations and experiments, they suggested a general mechanism that builds on preferential attraction/attachment to elongated structures. The proposed interactive particle model exhibits robust sprouting dynamics and results in patterns that are similar to native primordal vascular plexuses -- without any assumptions involving mechano-chemical signaling or chemotaxis.\par
\citet{WangCell} proposed a model for structures like the cytoskeleton, that is, for systems consisting of many interacting bio-macromolecules driven by energy-consuming motors. Readers interested in the models of active polar gels will find more details in \citep{RamOf2Julicher}.


\subsubsection{Swarming bacteria}\label{sec:BaciModels}

\begin{figure}
\centerline{\includegraphics[angle=0,width=.75\columnwidth]{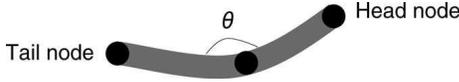}}
\caption{\label{figBendedRodCell}Each cell is represented by a string of $N$ nodes. In the simulations $N=3$, thus the cells consist of two segments, enclosing the angle $\theta$. The orientation is defined by the vector directed from the tail node to the head node. From \citet{MutansBacik}.}
\end{figure}

By using models and simulations, experimentally observed behaviors which are seemingly unintelligible might also be elucidated. Recently, as described in Sec. \ref{subsec:ExperimBaci}, bacteria belonging to \emph{Myxococcus xanthus} swarms were observed to reverse their gliding directions regularly, while the colony itself expanded \citep{ReverseBaci}. To compass this seemingly energy-wasting behavior, the authors simulated the observed phenomena using a cell-based model, taking into account only the contact-mediated, local interactions \citep{MutansBacik}. The individual cells are represented by a flexible string of $N$ nodes, consisting of $N-1$ segments, as depicted on Fig. \ref{figBendedRodCell} (basically a bendable rod, bended in $N-2$ points, being able to move in 2-D space). Each segment has the same length $r$. In the simulations $N$ was chosen to be 3, thus each cell had two segments, as the rod was blended in one point, in the middle. The orientations of the cells are defined by the vectors directed from the tail nodes to the head nodes. In order to keep the shapes of the cells within an interval that agrees with the observations, a Hamiltonian function was defined to characterize the certain node-configurations, as given by Eq. (\ref{eq:CellHamltn}).

\begin{equation}
H=\sum_{i=0}^{N-1}K_b(r_i-r_0)^2 + \sum_{i=0}^{N-2}K_{\theta}\theta_i^{2},
\label{eq:CellHamltn}
\end{equation}
where $r_i$ is the length of the $i$th segment, $r_0$ is its ``target length'' and $\theta_i$ is the angle enclosed by the neighboring segments $i$ and $i+1$. $K_b$ and $K_{\theta}$ are the stretching and the bending coefficients, respectively, defining the extent to which the length of the segments and the angles between them can vary. Both of them are dimensionless values, and are the same for all the segments and angles.\par
Regarding the active motion of the certain cells, first the head-node moves in a particular direction, followed by the other nodes which take positions so that the Hamiltonian function belonging to the new configuration is minimal. Since according to the observations, \emph{Myxococcus xanthus} cells do not have any kind of long-range communicating systems \citep{KaiserNRM}, the model takes the interactions only among neighboring cells into account.\par
The experimentally-observed reversals (sudden changes in the direction with $180^{o}$) are most probably regulated by an internal biochemical clock, which is independent of the actual interactions of the given cell. Therefore, the model takes into account these reversals by simply switching the roles of the head-nodes and the tail-nodes, according to an internal clock.\par
\begin{figure}
\centerline{\includegraphics[angle=0,width=1\columnwidth]{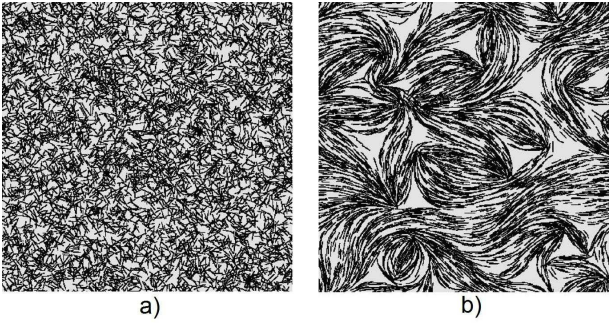}}
\caption{\label{figAlberSmltn}Simulational results of the bacteria motion and pattern formation deep inside the colony. (a) In the initial setup the cells are randomly distributed. (b) The inner area of the colony after 3 h of evolution. Adapted from \citet{MutansBacik}.}
\end{figure}

Simulations based on the above model did not result in swarming of the non-reversing cells in contrast to the simpler models by \citet{PeruaniDB06} and \citet{GinRods09}. On the other hand, it was found that the expansion rate of the colony depends on the length of the reversal period. Notably, the biggest expansion is obtained within the same time-period that was experimentally observed, that is $\approx 8$ min. The cellular motion and the emerged patterns deep inside the colony was also modeled. As it can bee seen on Fig. \ref{figAlberSmltn}, the considered social interactions result an enhanced order regarding the collective cellular motion. It should be noted, that in a very recent preprint \citet{PeruaniPreprint10} found signs of both ordering and clustering in experiments with a non-reversing, genetically modified mutant of a myxobacteria strain.\par

One of the earliest works on the collective motion of bacteria pointing out the reason why such models are important, is done by \citet{CmplxBactCln96}. The authors emphasized that the study of bacterial colonies can lead to interesting insights into the functioning of self-organized biological systems which rest on complex networks of regulation systems, since these are perhaps the simplest living systems exhibiting collective behavior, governed by interactions that are simple enough to be captured by mathematical tools.\par

In this paper the authors, on the one hand, reported on their experiments with \emph{Bacillus subtilis} (see also Sec. \ref{subsec:ExperimBaci} and Fig. \ref{figBSVortex}) and on the other hand introduced a step-by-step elaborated model, which is capable to describe the increasingly elaborated complex collective behavior. The simplest expression describes the \emph{collective migration} of the cells, which move with a fixed-magnitude velocity $v$ in the direction characterized by $\vartheta$, according to Eq. (\ref{eq:bcvm1})
\begin{equation}
\frac{d \vartheta_i}{dt}=\frac{1}{\tau} \left[ \left\langle \vartheta(t)\right\rangle_{i, \epsilon} - \vartheta_i(t)\right]+\zeta
\label{eq:bcvm1}
\end{equation}
where $\vartheta_i(t)$ is the direction of the $i$th bacterium at time $t$, $\tau$ is the relaxation time, which is related to the bacterial length to width ratio (the interaction is stronger for longer bacteria), and $\zeta$ indicates an uncorrelated noise. The term $\left\langle \vartheta(t)\right\rangle_{i, \epsilon}$ represents the average direction of the cells in the neighborhood of particle $i$, in the radius $\epsilon$.\par
For the simulations, a more simple, time-discretized form of Eq. (\ref{eq:bcvm1}) was used (Eq. \ref{eq:UjVi}), which is valid if the rotational relaxation time is fast compared to the change of the locations, that is, if $\tau << v^{-1}/\sqrt{\bar{\rho}}$. ($\bar{\rho}$ denotes the average bacterium density.) 


Eq. (\ref{eq:UjVi}) can be interpreted as a ``starting-point'' which is to be refined according to the specific systems. Here the noise takes values from the interval $[-\eta/2, \eta/2]$ randomly, with uniform distribution. The $x_i$ positions of particle $i$ is updated in each time-step according to Eq. (\ref{eq:UjXi}).\par
\begin{figure}
\centerline{\includegraphics[angle=0,width=0.55\columnwidth]{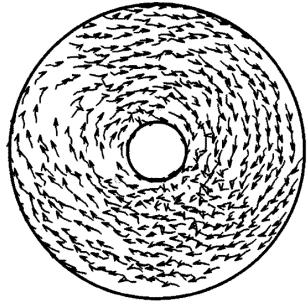}}
\caption{\label{figCBC96Fig6}A stationary state of the system characterized by short-range repulsion among the cells (defined by Eq. (\ref{eq:bcvm5})) using reflective circular walls. From \citet{CmplxBactCln96}.}
\end{figure}

Modifying the above model to be more system-specific, two changes were introduced: (i) the periodic boundary conditions were replaced with reflective circular walls, and (ii) a short-range ``hard-core'' repulsion was introduced, in order to prevent the cells to aggregate in a narrow zone. In other words, if the distance among cells decrease under a certain value $\epsilon^{*}$, then these cells will repel each other, and their direction of motion will be given by 

\begin{equation}
\vartheta_i(t+\Delta t) = \Phi \left( - \sum_{j \neq i, \left| \vec{x_j} - \vec{x_i} < \epsilon^{*} \right|} \vec{N}\left( \vec{x_j}(t) - \vec{x_i}(t) \right)\right),
\label{eq:bcvm5}
\end{equation}
where $\Phi(\vec{r})$ gives the angle $\vartheta$ between its argument vector and a predefined direction (for example the $x$ axis), and $\vec{N}=\vec{u}/|\vec{u}|$. Simulations with low noise and high density show correlated rotational motion (see Fig. \ref{figCBC96Fig6}), in which the direction of the vortices can be either clockwise or anti-clockwise, as it is selected by spontaneous symmetry breaking.\par

\begin{figure}
\centerline{\includegraphics[angle=0,width=0.64\columnwidth]{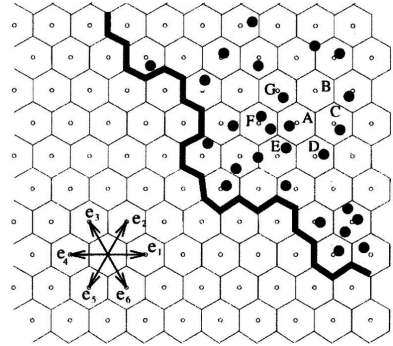}}
\caption{\label{figCBC96Fig7}The discretized concentration field: a hexagonal lattice defined by the lattice vectors $\vec{e_1}, \vec{e_2}, \ldots, \vec{e_6}$. The open circles in the middle of the hexagons are those points where the concentration level of the diffusing chemoattractants are calculated at each time-step. The thick line shows the boundary of the system, which reflects the particles (filled dots) which can move off-lattice. To define the average direction $\left\langle \vec{v} \right\rangle_{i, \epsilon}$ for the bacterium $i$ in lattice-cell $A$, the averaging involves all the particles in cells $A-G$. From \citet{CmplxBactCln96}.}
\end{figure}

The above constraint (reflective circular wall) is an externally imposed coercion to the bacterium colony. At the same time, in real colonies vortices often can be observed far from the boundaries as well, thus the confinement of the bacteria must be the resultant of some kind of interactions among the cells. Accordingly, the model can be further elaborated by adding ``chemoattractants'' to the system, which are interpreted in a broad sense: they can be reactions on ``passive'' physical forces as well (like surface tension, efficiency of the flagella-motors) which depend on the deposited extracellular slime. Cells slightly alter their propulsion forces according to the local concentration of the attractant, which results in a torque acting on the colony. To simulate the system that includes the above introduced attractants, the concentration field, $c_A$ (describing the concentration level of the secreted chemoattractants in each point of the field) is discretized by a hexagonal lattice (see Fig. \ref{figCBC96Fig7}). Supposing that a group of bacteria, a ``raft'', is held together by intercellular bonds, it can be treated as a rigid body of size $d$. In this case, the velocity difference $\Delta v$ at the opposite sides of the raft, in a linear approximation, is proportional to that component of $\nabla c_A$ which is orthogonal the velocity $\vec{v}$:

\begin{equation}
\Delta v \sim \frac{d}{v} | \vec{v} \times  \vec{\nabla} c_A |.
\label{eq:RaftDeltaV} 
\end{equation}

By neglecting the convective transport caused by the motion of bacteria, the chemoattractant field's time evolution can be written as

\begin{equation}
\frac{\partial c_A}{\partial t}=D_A \nabla^2 c_A + \Gamma_A \rho - \lambda_A c_A,
\label{eq:TimeEvolCA} 
\end{equation}
$\lambda_A$ is the constant rate of the decay, $\rho$ denotes the local density (number of particles in a unit area) and $\Gamma_A$ is the rate by which bacteria produce the chemoattractant material. The first term represents the diffusion.

\begin{figure}
\centerline{\includegraphics[angle=0,width=.8\columnwidth]{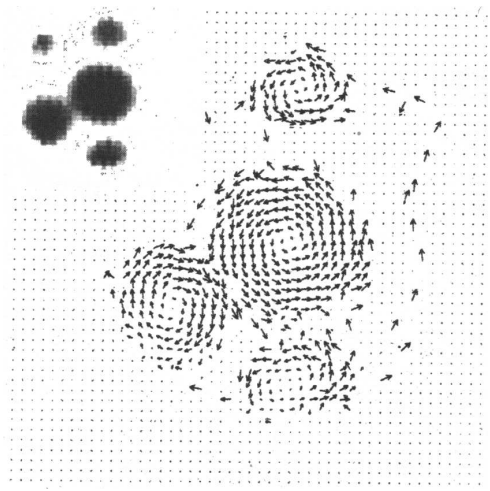}}
\caption{\label{figCBC96Fig8}Typical vortices formed by the model involving chemoattractants. The originally homogeneous particle distribution is destroyed by the positive feedback effect of the attractants. Arrows show the coarse grained velocity of the bacteria, while the corresponding distribution of the chemoattractant concentration is depicted in the upper left corner. From \citet{CmplxBactCln96}.}
\end{figure}

Figure \ref{figCBC96Fig8} depicts a typical snapshot of the simulations. The secretion of chemoattractants is a process with positive feedback effect, which breaks down the originally homogeneous particle distribution and results denser clusters in sparser regions.\par

\begin{figure}
\centerline{\includegraphics[angle=0,width=.85\columnwidth]{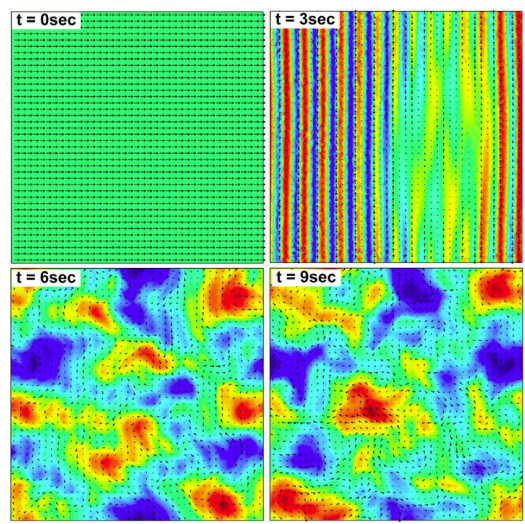}}
\caption{\label{figTurblntBct}(Color online) Four snapshots of the model reproducing the onset of the experimentally observed jets and vortices. The color map indicates the bacterial volume fraction, and the little arrows denote the fluid velocity field. According to the initial conditions ($t=0$), the bacteria are distributed uniformly and the fluid velocity field is directed (with a small random perturbation) along the $x$ axis. From \citet{TwoPhaseModBaci}.}
\end{figure}

In dense colonies of \emph{Bacillus subtilis} -- in which hydrodynamical effects (the effect of the medium through hydrodynamic interactions) play a significant role -- a surprising behavior can be observed: in regions of high bacterium concentration (having at least $10^{9}$ cells per $cm^{3}$) transient jet-like patterns and vortices appear. The latter ones persist for timescales of $\approx 1 s$ \citep{MBWAW99, ZurosBaciC}. The speed of the observed jets are typically larger than that of the individual bacteria. To elucidate these observations, \citet{TwoPhaseModBaci} developed a two-phase model in which the fluid and the bacteria were modeled by two independent, but interpenetrating continuum phases. Since their propulsive motors (the flagella) do not act on the center of mass, the rod-shaped bacteria exert a dipole force on the fluid. For reasonable parameter-values, the model (a system of partial differential equations) reproduces the observed behavior qualitatively. Figure \ref{figTurblntBct} represents the onset of the observed turbulent behavior with the jets and vortices.\par

\begin{figure}
\centerline{\includegraphics[angle=0,width=.95\columnwidth]{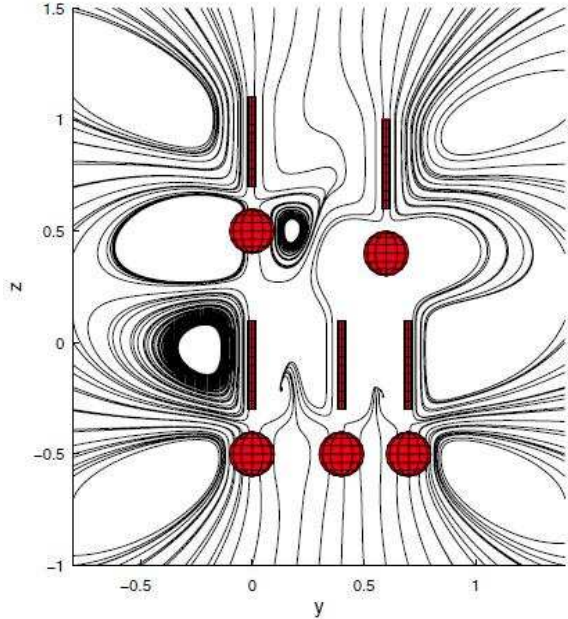}}
\caption{\label{figBctMdWall}(Color online) Streamlines of the fluid velocity field surrounding a group of five bacterium near to the walls. As it can be seen, there is little front-to-end penetration of the fluid into the group. Remarkably, as the authors point it out, this circumstance can lead to the split-up of a group because, as it follows, the oxygen supply for the organisms within a phalanx consisting of many bacteria will be insufficient, thus the inner cells will alter their velocity according to the gradient of oxygen concentration. From \citet{BacterialFluidMech}.}
\end{figure}

The interaction between these organisms under similar circumstances (namely, closely packed populations of \emph{Bacillus subtilis}) with one other and with the boundaries (walls) is in the focus of \citet{BacterialFluidMech}. Their model swimmer consist of a sphere (which is the ``body'' of the cell) and a cylinder representing the rotating bundle of helical flagella (see Fig. \ref{figBctMdWall}). The occurrence of the turbulent states at small Reynolds numbers (at $Re << 1$) is explained by the energy that the bacteria insert into the fluid as they swim in it. \par

\citet{CzirMatVics} used coupled differential equations to describe experimentally observed patterns of bacterial colonies. With such a method, they captured the periodic growth of \emph{Proteus mirabilis} colonies (see Fig. \ref{figBaciKorok} in Sec. \ref{subsec:ExperimBaci}). \citet{BioMechBaciOrder} emphasized the role of bio-mechanical interactions arising from the growth and division of the cells (see also Fig. \ref{figBaciMechOrdr} in Sec. \ref{subsec:ExperimBaci}), and developed a continuum model based on equations for local cell density, velocity and the tensor order parameter. \par
Readers interested in this field will find more details in \citep{HydrodymOf09} and in \citep{RamOf1Ishikawa}.

\subsection{Models with segregating units}\label{sec:CellSorting}

\begin{figure}
\centerline{\includegraphics[angle=0,width=.85\columnwidth]{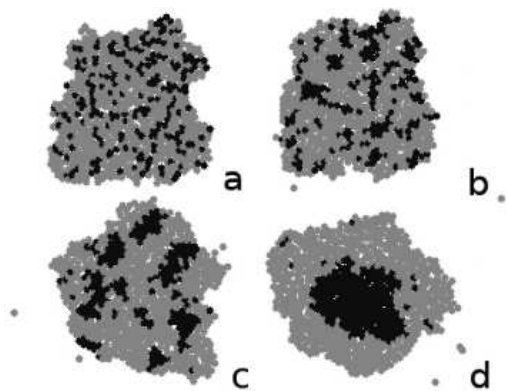}}
\caption{\label{figChtCellSrt}Cell sorting of 800 cells. The endodermic cells are represented by black, and the ectodermic ones by gray circles, respectively. (a) The initial cluster with mixed cell types. (b) the cluster after 3000 time step, and (c) is taken at $t=3 \times 10^{5}$. Clusters of endodermic cells form and grow as time passes by. (d) $t=2 \times 10^{6}$. A single endodermic cluster is formed, but some isolated cells remain within the ectoderm tissue, in agreement with the experiments of \citet{RieuHydra98}. From \citet{ChateSellSortingt}}
\end{figure}

Cell sorting denotes a special type of collective motion during which an originally heterogeneous mixture of cells segregate into two (or more) homogeneous cell clusters without any kind of external field. This can be observed, for example, during the development of organs in an embryo or during regeneration after tissue dissociation. To simulate this phenomena, \citet{ChateSellSortingt} considered two kinds of cells, differing in their interaction intensities. According to the model, $N$ particles move in a two-dimensional space with constant $v_0$ velocity. The velocity and the angle of the orientation of particle $n$ at time $t$ is denoted by $\vec{v}_{n}^{t}$ and $\theta_{n}^{t}$, respectively. The new orientation $\theta_{n}^{t+1}$ of particle $n$ is
\begin{equation}
\theta_{n}^{t+1}=arg \left[ \sum_{m} \left( \alpha_{nm}\frac{\vec{v}_{m}^{t}}{v_0}+\beta_{nm}f^{t}_{nm}\vec{e}^{t}_{nm} \right)+ \vec{u}_{n}^{t}\right],
\label{eq:BlmTh}
\end{equation}
where the summation refers to those particles ($m$) which are within a radius $r_0$. These `cells' exert a force $f^{t}_{nm}\vec{e}^{t}_{nm}$ on $n$, along the direction $\vec{e}^{t}_{nm}$. The noise is taken into account by $ \vec{u}_{n}^{t}$, which is a random unit vector with uniformly distributed orientation. $\alpha_{nm}$ and $\beta_{nm}$ are the control parameters: $\alpha$ controls the relative weights of the alignment interaction, and $\beta$ shows the strength of the radial two-body forces $f_{nm}$, which is defined as

\begin{equation}
\label{eq:Fnm}
f_{nm} = \left\{
	\begin{array}{lll}
		\infty & \mbox{if $r_{nm}<r_c$,}\\
		1-\frac{r_{nm}}{r_e} & \mbox{if $r_c < r_{nm}<r_0$,}\\
		0 & \mbox{if $r_{nm}>r_0$,}
	\end{array}	
\right.
\end{equation}
that is, for distances smaller than a core radius $r_c$, it is a strong repulsive force, around the equilibrium radius $r_e$ it is a harmonic-like interaction, and for distances bigger than the interaction range $r_0$ it is set to zero. For modeling the observations regarding \emph{Hydra} cells \citep{RieuHydra}, the authors defined two kinds of particles, ``endodermic'' and ``ectodermic'', denoted by 1 and 2, respectively. Accordingly, $\beta_{11}$ and $\beta_{22}$ stand for the cell cohesion \emph{within} the two cell-types, while $\beta_{12}$ and $\beta_{21}$ account for the \emph{inter-cell}-type interactions. These latter ones are assumed to be symmetric, that is, $\beta_{12}=\beta_{21}$. For the sake of simplicity, all the cells have the same $\alpha$ value. Figure \ref{figChtCellSrt} shows how a group of 800 cells evolve in time. The proportion is 1:3 of endodermic (black) to ectodermic (gray) cells, and $\alpha$=0.01.\par

Segregation occurs in various 3D systems as well, such as in flocks of birds or schools of fish. Mostly, models assume \emph{identical} particles to simulate collective motion. At the same time, those simulations which suppose diverse particles, exhibit sorting \citep{Romey96, CouzinKrause03}. This means that behavioral and/or motivational differences among animals effect the structure of the group, since individuals change their positions relative to the others according to their actual inner state. This involves that if the individual variations are persistent then the group will reassemble to its' original state after perturbations \citep{CouzinCollMem02}. The sorting phenomenon depends primarily on the \emph{relative} differences among the units.\par
\begin{figure}
\centerline{\includegraphics[angle=0,width=1\columnwidth]{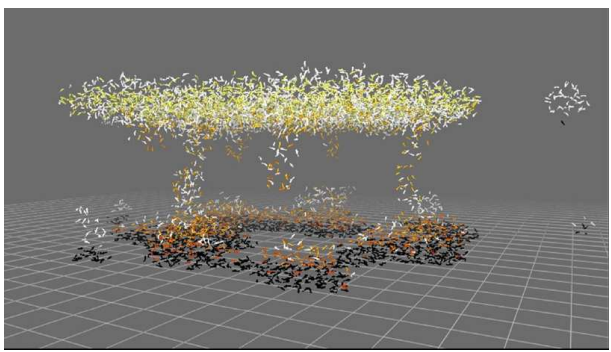}}
\caption{\label{figVaboHrng}(Color online) Simulational results, in which the motivational level of the individuals are taken into account. Large herring populations with small motivational synchronization tend to form multi-layered schools in which the layers are connected by cylindrical shaped ``bridges''. The motivational level depends primarily on the age of the fish: mature herrings are denoted with yellow color (online), ovulating ones are orange to red, spawning individuals are black and white color registers spent herring. Adapted from \citet{VaboHerring}.}
\end{figure}

In a similar spirit, \citet{VaboHerring} showed that differences in the \emph{motivational level} can cause segregation within a school of spawning herrings. They used an individual based model in which the parameters describing the states of the individuals were varied and they measured a range of parameters at system level. The motion of each individual was determined by the combination of five behavioral rules: (1) avoiding boundaries, (2) social attraction, (3) social repulsion, (4) moving towards the bottom to spawn and (5) avoiding predation. The motivational level was controlled by a parameter. To capture how the system as a whole reacts on changes in the individual level, various metrics were recorded, like the size and age of the school, its vertical and horizontal extension, etc. By varying the size of the population and the level of the motivational synchronization, different systems emerged regarding its morphology and dynamics. 
Similar motivational levels resulted in an integrated school, whereas diverse inner states produced a system with frequent split-offs. More complex structure appeared by an intermediate degree of synchronization characterized by layers connected with vertical cylindrical shaped schools (see Fig. \ref{figVaboHrng}) allowing ovulating and spent herring to move across the layers, in agreement with the observations \citep{HerringObs}. These findings suggest that the level of motivational synchronization among fish determines the unity of the school. Furthermore, this study also demonstrates that larger populations can exhibit such emergent behaviors that smaller ones can not (for example the cylindrical bridges mentioned above).\par
More general simulations also support these results. \citet{YouSegr09} investigated the behavior of two-component swarms, consisting of two different kinds of particles, varying in their parameters, such as mass, self-propelling strength or preferences for shelters. The units, having different parameter-set, were observed to segregate from each other.

Other experiments focus on the emergent patterns of particles having different kinetic parameter settings (preferred speed, the range of perception in which a particle perceives the velocity vectors of other particles, etc.).  The study of \citet{SwarmChemistry} was prompted by an in-class experiment aiming to test a new version of a software called ``Swarm Chemistry'',\footnote{it can be downloaded from the author's website, http://bingweb.binghamton.edu/\~{ }sayama/SwarmChemistry/} which is an interactive evolutionary algorithm.\par
\begin{figure}
\centerline{\includegraphics[angle=0,width=0.93\columnwidth]{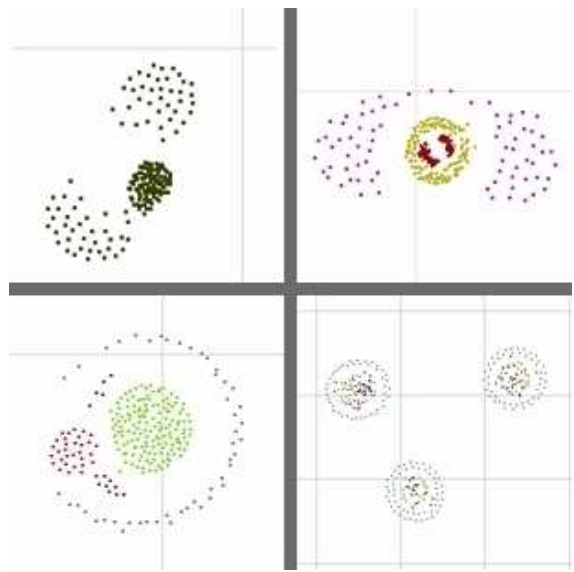}}
\caption{\label{figSwrmChm}(Color online) Some snapshots of the emergent patterns that particles with different parameter-set (preferred speed, range of perception and strength of the kinetic rules) can produce. From \texttt{http://bingweb.binghamton.edu/\~{ }sayama/} \texttt{SwarmChemistry/\#recipes}.}
\end{figure}

 
The software assumes that the particles move in an infinite two-dimensional space according to kinetic rules resembling to the ones introduced by \citet{Reynolds87} (see Sec. \ref{sec:OrigSPPMdl}). The strength of these kinetic rules, as well as the preferred speed and the local perception range, \emph{differ} from particle to particle. Those units that (accidentally) share the same parameter-set, are considered to be of the same type. Some snapshots of the emergent dynamic patterns that these particles produce with their various parameter-sets, can be seen on Fig. \ref{figSwrmChm}.

According to the simulations, these mixtures of multiple type units usually spontaneously undergo to some kind of segregation process, often accompanied by the appearance of multilayer structures. Furthermore, the formed clusters may exhibit various dynamic macroscopic behaviors, such as oscillation, rotation or linear motion.\par
Interestingly, simulations of \emph{hunting} showed segregation as well \citep{ChasingNtr}. Here the two kinds of particles were the chasers (or predators) and the targets (or preys) which differed in their behavior.

\subsection{Models inspired by animal behavioral patterns}

\subsubsection{Insects}

\begin{figure}[t]
	\centerline{\includegraphics[angle=0,width=1\columnwidth]{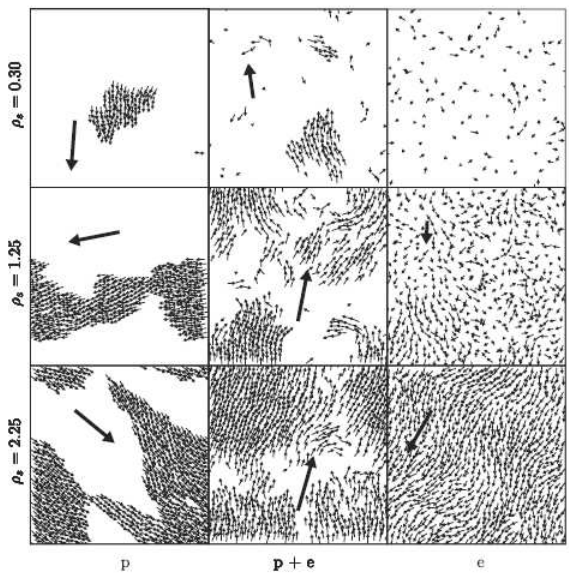}}
  \caption{\label{figRomanczukCnbl}The global collective motion emerging from escape and pursuit behaviors. $\rho_s$ is the \emph{rescaled density}, which is defined as $\rho_s=N l_s^{2}/L^{2}$, where $N$ is the total number of the individuals, $l_s$ is their interaction range and $L$ is the size of the simulation field. The simulations were carried out by using periodic boundary conditions. The column denoted by $p$ shows the typical spatial configurations for the pursuit-only case, $e$ for the escape-only case, and $p+e$ when both interactions are present. The large arrows indicate the direction and speed of the mean migration. As it can be seen on the top row, at low rescaled densities the emergent patterns strongly vary according to the strength of the escape and pursuit interactions. From \citet{Romanczuketal2009}.}
\end{figure}

As mentioned in Sec. \ref{subsec:ExperimInsects}, Mormon crickets and Desert locusts tend to exhibit cannibalistic behavior in case of the depletion of nutritional resources \citep{CannibalCrickets, CannibalLocusts}. Motivated by these observations, it can be shown that individuals with \emph{escape} and \emph{pursuit} behavior-patterns (which special kind of repulsive and attractive behaviors can be correlated with cannibalism) exhibit collective motion. \citep{Romanczuketal2009}. The escape reaction is triggered in an individual if it is approached from behind by another one; in this case the escaping animal increases its velocity in order to prevent being attacked from behind. In contrast, if the insect perceives one of its mates moving away, it increases its velocity in the direction of the escaping one; this is the pursuit behavior. Other cases do not trigger any response. According to the simulations, at moderate noise intensity and high particle density, these interactions (pursuit and escape) lead to global collective motion, irrespective of the detailed model parameters (see Fig. \ref{figRomanczukCnbl}). Both interaction-types lead to collective motion, but with an opposite effect on the density distribution. Whereas pursuit leads to density-inhomogeneities (that is, to the appearance of clusters, as it can be seen on the first column on Fig. \ref{figRomanczukCnbl}), escape calls to forth homogenization. Thus, the collective dynamic in which both behavior-types are present, is a competition between the two opposite effects.\par

Another often observed phenomena regarding collective locust motion is their sudden, coherent switches in the direction of motion. \citet{YatesPnas09} suggested to use Fokker-Plank equations in order to describe these observations. They found a seemingly counterintuitive result, namely that the individual locusts \emph{increased their motional randomness} as a reaction for a \emph{loss} of alignment in the group. This reaction thought to facilitate the group to find a highly aligned state again. They also found that the mean switching time increased exponentially with the number of individuals. Recently \citet{ErgodicDirSw} suggested that these ergodic directional switches might be the resultants of the small errors that the insects make when trying to adopt their motion to that of their neighbors. These errors usually cancel each other out, but over exponentially long time periods they have the possibility of accumulating and producing a switch.


\subsubsection{Moving in three dimensions -- fish and birds}\label{sec:ThreeDModels}

The main goal of the first system-specific models aiming to simulate the motion of animals moving in 3 dimensions (primarily birds and fish) was to produce realistically looking collective motion \citep{Reynolds87}, to give system-specific models taking into account many parameters \citep{OkuboPrsKonyv}, or to create systems in which some characteristics (for example nearest neighbor distance or density) resembles to an actual biological system \citep{HuthWissel94}. Later various other aspects and features were also studied, such as the function and mechanism of line versus cluster formation in bird flocks \citep{BHeppnerCikk}, how the fish size and kinship correlates with the spatial characteristics (e.g., animal density) of fish schools \citep{HemelrijkK05}, the effect of social connections (``social network'') on the collective motion of the group \citep{BodeSocNetwork2011}, the cohort departure of bird flocks \citep{HeppnerPrsBook}, the collective behavior in an ecological context (in which not only the external stimuli, but the internal state of the individuals are also taken into account) \citep{VaboHerring}, or the effect of perceived threat \citep{BodeFishThreat}. This latter circumstance was taken into account by relating it to higher updating frequency, and it resulted a more synchronized group regarding its speed and nearest neighbor distribution. The book of \citet{AnimalGrIn3D} provides an excellent review on this topic.\par

Models proposed by biologists tend to take into account many of the biological details of the modeled animals. A good example for this kind of approach is the very interesting work of \citet{heppner90}, who proposed a system of stochastic differential equations with 15 parameters. \citet{hdofHW92} used a similar approach for schools of fish. Some other models included a more realistic representation of body size and shape \citep{hdofKH03}.\par

Regarding the methodology of most of the simulations, the \emph{agent-based} (or \emph{individual-based}) approach proved to be very popular (although there are alternative approaches as well, e.g., \citet{JapiGeese}). The reason behind this is that this approach provides a link between the behavior of the individuals and the emergent properties of the swarm as a whole, thus appropriate to investigate how the properties of the system depend on the actual behavior of the individuals. The most common rules applied in these models are: (i) short-range repulsive force aiming to avoid collision with mates and with the borders, (ii) adjusting the velocity vector according to the direction of the neighboring units, (iii) a force avoiding being alone, e.g., moving towards the center of the swarm's mass, (iv) noise, (v) some kind of drag force if the medium is considered in which the individuals move (which is usually \emph{not} taken into account, in case of birds and fish). Then, the concrete models differ in the rules they apply (usually most of the above ones), in their concrete form and in the system parameters.\par
Some biologically more realistic, yet still simple individual-based models were also suggested \citep{CouzinCollMem02, HemHildFish08}. In a computer model called `StarDisplay', \citet{HildStarlingModel} combined the above mentioned ``usual rules'' with system specific ones in order to generate patterns that resemble 
to the aerial displays of starlings, recorded by the Starflag project (see Sec. \ref{subsec:ExperimBirds}).\par
\begin{figure}
  \centerline{\includegraphics[angle=0,width=.6\columnwidth]{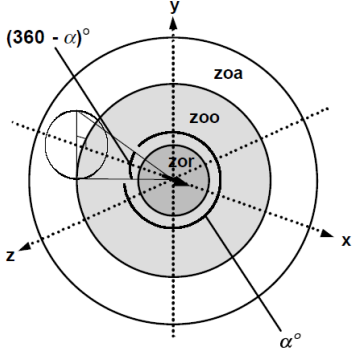}}
  \caption{\label{figCouzinGen3DZones}The interaction zones, centered around each individual. The inner-most sphere with radius $R_r$ is the \emph{zone of repulsion} (``zor''). If others enter this zone, the individual will response by moving away from them into the opposite direction, that is, it will head towards $-\sum_{j \neq i}^{n_r} (\vec{r}_j-\vec{r}_i) / |\vec{r}_j-\vec{r}_i| $, where $n_r$ in the number if individuals being in the \emph{zor}. The interpretation of this zone is to maintain a personal space and to ensure the avoidance of collisions. The second annulus, ``zoo'', represents the \emph{zone of orientation}. If no mates are in the `zor', the individual aligns itself with neighbors within this `zoo' region. The outermost annulus, ``zoa'', is the \emph{zone of attraction}. The interpretation of this region is that group-living individuals continually attempt to join a group and to avoid being alone or in the periphery. From \citet{CouzinCollMem02}.}
\end{figure}

\citet{CouzinCollMem02} categorized the emergent collective motions as the function of the system parameters. In this framework, the individuals obey to the following basic rules: (i) they continually attempt to maintain a certain distance among themselves and their mates, (ii) if they are not performing an avoidance manoeuvre (described by rule $i$), then they are attracted towards their mates, and (iii) they align their direction to their neighbors. Their perception zone (in which they interact with the others) is divided into three non-overlapping regions, as illustrated in Fig. \ref{figCouzinGen3DZones}. The radius of these spheres (zone of repulsion, zone of orientation and zone of attraction) are $R_r$, $R_o$ and $R_a$, respectively. Thus, the width of the two outer annulus are $\Delta R_o=R_o-R_r$ and $\Delta R_a=R_a-R_o$. $\alpha$ denotes the field of perception, thus, the ``blind volume'' is behind the individual, with interior angle $(360-\alpha)^{}$.

In order to explore the global system behavior, the authors analyzed the consequences of varying certain system parameters, like the number of individuals, preferred speed, turning rate, width of the zones, etc. For every case, the following two global properties were calculated:

\begin{equation}
 \varphi(t)=\frac{1}{N}\left| \sum_{i=1}^{N} \vec{v_i^u}(t) \right|,
\label{eq:GrpPolarization}
\end{equation}
where $N$ is the number of individuals within the group, $(i=1, 2, \ldots, N)$, and $\vec{v_i^u}(t)$ is the unit direction vector of the $i$th animal at time $t$. (Since $\vec{v_i^u}(t)$ is a \emph{unit} vector, the expression defined by Eq. (\ref{eq:GrpPolarization}) is equivalent with the order parameter defined by Eq. (\ref{eq:NormV}).)\par
The other measure, \emph{group angular momentum}, is the sum of the angular momenta of the group-members about the center of the group, $\vec{r}_{Gr}$. This expression measures the degree of rotation of the group about its center

\begin{equation}
 m_{Gr}(t)=\frac{1}{N}\left| \sum_{i=1}^{N} \vec{r}_{i-Gr}(t) \times \vec{v_i^u}(t) \right|,
\label{eq:GrpMomentum}
\end{equation}
where $\vec{r}_{i-Gr} = \vec{r}_i - \vec{r}_{Gr}$, is the vectorial difference of the position of individual $i$, $\vec{r_i}$, and the position of the group-center, $\vec{r}_{Gr}$.

\begin{equation}
 r_{Gr}(t)=\frac{1}{N} \sum_{i=1}^{N} \vec{r_i}(t) 
\label{eq:GrCenter}
\end{equation}

\begin{figure}
  \centerline{\includegraphics[angle=0,width=1\columnwidth]{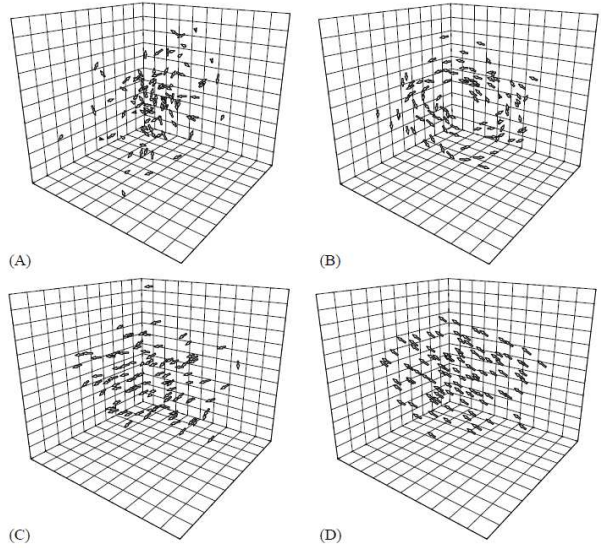}}
  \caption{\label{figCouzinGen3DReslt}The ``basic types'' of collective motions exhibited by the model, according to the various parameter setups. The denominations are: (a) swarm, (b) torus, (c) dynamic parallel group, (d) highly parallel group. Adapted from \citet{CouzinCollMem02}.}
\end{figure}

Figure \ref{figCouzinGen3DReslt} summarizes the four ``basic types'' of collective motions emerged according to the various parameter setups.\par
``\emph{Swarm:}'' Both the order parameter ($\varphi$) and the angular momentum ($m_{Gr}$) are small, which means little or no parallel orientation. (Fig. \ref{figCouzinGen3DReslt}, (a) sub-picture.)\par
``\emph{Torus}'' or ``milling'': Individuals rotate around an empty core with a randomly chosen direction. The order parameter ($\varphi$) is small, but the angular momentum ($m_{Gr}$) is big. This occurs when $\Delta r_a$ is big, but $\Delta r_o$ is small. (Fig. \ref{figCouzinGen3DReslt}, (b) sub-picture.)\par
``\emph{Dynamic parallel group:}'' Occurs at intermediate or high values of $\Delta r_a$ and intermediate values of $\Delta r_o$. This formation is much more mobile than either of the previous ones. The order parameter ($\varphi$) is high but the angular momentum ($m_{Gr}$) is small. (Fig. \ref{figCouzinGen3DReslt}, (c) sub-picture.)\par
``\emph{Highly parallel group:}'' By increasing $\Delta r_o$, a highly aligned formation emerges characterized by very high order parameter ($\varphi$) and low angular momentum ($m_{Gr}$). (Fig. \ref{figCouzinGen3DReslt}, (d) sub-picture.)\par

However, the approach of individual (or agent) based modeling has its own limitations or ``traps'' -- as pointed out by \citet{bla1}. Namely, that different combinations of rules and parameters may provide the same (or very similar) patterns and collective behaviors. Accordingly, in order to prove that the emergent behavior of a certain biological system obeys given principles, it is not enough to provide a rule and a parameter set (modeling these principles) and demonstrate that they reproduce the observed behavior. (On the other hand, \citet{CouzinCollMem02} demonstrated that -- vice versa -- the same rule and parameter set may result in different collective behavior in the very same system, depending on its recent past, ``history''.) \citet{YatesPnas09} note that many models have weak predictive power concerning the various relevant aspects of collective motion, including, for example, the rate at which groups suddenly change their direction.

\subsection{The role of leadership in consensus finding}\label{sec:LdrshpModels}

Animals traveling together have to develop a method to make collective decisions regarding the places of foraging, resting and nesting sites, route of migration, etc. By slightly modifying (typically extending) models (such as the one described in Sec. \ref{sec:ThreeDModels}), a group of individuals can get hold of such abilities. Accordingly, \citet{CouzinNtrCollDcsn} suggested a simple model in which individuals were not required to know how many and which individuals had information, they did not need to have a signaling mechanism and no individual recognition was required from the group members. Informed individuals were not necessitated to know anything about the information-level of their mates either and that how the quality of their information was compared to that of others.
The model looks as follows:\par
$N$ individuals compose the group. The position of the $i$th particle is described by the vector $\vec{r}_i$, and it is moving in the direction $\vec{v}_i$. The group members endeavor to continually maintain a minimum distance, $\alpha$, among themselves, by turning away from the neighbors $j$ which are within this range towards the opposite direction, described by the desired direction $\vec{d}_i$

\begin{equation}
 \vec{d}_i(t+\Delta t)= -\sum_{j \neq i} \frac{\vec{r}_{j}(t)-\vec{r}_i(t)} {\left| \vec{r}_{j}(t)-\vec{r}_i(t)  \right|}
\label{eq:LdrspMinDstnc}
\end{equation}

This rule has the highest priority. If there are no mates within this range, than the individual will attempt to align with those neighbors $j$, which are within the interaction range $\rho$. If so, the desired direction is defined as

\begin{equation}
 \vec{d}_i(t+\Delta t)= \sum_{j \neq i} \frac{\vec{r}_{j}(t)-\vec{r}_i(t)} {\left| \vec{r}_{j}(t)-\vec{r}_i(t)  \right|} + \sum_{j \neq i} \frac{\vec{v}_j(t)}{\left| \vec{v}_j(t)\right|}.
\label{eq:LdrspRule2}
\end{equation}

We will use the corresponding unit vector, $\hat{d}_i(t)=\vec{d}_i(t) / |\vec{d}_i(t)|$.\par
Until this point the algorithm is very similar to the one described in Sec. \ref{sec:ThreeDModels}. In order to study the influence of informed individuals, a portion of the group, $p$, is given information about a preferred direction, described by the unit vector $\vec{g}$. The rest of the group is naive, without any preferred direction. Informed individuals balance their social alignment and their preferred direction with the weighting factor $\omega$

\begin{equation}
 \vec{d}_i(t+\Delta t)= \frac{\hat{d}_{i}(t +\Delta t) + \omega \vec{g}_i } {\left|   \hat{d}_{i}(t +\Delta t) + \omega \vec{g}_i   \right|}.
\label{eq:LdrspRule3}
\end{equation}
($\omega$ can exceed 1; in this case the individual is influenced more heavily by its own preferences than by its mates.) The \emph{accuracy of the group} (describing the quality of information transfer) is characterized by the normalized angular deviation of the group direction around the preferred direction $\vec{g}$, similarly to the term given in Eq. \ref{eq:GrpMomentum}.\par
\begin{figure}
  \centerline{\includegraphics[angle=0,width=1\columnwidth]{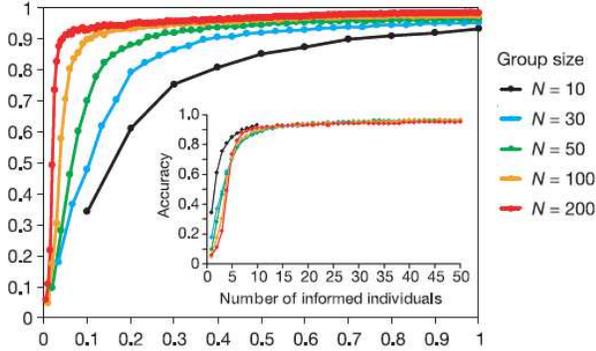}}
  \caption{\label{figGrAccrc}(Color online) SPPs following simple rules can compose systems in which a few informed individual is capable to guide the entire group towards a preferred direction. The accuracy of the group (following the rules given by Eqs. \ref{eq:LdrspMinDstnc}, \ref{eq:LdrspRule2} and \ref{eq:LdrspRule3}) increases asymptotically as the portion of the informed individuals increases. Adapted from \citet{CouzinNtrCollDcsn}.}
\end{figure}

The authors found that for fixed group size, the accuracy increases asymptotically as the portion $p$ of the informed members increased, see Fig. \ref{figGrAccrc}. This means, that the larger the group, the smaller the portion of informed members is needed, in order to guide the group towards a preferred direction.\par

\begin{figure}
  \centerline{\includegraphics[angle=0,width=1\columnwidth]{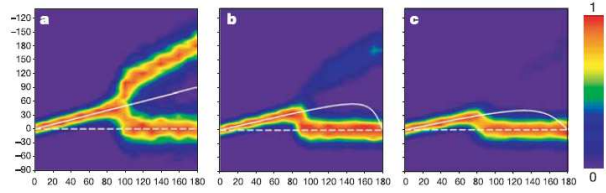}}
  \caption{\label{figCouzin05Lila}(Color online) Collective group direction when two groups of informed individuals differ in their preferences. The vertical axis shows the degree of the  most probable group motion. The first group (consisting of $n_1$ informed individuals) prefers the direction characterized by 0 degrees (dashed line), while the second group (consisting of $n_2$ informed individuals) prefers a direction between $0 - 180$ degrees (horizontal axis). The group consists of 100 individuals altogether, of which the numbers of informed individuals are (a) $n_1=n_2=5$, (b)  $n_1=6$ and $n_2=5$ (c) $n_1=6$ and $n_2=4$. Adapted from \citet{CouzinNtrCollDcsn}.}
\end{figure}

However, informed individuals might also differ in their preferred direction. 
If the number of individuals preferring one or another direction is equal, than the group direction will depend on the degree to which these
preferred directions differ from each other: if these preferences are similar, then the group will go in the average preferred direction of all informed individuals. As the differences among the preferred directions increase, individuals start to select randomly one or another preferred direction. If the number of informed individuals preferring a given direction increases, the entire group will go into the direction preferred by the majority, even if that majority is small (see Fig. \ref{figCouzin05Lila}).\par

\citet{FreemanBiroLdrsp} extended this model by including a ``social importance factor'', $h$, describing the \emph{strength of the effect} of a given individual on the group movement. That is, $h$ varies with each agent, and the higher this value is, the bigger influence the given unit exerts on the group. Equation \ref{eq:LdrspRule2} is modified accordingly
\begin{equation}
 \vec{d}_i(t+\Delta t)= \sum_{j \neq i} h_j \frac{\vec{r}_{j}(t)-\vec{r}_i(t)} {\left| \vec{r}_{j}(t)-\vec{r}_i(t)  \right|} + \sum_{j \neq i} h_j \frac{\vec{v}_j(t)}{\left| \vec{v}_j(t)\right|}
\label{eq:LdrspBiro}
\end{equation}


These models show that leadership might emerge from the differences of the level of information possessed by the group members. Importantly, since the information can be transient and different group members may have pertinent information at different times or in different contexts, leadership can be transient and transferable as well. Other studies also support these results. \citet{quera2010} used an other kind of rule-set by which their agents moved, and observed the same: certain agents did become leaders without anything in the rule-set or in the initial conditions that would have prompted or predicted it.\par

In addition, even simpler models can lead to consensus decisions. For example, the severe quorum rule (in which the probability that an individual follows a given option, sharply increases when the number of other group members making that very decision reaches a threshold) resulted accurate group decisions as well \citep{SumpterConsFish}.\par


Despite many attempts, the research of ``human decision making'' is still in its infancy. \citet{SocStatFizOlasz} give a nice review of the state of the art regarding the physical and mathematical models which have been proposed throughout the years in the field of social dynamics. \citep{RamOf4Ramaswamy} and \citep{RamOf5TTR} provide general informative reviews of the models and approaches of collective motion.

\subsection{Relationship between observations and models}\label{sec:ModelMegfigyKapcs}

Throughout this review, we have discussed two kinds of models: one of them, reviewed in Sec. \ref{sec:Models}, constitutes the ones addressing the basic laws and characteristics of flocking. In fact, the aim of these models is to give a \emph{minimal model}, i.e., a description, including those and only those rules which are inevitable for the emergence of collective motion. All other aspects are neglected, by design. The question these studies aim to clarify is the following: What are the necessary conditions for collective motion to appear and how the emergent behavior is affected by these terms?\par
On the other hand, Sec. \ref {sec:SystmSpecModels} comprises models which aim to apprehend a well-described observation as adequately as possible. In fact, a model is considered to describe a phenomenon properly if it is capable of giving predictions, that is, the behavior of the observed system can be foretold by applying the model. Since models ensure a deep insight into the observed phenomenon, many authors prefer to propose straight away a model accompanying their experimental findings (\citep{KellerSegelM, CmplxBactCln96, ReverseBaci, MutansBacik, CzirMatVics, RotatingSlimeMold, SzaboEtAl, AntLaneCF, CouzinScience, SaintillanGrMElectr})
while other models give account for previously reported observations. As such, \citet{TwoPhaseModBaci} developed a two-phase model in order to describe the jet-like patterns and vortices appearing in dense colonies of Bacillus subtilis (\citep{MBWAW99, ZurosBaciC}), \citet{ChateSellSortingt} proposed a model which gives account for the cell-sorting phenomenon observed in Hydra cells by \citet{RieuHydra}, or we can mention \citet{VaboHerring} who modeled the unique forms of  herring schools observed by Axelsen et al. in 2000.\par
Sometimes the theoretical calculations come before the observations, i.e., the given model predicts a previously unreported phenomenon. As an example, Toner and Tu predicted the phenomenon of giant number fluctuation in their 1995, 98 paper, which was observed more than a decade later by \citet{LongLivedFlucN}. These cases lend strong support for the models giving the predictions.\par
Overall, we conclude that a considerable number of evidences have accumulated each demonstrating that both qualitative and quantitative agreements between observations (of collective motion) and their theoretical/modeling interpretations can be established.


\section{Summary and conclusions}\label{sec:summ}

From the continuously growing number of exciting new publications on flocking we are tempted to conclude that collective motion can be regarded as an emerging field on the borderline of several scientific disciplines. Thus, it is a multidisciplinary area with many applications, involving statistical physics, technology and branches of life sciences. Because of the nature of the problem (treating many similar entities) studies in this field make quantitative comparison with observations possible even for living systems and there is a considerable potential for constructing theoretical approaches.\par

The results we have presented support a deep analogy with equilibrium statistical physics. The essential deviation from equilibrium is manifested in the ``collision rule'': since the absolute velocity of the particles is preserved and in most cases an alignment of the direction of motion after interaction is preferred, the total momentum increases both during individual collisions and, as a result, gradually in the whole system of particles as well.\par

The observations we have discussed can be successfully interpreted in terms of simple simulational models. Using models based on simplified units (also called particles) to simulate the collective behavior of large ensembles has a history in science, especially in statistical physics, where originally particles represented atoms or molecules. With the rapid increase of computing power and a growing appreciation for `understanding through simulations', models based on a plethora of complex interacting units, nowadays widely called agents, have started to emerge. Agents, even those that follow simple rules, are more complex entities than particles  because they have a goal they intend to achieve in an optimal way (for example, using as little amount of resources as possible).\par

As a rule, models of increasing complexity are bound to be born in order to account for the interesting variants of a fundamental process. However, there is a catch. A really good model must both reproduce truly life-like behavior and be as simple as possible. If the model has dozens of equations and rules, and correspondingly large numbers of parameters, it is bound to be too specific and rather like an `imitation' than a model that captures the few essential features of the process under study.\par

Thus, if a model is very simple, it is likely to be applicable to other phenomena for which the outcome is dominated by the same few rules. On the other hand, simplification comes with a price, and some of the exciting details that distinguish different phenomena may be lost. But, as any fan of natural life movies knows, collective behavior in nature typically involves sophisticated, occasionally amazing techniques aimed at coordinating the actions of the organisms to maximize success. The art of designing models of reality is rooted in the best compromise between oversimplification and including too many details (eventually preventing the location of the essential features).\par

On the basis of the numerous observations and models/simulations we have discussed above, the following conclusions can be made concerning the general features of systems exhibiting collective motion:
\begin{itemize}
\item Most patterns of collective motion are universal (the same patterns occur in very different systems)
\item Simple models can reproduce this behavior
\item A simple noise term can account for numerous complex deterministic factors
\item Global ordering is due to non-conservation of the momentum during individual collisions/interactions of pairs of units
\end{itemize}

The universally occurring patterns can be divided into a few classes of motion patterns:
\begin{itemize}
\item Disordered (particles moving in random directions)
\item Fully-ordered (particles moving in the same direction)
\item Rotational (within a rectangular or circular area)
\item Critical (flocks of all sizes moving coherently in different  directions. The whole system is very sensitive to perturbations)
\item Quasi-long range velocity correlations and ripples 
\item Jamming
\end{itemize}
A few further, less widely-occurring patterns are also possible, for example in systems made of two or more distinctively different types of units.

The following types of transitions between the above collective motion classes are possible:
\begin{itemize}
\item Continuous (second order, accompanied with large fluctuations and algebraic scaling)
\item Discontinuous (first order)
\item No singularity in the level of directedness
\item Jamming (transition to a state in which mobility is highly restricted)
\end{itemize}

The above transitions usually take place as i) a function of density, or, ii) the changing magnitude of perturbations the units are subject to. The role of noise is essential; all systems are prone to be strongly influenced by perturbations. In some cases noise can have a paradoxical effect and, e.g., facilitate ordering. This could be understood, for example, as a result of perturbations driving the system out from an inefficient deterministic regime (particles moving along trajectories systematically (deterministically) avoiding each other) into a more efficient one, characterized by an increased number of interactions.\par

After reviewing the state of the art regarding collective motion, we think that some of the most exciting challenges in this still emerging field can be summarized as: i) Additional, even more precise data about the positions and velocities of the collectively moving units should be obtained for establishing a well defined, quantitative set of interaction rules typical for most of the flocks. ii) the role of leadership in collective decision making should be further explored. Is it hierarchical? Can a leadership-driven decision-making mechanism be scalabe up to huge group sizes? iii) The problem of a coherently-moving, self-organized flock of unmanned aerial vehicles is still unsolved in spite of its very important potential applications, iv) and last, but far from being the least, the question about the existence of some simple underlying laws of nature (such as, e.g., the principles of thermodynamics) that produce the whole variety of the observed phenomena we discussed is still to be uncovered.\par

We have seen that using methods common in statistical physics has been very useful for the quantitative description of collective motion. Theories based on approaches borrowed fluid dynamics, data evaluation techniques making use of correlations functions and many particle simulations all have led to a deeper insight into flocking phenomena. This is all in the spirit of going from a qualitative to a more quantitative interpretation of the observations: a widely preferred direction in life sciences these days.\par

A quantitative frame for describing the behavior of a system enables important, highly desirable features of treating actual situations. For example, prediction of the global displacement of huge schools of fish may have direct economic advantages. Understanding the collective reaction of people to situations including panic may lead to saving lives. Using computer models to simulate migration of birds or mammals can assist in preserving biodiversity. The list of potential applications is long, and likely to get longer, especially if we take into account the rapidly increasing interest in collective robotics.

\subsection*{Acknowledgments}
T.V. thanks collaboration and numerous helpful exchanges of ideas with a long list (too long to reproduce here, but cited in the text) of outstanding colleagues over the past 15 years. Writing this review was made possible in part by a grant from EU ERC No. 227878-COLLMOT. A.Z. is very grateful to Elias Zafiris for his help during the preparation of the manuscript.


\bibliographystyle{plainnat}
\bibliography{refek}

\end{document}